\documentclass[11pt,letterpaper]{article}
\usepackage{fullpage}
\usepackage[top=0.5in, bottom=0.5in, left=0.5in, right=0.5in]{geometry}
\usepackage{amsmath,amsthm,amsfonts,amssymb,amscd}
\usepackage{enumerate}
\usepackage{graphicx}
\usepackage{listings}
\usepackage{float}
\allowdisplaybreaks
\usepackage{subcaption}

\usepackage{lipsum}
\usepackage{epsfig}
\usepackage{bm}
\usepackage{multicol}
\usepackage{url}
\usepackage[utf8]{inputenc}
\usepackage{mathtools}
\usepackage{extarrows}

\usepackage{color}
\definecolor{dkgreen}{rgb}{0,0.6,0}
\definecolor{gray}{rgb}{0.5,0.5,0.5}
\definecolor{mauve}{rgb}{0.58,0,0.82}

\usepackage{booktabs}
\usepackage{longtable}
\usepackage{array}
\usepackage{multirow}
\usepackage{wrapfig}
\usepackage{colortbl}
\usepackage{pdflscape}
\usepackage{tabu}
\usepackage{threeparttable}
\usepackage{threeparttablex}
\usepackage[normalem]{ulem}
\usepackage{makecell}
\usepackage{xcolor}
\usepackage{natbib}
\usepackage{placeins}
\usepackage{hyperref}
\usepackage{bbm}

\graphicspath{{figures/}}

\usepackage[utf8]{inputenc}

\title{
Unified calibration
and spatial mapping of fine particulate matter data from multiple low-cost air pollution sensor networks in Baltimore, Maryland}
\author{
  Claire Heffernan\thanks{Department of Biostatistics, Johns Hopkins University, $^1$\texttt{cmheff96@gmail.com}, $^5$\texttt{abhidatta@jhu.edu}} $^1$,
  Kirsten Koehler\thanks{Department of Environmental Health and Engineering, Johns Hopkins University, $^2$\texttt{kkoehle1@jhu.edu}} $^2$,
  Drew R. Gentner\thanks{Department of Chemical \& Environmental Engineering, Yale University, $^3$\texttt{drew.gentner@yale.edu}} $^3$,
  Roger D. Peng\thanks{Department of Statistics and Data Sciences, University of Texas, Austin, $^4$\texttt{roger.peng@austin.utexas.edu}} $^4$,\\
  Abhirup Datta\footnotemark[1] $^5$
}

\date{}

\newcommand{\given}{\;|\;}

\newcommand{\calW}{\mathcal{W}}

\newcommand{\bx}{\mathbf{x}}
\newcommand{\by}{\mathbf{y}}

\newcommand{\ba}{\mathbf{a}}
\newcommand{\bs}{\mathbf{s}}
\newcommand{\bS}{\mathbf{S}}
\newcommand{\boldm}{\mathbf{m}}
\newcommand{\bM}{\mathbf{M}}

\newcommand{\bX}{\mathbf{X}}

\newcommand{\bB}{\mathbf{B}}
\newcommand{\bC}{\mathbf{C}}

\newcommand{\bz}{\mathbf{z}}
\newcommand{\bD}{\mathbf{D}}
\newcommand{\bZ}{\mathbf{Z}}

\newcommand{\st}{(\bs,t)}

\newcommand{\sot}{(\bS_0,t)}
\newcommand{\slt}{(\bS_1,t)}

\newcommand{\sat}{(\bS_A,t)}
\newcommand{\sbt}{(\bS_B,t)}

\newcommand{\Sn}{\bS_{new}}

\newcommand{\blue}[1]{\leavevmode{\color{black}#1}}

\newcommand{\bbeta}{ {\boldsymbol \beta} }

\newcommand{\beps}{ {\boldsymbol \epsilon} }

\newcommand{\bmu}{ {\boldsymbol \mu} }

\usepackage{titlesec}

\titlespacing{\section}{0pt}{\parskip}{-\parskip}
\titlespacing{\subsection}{0pt}{\parskip}{-\parskip}

\usepackage{setspace}
\usepackage{xr}
\makeatletter
\newcommand*{\addFileDependency}[1]{
  \typeout{(#1)}
  \@addtofilelist{#1}
  \IfFileExists{#1}{}{\typeout{No file #1.}}
}
\makeatother
\newcommand*{\myexternaldocument}[1]{
    \externaldocument{#1}
    \addFileDependency{#1.tex}
    \addFileDependency{#1.aux}
}
\myexternaldocument{supplement}

\begin{document}
	\setlength{\abovedisplayskip}{0pt}
	\setlength{\belowdisplayskip}{4pt}
	\setlength{\abovedisplayshortskip}{0pt}
	\setlength{\belowdisplayshortskip}{0pt}
\clearpage\maketitle

\begin{abstract}
Low-cost air pollution sensor networks are increasingly being deployed globally, supplementing sparse regulatory monitoring with localized air quality data. In some areas, like Baltimore, Maryland, there are only few regulatory (reference) devices but multiple low-cost networks. There are many available methods to calibrate data from each network individually, \blue{including the recently proposed Gaussian process filter (GP filter) method, that mitigates underestimation issue of other calibration methods, models spatial correlation, and yields a dynamic and probabilistic (Bayesian) calibration equation.} However, separate calibration of each network \blue{using GP filter or any other calibration approach} leads to conflicting air quality predictions. \blue{In this manuscript, we extend the GP filter 
to jointly model} data from multiple \blue{low-cost} networks and reference devices. \blue{The approach provides} dynamic calibrations (informed by the latest reference data) and unified predictions (combining information from all available \blue{low-cost and reference} sensors) for the entire region. This method accounts for network-specific bias and noise, as different networks can use different types of sensors, and uses a Gaussian process to capture spatial correlations. We apply the method to calibrate PM$_{2.5}$ data from Baltimore in June and July 2023 -- a period including days of hazardous concentrations due to wildfire smoke. Our method helps mitigate the effects of preferential sampling of one \blue{low-cost sensor} network in Baltimore, resulting in better predictions and  \blue{more precise credible} intervals. 
Our approach can be used to calibrate low-cost air pollution sensor data in Baltimore and other areas with multiple low-cost networks.
\end{abstract}

\textbf{Keywords: } Gaussian process, Bayesian statistics, Kalman filtering, low-cost networks, air pollution, spatial statistics

\newpage

\section{Introduction}

Air pollution is a major concern worldwide. 
\blue{A recent study estimated that} 6.5 million deaths per year worldwide are caused by air pollution \citep{fuller2022pollution}. 
The U.S. Environmental Protection Agency (EPA) has established guidelines for concentrations of major pollutants and has recently lowered the annual standard for fine particulate matter (PM$_{2.5}$) from 12 $\mu g/m^3$ to 9 $\mu g/m^3$ \citep{NAAQS_new}. The World Health Organization (WHO) 
estimates that 99\% of people are exposed to higher concentrations of pollution than the guidelines \citep{WHO_airpollution}.


In Baltimore, there is evidence that air pollution is not evenly distributed across the city \citep{boone2014long}. Therefore, it is important to understand the air quality throughout the entire city at a fine spatial scale. 
Devices that meet the EPA's standards, the Federal Reference Method (FRM) or Federal Equivalent Method (FEM) are the gold standard of air quality measurement. We call these devices \textit{reference devices}. 
States deploy networks of reference devices to record these high-quality measurements. However, these networks are often sparse; for example, Maryland only has 26 devices in the state, and only one device in Baltimore \blue{measures PM$_{2.5}$ on an hourly scale.} \citep{MDE_2023}. Thus, reference devices alone are not enough to give spatial information about air quality in Baltimore. 

\textit{Low-cost sensors} (LCS) are a solution to the sparsity of reference devices. These sensors usually cost a few hundred dollars, considerably less than reference devices, and can be installed in many more locations within a city. Such networks exist in many areas, including many major  cities in the United States \citep{kim2018berkeley,desouza2022calibrating,esie2022neighborhood}. The PurpleAir network is one example of a low-cost air pollution network. It is a community science network where individuals can purchase a sensor to place outside their homes and record air quality. PurpleAir data is publicly available for those who agree to share their data and sensors have been installed in many areas of the United States. 

Low-cost sensors suffer from having bias and noise in the measurements, and thus they must be calibrated before being used. 
One approach to calibration is field-calibration, 
where sensors are deployed and calibration equations are created based on the real-world performance of the sensors. To fit the calibration equations, some high-quality air pollution measurements are needed. Thus, some low-cost sensors are placed in the same location as reference devices. These locations, with both a reference and low-cost sensor, are called \textit{collocated sites}. The the measurements at these sites are used to estimate calibration equations 
based on various types of regression models with the reference measurement as the outcome and the low-cost measurement and other meteorological variables as predictors \citep{barkjohn2021development,datta2020statistical,Patton2022,levy2022evaluating,levy2023identifying,bigi2018performance,bi2020incorporating,ardon2020measurements,romero2020development}.

Recently, \cite{heffernan2023dynamic} showed that regression-based field-calibration equations systematically underestimate high concentrations. Additionally, many of these approaches are not spatially informed, \blue{i.e., do not leverage correlation between air pollution concentrations at nearby locations. Finally, calibration equations are often not dynamic, as they are based on one time (or periodic) training, and are not informed by the latest available reference data in the area.} To mitigate these issues, \citep{heffernan2023dynamic} proposed a spatial filtering model called the Gaussian Process Filter (GP filter), reviewed in Section \ref{sec:gpf}. The observation model part of GP filter switches the roles of outcome and predictor variables from standard regression calibration, and  models the low-cost measurements as a noisy and biased version of the true concentrations, as measured by the reference devices. The state-space model accounts for spatial structure in true pollutant levels and is specified as a conditional Gaussian process regression for the true concentrations given the reference data at a 
handful of locations. 
The GP filter thus uses all the low-cost data available and the contemporary reference data to make dynamic, spatially informed predictions. 
The approach was used to calibrate PM$_{2.5}$ data from $\sim 45$ sensors of the SEARCH (Solutions to Energy, Air, Climate, and Health Center) low-cost network in Baltimore. 

Besides the SEARCH network, the PurpleAir network also has considerable presence in Baltimore, with  roughly 30 PM$_{2.5}$ sensors in the area. \blue{This data was not utilized in \cite{heffernan2023dynamic}.} Utilizing this additional data can enrich our knowledge of intra-urban variations of PM$_{2.5}$ in Baltimore. 
However, we will illustrate that there is evidence of preferential sampling in the PurpleAir network, which would occur when sensors are more concentrated in areas with either higher or lower concentrations. Studies \citep{shaddick2014case,lee2015impact} have shown that preferential sampling can lead to biased estimates of city-level or regional air quality. 
Thus, the PurpleAir network does not give an accurate assessment of air quality as a whole, and this data alone may not be the best approach for estimating air quality in Baltimore. 
Preferential sampling is less of a concern for the SEARCH network since sites were selected by using a weighted random sample to select a representative set of locations. However, the SEARCH network is more spread out, leaving large gaps in the city where predictions are imprecise. 

Leveraging data from both the SEARCH and the PurpleAir networks in Baltimore has the potential to improve predictions over using either network individually. 
However, multiple low-cost networks in an area can have different biases and noise levels, since different brands of sensors from different manufacturers are made differently. For example, the PurpleAir network uses Plantower PMS5003 sensors, while the SEARCH network uses Plantower A003 sensors, two different types of devices from the same manufacturer whose inlets are designed differently. Therefore, the measurements made by the different networks are not directly comparable, and they should be treated as separate networks for the purpose of calibration, with the biases and noises of each network addressed separately. However, most of the existing calibration methods for LCS networks, \blue{including the GP filter method of \cite{heffernan2023dynamic},} do not address the setting of multiple low-cost networks within an area. A simple approach would be to calibrate each network separately. However, na\"ive network-specific calibration followed by spatial interpolation would lead to different predicted air quality maps for the region using each network, with potentially conflicting predictions of air-quality from each network. Therefore some ad-hoc way to merge the two or more sets of predictions to create unified maps would be required. It is more desirable to develop new methodology for coherent fusion of air pollution data coming from multiple networks. 


\blue{Our contributions in this manuscript are as follows. }
We develop a principled \blue{extension of the GP filter method to combine and jointly model data from multiple LCS networks with overlapping geographical coverage. Methodologically, the extension is straightforward. Leveraging the Bayesian formulation of the filtering, we include multiple observation models, one corresponding to LCS data from each new network, modeled as a biased and noisy version of the true pollutant concentrations.  The second part of the model remains unchanged --- a  Gaussian process state-space model for the true pollutant surface accounts for the spatial correlation in the concentrations, informed by the available reference data at one or few locations. Like the single network GP filter, calibrations obtained from our method will be dynamic, i.e., informed by the latest reference data in the region. 

However, this simple methodological extension considerably broadens the scope of the GP filter approach. 
It enables leveraging all available data on the pollutant concentrations, from} 
multiple LCS networks as well as the 
from the sparse reference network, sharing information across all the networks and locations. 
\blue{It accounts for} \blue{ network-specific biases and noises.} 
 Our method offers three main advantages over calibrating each network individually. 
 \begin{enumerate}
 \item It offers a principled approach to obtain unified set of spatial predictions of air quality at any location in the region that is informed by all available data (LCS networks as well as any available reference data), \item It improves prediction accuracy, especially when some networks have preferential sampling. 
 \item It generally reduces uncertainty around the predictions by using data from a denser set of locations by combining multiple networks. 
 \end{enumerate}
 \blue{Together, these advantages make our approach robust and comprehensive for regional air quality assessment. The rest of the manuscript is organized as follows. In Section \ref{sec:bmoredata}, we introduce the low-cost and reference networks for fine particulate matter (PM$_{2.5}$) in Baltimore, which motivates our methods development. In Section \ref{sec:met} we review the single-network GP filter (Section \ref{sec:gpf}), and present the extension to use multiple LCS networks (Section \ref{sec:mgpf}),  
 alongwith parameter estimation strategies (Section \ref{sec:est}), details about the advantages of the multi-network approach (Section \ref{sec:goals}), and an extension to model heteroscedasticity, motivated by our data application 
 (Section \ref{sec:het_obs_model}). Section \ref{sec:gensim_main} presents performance assesment of our method using simulation studies. 
Section \ref{sec:baltimore} applies the method to calibrate and map PM$_{2.5}$ in Baltimore using data from two LCS networks and one reference device.} 

\section{Baltimore low-cost sensor air quality data networks}\label{sec:bmoredata}

\begin{figure}[h]
\centering
\includegraphics[width=4.5in,trim={0 0.5in 0 0.5in},clip]{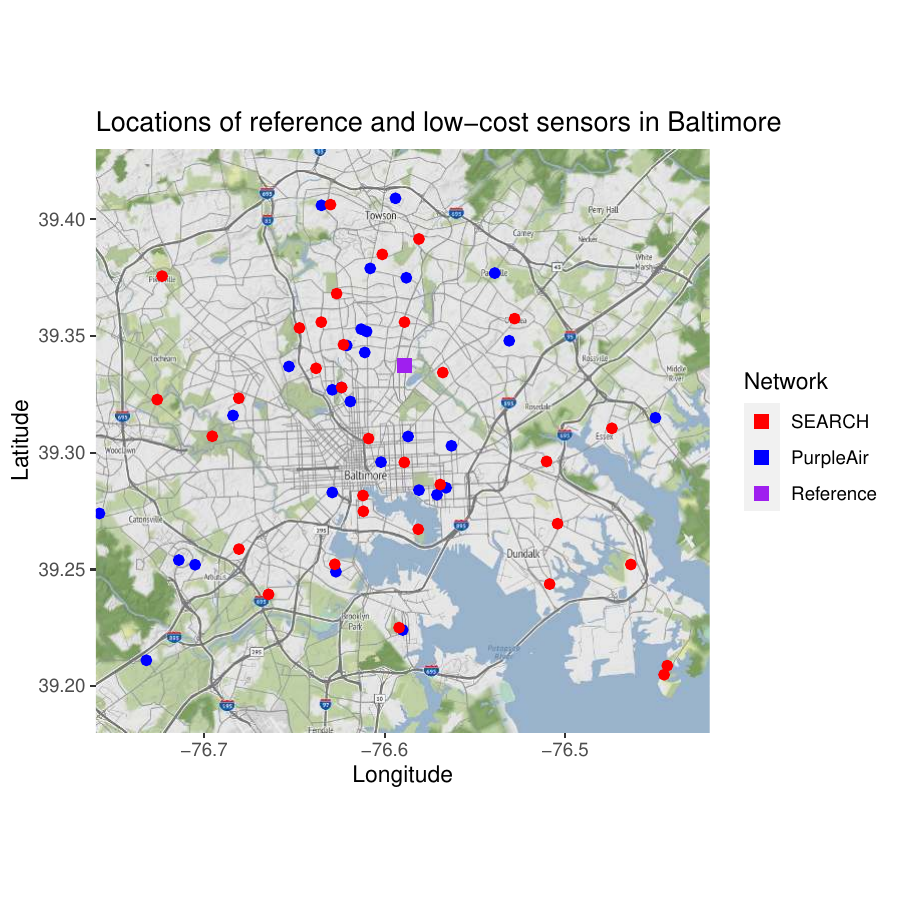}
\caption{Low-cost sensor networks in Baltimore in 2023. The red sites are the SEARCH low-cost network. The blue sites are the PurpleAir low-cost network. The purple site is the reference device at Lake Montebello, which also has two SEARCH sensors at the same location. }\label{fig:locs} 
\end{figure}

In Baltimore, Maryland, the PurpleAir network \citep{barkjohn2021development} and the Solutions to Energy, Air, Climate, and Health (SEARCH) Center \citep{levy2018field,heffernan2023dynamic} are low-cost networks that measure PM$_{2.5}$, the mass concentration of fine particulate matter in the air. Additionally, there is one reference device with high quality measurement of PM$_{2.5}$ 
at Lake Montebello in 2022. 
A map of the networks is shown in Figure \ref{fig:locs}. Note that the networks have different geographic distributions, with the PurpleAir network being concentrated closer to the center and north of the city and having no sites in the southeast corner of the city. 
This indicates that there may be stronger preferential sampling in the PurpleAir network. Section \ref{sec:preferential} will explore this possibility further. 

We wish to combine the information from the two LCS networks along with the data from the single reference device to make predictions of the spatial distribution of PM$_{2.5}$ in Baltimore. Since the two networks use different types of sensors, they may have different biases, and thus they cannot be treated as one larger network. Given the severe lack of spatially-resolved PM$_{2.5}$ data in Baltimore, we seek a model that 
offers unified predictions of PM$_{2.5}$ combining data from these three sources. Contingent upon adequate model validation under different conditions, the model can then be used to make spatially-detailed map of air quality in Baltimore as new data is collected, which in turn 
can be used in downstream health association studies or policy evaluations to assess the burden of air quality in Baltimore.


For this study, we consider the period of June and July 2023, an unusual time for air quality in Baltimore because there were several days of extremely high and unprecedented PM$_{2.5}$ concentrations due to wildfire smoke, interspersed in days of typical concentrations. \cite{heffernan2023dynamic} showed that the single-network spatial filtering method theoretically performs well at high concentrations, but there were only very limited days of poor air quality in the study period of \cite{heffernan2023dynamic}. 
It is important to check the performance of any low-cost sensor calibration method at such high concentrations, especially to assess impact of potential model misspecification as the training window for these models will not often include high concentrations. The chosen study period thus provides the unique opportunity to validate our proposed approach  across a wider range of concentrations. 

\section{Methods}\label{sec:met}

\subsection{Overview of the single network GP filter}\label{sec:gpf}

\begin{figure}[t]
\centering
\includegraphics[width=6in]{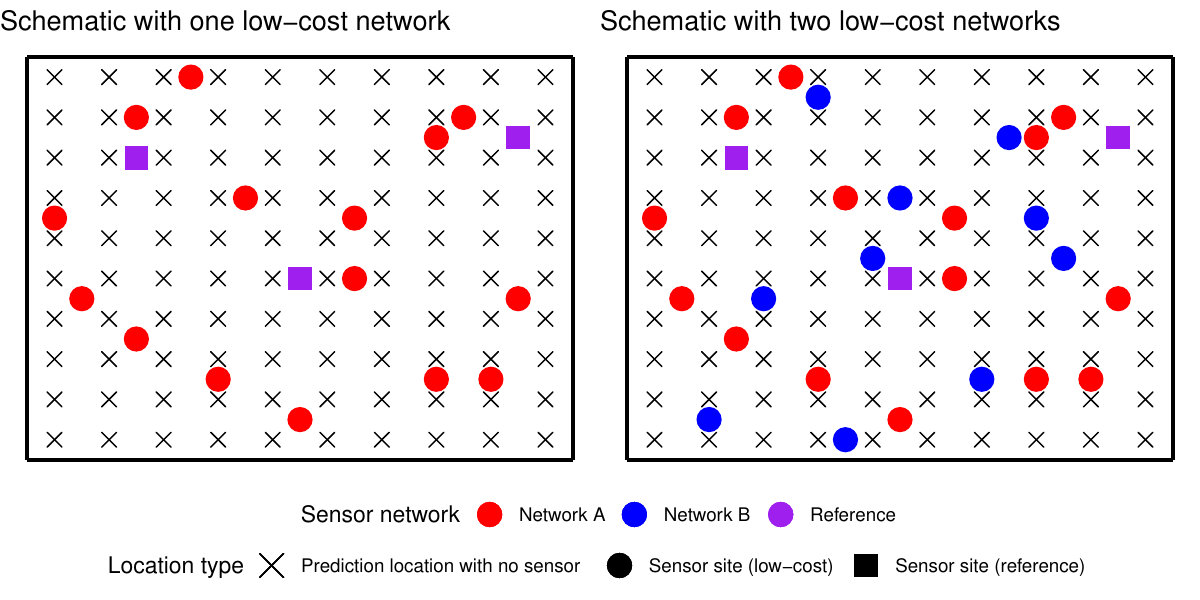}
\caption{Illustrations of multiple networks operating in a same region: with (left) one or (right) two low-cost network(s) and 3 reference sites. }\label{fig:schematic} 
\end{figure}

We first briefly review the single network filtering approach, GP filter, of 
\cite{heffernan2023dynamic}. 
This method considers a low-cost sensor (LCS) network and a sparse 
reference network (typically with one or very few sites). 
A schematic of such a low-cost network is shown in Figure \ref{fig:schematic} (left), with the purple sites being reference sites and the red sites being the low-cost network sites.

\cite{heffernan2023dynamic} proposed the following two-stage model to combine data from the LCS network $y$ and the reference devices $x$ to predict the true concentrations across the region: 
\begin{align}\label{eq:gpf}
\begin{split}
    \text{Observation model: }&y\st=\beta_0+\beta_1x\st+\bbeta_2\bz\st+\bbeta_3x\st\bz\st+\epsilon\st\\
    \text{State space model: }&(\bx\sot,\bx\slt)\sim N(\blue{\bmu_t},\bC_t)
\end{split}
\end{align}
where $x\st$ denotes the true pollutant concentrations as would be measured by a reference device at location $\bs$ and time $t$, $y\st$ denotes the measurement from the low-cost sensor, and $\bz\st$ denotes other covariates (detailed later) that influences the bias of the low-cost data. The set of locations $\bS_0$ are the locations with reference sites, the set of locations $\bS_1$ denotes locations with low-cost devices but no reference device, $\bx(\bS_0,t)$ denotes $x(s,t)$ for $s \in \bS_0$ and other quantities are similarly defined. The state space model accounts for the spatial correlation in the true air pollution levels by assuming that the true air pollution surface $\bx_t=\{x\st,\,\forall \bs\}$ follows a Gaussian process $\bx_t\sim GP(\mu_t,C_t)$, with 
time-specific mean \blue{function $\mu_t(s) = E(x(s,t))$ and covariance function $C_t(s,s')=Cov(x(s,t),x(s',t))$.} At a finite set of locations $\bS= \bS_0 \cup \bS_1$, the Gaussian Process is a multivariate normal distribution, as shown in the second equation of Equation (\ref{eq:gpf}), \blue{where the mean $\bmu_t = (\mu(s,t))_{s \in \bS}$ is the vector stacking up $\mu(s,t)$ for $s \in \bS$, and the matrix $\bC_t = \Big(C_t(s,s')\Big)_{s,s' \in \bS}$ is the covariance between sites,} which depends on the site locations and on covariance parameters. The observation model is a linear regression equation with errors $\epsilon\st$. \cite{heffernan2023dynamic} modelled the errors as i.i.d with a Gaussian distribution. Note that at the reference sites (purple sites), we say that the true concentration $x$ is observed, since the reference measurement is the gold standard for measurement, while at the low-cost sites, only $y$ is observed and we wish to predict $x$. The observation model is a natural way to model data from low-cost sensors because the low-cost measurement is modelled as the noisy version of the true concentration $x$, \blue{which reflects the reality.} The state space model assumes that air pollution concentrations form an underlying smooth spatial surface, which is another realistic modelling choice. 

The GP filter presents several advantages over typical field-calibration approaches. First, the use of the observation model with $x$ as the independent variable means that $x$ is not systemically underestimated when it is large, which is a limitation of many regression-based calibration techniques that typically regress the true concentrations $x$ on the low-cost data $y$ \citep{heffernan2023dynamic}. Additionally, GP filter incorporates spatial information, both from other network sites and from available reference devices, into prediction, while common calibration approaches only use a particular site's measurement for calibration. 

The spatial filtering has two modules: i) training of the observation model, and ii) simultaneous training of the state-space model and spatial filtering. The parameters of the observation model, both the regression coefficients and the error model variance, are typically estimated before filtering is performed -- as is common in Kalman-filtering or other filtering applications. For this, low-cost sensors are collocated with one or more of the reference sites permanently or for a long period of time. Then, the observation model can be trained on this abundant data, yielding highly precise estimates of the parameters which are subsequently held fixed at these estimated values for the filtering module. 
During the filtering part, the parameters of the state space model and the latent true concentrations are jointly estimated using the hierarchical model (\ref{eq:gpf}) in a Bayesian implementation. Priors are added to \blue{the parameters specifying the mean and covariance functions of the GP,} 
and the parameters and $\bx\slt$ are estimated simultaneously using Markov Chain Monte Carlo (MCMC). 

\subsection{Multinetwork Gaussian process filter \blue{model}}\label{sec:mgpf}

A schematic of a region with multiple low-cost sensor networks, such as Baltimore, is shown in Figure \ref{fig:schematic} (right). The first low-cost network (in red) is still present, but the second low-cost network (in blue) has now been added. The Bayesian filtering formulation of the GP filter lends itself naturally to multiple low-cost networks in a region, each with a different observation model, tied to the same state-space model for the true concentrations. Thus, data from multiple networks can be used together to make unified predictions both at the network sites and across the entire region. We now present this extension of the GP filter, which we call the {\em Multi-network Gaussian Process Filter (MGPF)}. 

Let $A$ and $B$ denote the two LCS networks, $\bS_A$ and $\bS_B$ denote the respective set of LCS locations for the networks, and $\bS_0$ denote the reference network (which can even be only a single site, as in Baltimore).  
We extend the model in (\ref{eq:gpf}) naturally to include two networks. Continuing to use the Gaussian Process $\bx_t\sim GP({\mu_t},C_t)$ as the underlying air pollution distribution, we can write the model \blue{for the data} as follows:
\begin{align}\label{eq:mgpf}
\begin{split}
    & \text{Observation model A: }\by\sat=\\
    & \qquad \beta_{A,0}+\beta_{A,1}\bx\sat+\bbeta_{A,2}\bz\sat+\bbeta_{A,3}\bx\sat\bz\sat+\epsilon_A\sat\\
    & \text{Observation model B: }\by\sbt=\\
    & \qquad \beta_{B,0}+\beta_{B,1}\bx\sbt+\bbeta_{B,2}\bz\sbt+\bbeta_{B,3}\bx\sbt\bz\sbt+\epsilon_B\sbt\\
    & \text{State space model: }(\bx\sot,\bx\sat,\bx\sbt)\sim N(\blue{\bmu_t},\bC_t), \\
    & \qquad \blue{\mbox{ where } \bmu_t =(\mu(s,t))_{s \in \bS} \mbox{ and } \bC_t = \Big(C_t(s,s')\Big)_{s,s' \in \bS}, \mbox{ with } \bS=\bS_0 \cup \bS_1 \cup \bS_2, }
\end{split}
\end{align}
and $\epsilon_A$ and $\epsilon_B$ are independent errors. Rather than assuming i.i.d. errors as was done in \cite{heffernan2023dynamic}, we allow for heteroscedastic errors (details in Section \ref{sec:het_obs_model}).  \blue{Our approach and the single network GP filter of \cite{heffernan2023dynamic} are flexible in terms of choice of the mean and covariance parameters. The means $\blue{\bmu_t}$ can be modeled as a regression on land-use variables. Here, we model them as time-specific constants, i.e., $\bmu_t = \mu_t \mathbbm{1}$. This is due to there being limited number of total locations with a measurement to meaningfully learn the parameters of land-use regression. Similarly,} our methodology can be used with any valid Gaussian process covariance function $C_t$. In our application in Baltimore, the exponential covariance function will be used. 

\begin{figure}[t]
\centering
\includegraphics[width=4.5in]{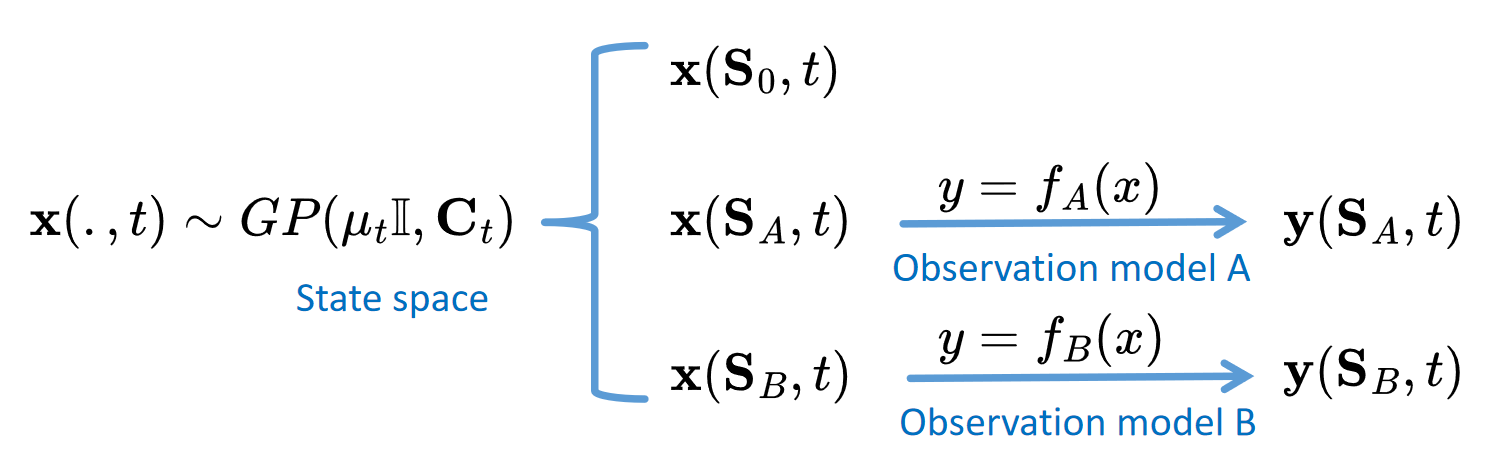}
\caption{Schematic of the MGPF model}\label{fig:model}
\end{figure}

\blue{A schematic of this model is shown in Figure \ref{fig:model}.}
We note that while we primarily focus on two LCS networks as is the setting for Baltimore, this model can be written more generally to include $K$ networks as follows. 
\begin{align}\label{eq:mgpf_K}
\begin{split}
    K \text{ observation models: }\by(\bS_k,t)&= \beta_{k,0}+\beta_{k,1}\bx(\bS_k,t)+\bbeta_{k,2}\bz(\bS_k,t) \\
    & \quad +\bbeta_{k,3}\bx(\bS_k,t)\bz(\bS_k,t)+\epsilon_k(\bS_k,t)
    \quad \text{for }k\in\{1,\dots K\}\\
    \text{State space model: } & (\bx\sot,\bx(\bS_1,t),\dots\bx(\bS_K,t))\sim GP(\blue{\bmu_t},\bC_t)
\end{split}
\end{align}

\noindent \blue{{\em Misaligned or missing data:} The MGPF allows each network to have its own set of sites $\bS_k$. Although we have denoted each network as a static set of locations $\bS_k$, in practice the locations with low-cost data can vary between time-points, due to factors such as new sensors being added (misalignment over time) or a mechanical issue with a sensor for a period of time (resulting in missing data). 
The MGPF allows for each time-point to have a different number of low-cost sites active per network with no issues. Formally, we can assume at time $t$, the set of sites offering data for network $k$ to be $\bS_{k,t}$. Missingness of low-cost data is usually unrelated to the concentration levels of the pollutant and typically related to network connectivity or other external issues. 
Hence, the missingness pattern is not expected to be informative and need not be modeled. 
All the estimation strategies outlined in the next Section remains valid for low-cost data with time- and network-specific missingness, by simply modeling the low-cost data at the set of active locations $\bs_{k,t}$ in the filtering part. Similarly, the set of locations with reference data can also vary with time. 

In other applications, if missingness is informative of the underlying true concentration value, we can, e.g., replace the observation model for a missing sensor-time-point pair can be replaced with a logistic model for missingness of the low-cost observation based on the unobserved $PM_{2.5}$). Thus our Baysian hierarchical formulation of the filtreting problem can easily model missingness mechanism via another hierarchy. Information about the distribution of missingness in our data application is in Supplement \ref{sec:supp_missing}.}

\subsection{\blue{Parameter estimation and prediction}}\label{sec:est}

\blue{We now discuss different strategies for parameter estimation and prediction of the true pollutant surface in the MGPF model.} In Equation (\ref{eq:mgpf_K}), the only known quantities are the LCS data $\by(\bS_1,t)$, $\by(\bS_2,t), \ldots  \by(\bS_K,t)$ at the respective network sites  and the true concentrations $\bx(\bS_0,t)$ at a handful of reference sites $\bS_0$. The goal is to estimate the true concentrations $\bx(\bs,t)$ at each LCS location $\bs \in \bS_1 \cup \bS_2 \cup  \ldots \cup \bS_K$ \blue{as well as predict $\bx(\bs,t)$ at any location $\bs$ without a sensor.}

\noindent\blue{{\em Pre-estimation of observation model parameters:}} 
\blue{The observation model is only informed by data at the collocated sites which have both a low-cost sensor and a reference site. Let $\bs_k$ be the site of collocation for network $k$, $(x(\bs_k,t),y(\bs_k,t))$ denote the collocated timeseries for that network at $\bs_k$ for a timewindow $t \in \calW$. Let $\by(s_k,\calW)$ be the vector stacking up the low-cost observations $y(\bs_k,t)$ for $t \in \calW$, and define $\bx(s_k,\calW)$ for the true concentrations, $\bz(s_k,\calW)$ for the covariates, and $\beps(s_k,\calW)$ for the errors similarly. Then the observation model for the network for this location and timewindow can be expressed as 
\begin{align}\label{eq:obstrain}
    \by(\bs_k,\calW) &= \bX(\bs_k,\calW)\bbeta_k + \beps(\bs_k,\calW), \epsilon(s_k,t) \sim N(0,\sigma^2_{k,t}) \nonumber \\
    \mbox{ where } \bX(\bs_k,\calW) &=(\mathbbm{1}, \bx(\bs_k,\calW),\bz(\bs_k,\calW),\bx(\bs_k,\calW) \odot \bz(\bs_k,\calW)) \mbox{, and }\\ 
    \bbeta_k &= (\beta_{k,0}, \beta_{k,1}, \beta_{k,2}, \beta_{k,3})'. \nonumber
\end{align}
Here $\odot$ is the Hadamard (elementwise) product used to write the interaction terms in vector form. The parameters $\bbeta_k$ and $\sigma^2_{k,t}$ 
can then simply be estimated using least squares. 

While the observation model can be trained jointly with the filtering part in a fully Bayesian setup, \cite{heffernan2023dynamic} recommended pre-estimating
the parameters of each observation model. This is because after an adequate period of collocation (as is the case for the SEARCH network in Baltimore), there is usually enough data available to estimate these parameters in (\ref{eq:obstrain})  very high precision. 
Another scenario is that there maybe no collocated sites for a LCS network in the area of study (as in the case of the PurpleAir network in Baltimore). However, the estimates of the observation model parameters already available based on prior studies using collocated or near-collocated data from another region, e.g., the US-wide calibration equation for PurpleAir PM$_{2.5}$ sensors derived  \citep{barkjohn2021development}. In either case, it is reasonable to plug in the parameter estimates from the pretrained observation model into the subsequent filtering part described below.}  

\noindent\blue{{\em Spatial parameter estimation and filtering:}} \blue{Given the observation model parameter estimates $\widehat\bbeta_k$ and $\widehat \sigma^2_{k,t}$, we estimate 
the true concentrations $\bx(\bS_1,t),\bx(\bS_2,t),\ldots,\bx(\bS_K,t)$ 
jointly with the mean $\blue{\mu_t}$ and the parameters $\theta_t$ of the covariance function $C_t :=C(\cdot,\cdot \given \theta_t)$  in the Bayesian filtering model. To explain the algorithm, we first define the following quantities. Let $\bx(\bS^*,t)$ denote the vector created by stacking up $\bx(\bS_k,t)$ for $k=1,\ldots,K$. At a time point $t$, where we wish to estimate $\bx(\bS^*,t)$, we observe the low-cost data $\by(\bS^*,t)$ (defined similarly as $\bx(\bS^*,t)$) and the reference data $\bx(\bS_0,t)$ at the reference site(s) $\bS_0$. 
Let $\ba(\bS^*,t)$ denote the $|\bS^*| \times 1$ vector created by stacking the members of the set $\{\hat\beta_{k,0} + z(s_{k},t)\hat\beta_{k,2}: k=1,\ldots,K, s_k \in \bS_k\}$. Similarly, define $\bB(\bS^*,t)$ denote the diagonal matrix created from members of the set $\{\hat\beta_{k,1} + z(s_{k},t)\hat\beta_{k,3}: k=1,\ldots,K, s_k \in \bS_k\}$. 
Finally, the measurement error covariance is represented by the block diagonal matrix \(\mathbf{D}(\bS^*,t) = \mathrm{blockdiag}\!\left(\sigma_{1,t}^2 I_{|\bS_1|}, \dots, \sigma_{K,t}^2 I_{|\bS_K|}\right)\).

We can now rewrite (\ref{eq:mgpf_K}) as
\begin{equation}\label{eq:mgpf_mat}
\begin{aligned}
    \by(\bS^*,t) &= N \left(\ba(\bS^*,t) + 
    \bB(\bS^*,t)
    \bx(\bS^*,t), \bD(\bS^*,t)\right) \\
    \bx(\bS_0,t) &\sim N(\mu_t \mathbbm{1},C(\bS_0,\bS_0 \given \theta_t)), \\
    \bx(\bS^*,t) \given \bx(\bS_0,t) & \sim N\left(\bmu_{t,\bS^* \given \bS_0} ,  \bC_{t,\bS^* \given \bS_0}\right), \mbox { where } \\
    \bmu_{t,\bS^* \given \bS_0} & = \mu_t \mathbbm{1} + C(S,\bS_0 \given \theta_t)C(\bS_0,\bS_0 \given \theta_t)^{-1}(\bx(\bS_0,t) - \mu_t \mathbbm{1}), \mbox{ and } \\
    \bC_{t,\bS^* \given \bS_0} & = \bC(\bS^*,\bS^* \given \theta_t) - \bC(\bS^*,\bS_0 \given \theta_t)\bC(\bS_0,\bS_0 \given \theta_t)^{-1}\bC(\bS_0,\bS^* \given \theta_t). 
\end{aligned}
\end{equation}

In (\ref{eq:mgpf_mat}), the only unknowns are $\bx(S,t)$, $\mu_t$, and $\theta_t$ as all the other quantities are functions of observed data $\by(S,t)$, $\bZ(S,t)$, $\bx(\bS_0,t)$ or pre-estimated parameter $\widehat \bbeta_k$, $k=1,\ldots, K$. For a fixed value of $\mu_t$ and $\theta_t$, the posterior of $\bx(S,t)$ is available as a closed form as
\begin{equation}\label{eq:kalman update}
\begin{aligned}
    \bx(S,t) \given \cdot & \sim N(\bM_t^{-1} \boldm_t , \bM_t^{-1}) \mbox{ where }\\
    \boldm_t & = 
    \bB(\bS^*,t)
    '\bD(\bS^*,t)^{-1}(\by(\bS^*,t) - \ba(\bS^*,t) 
    + \bC_{t, S \given \bS_0}^{-1}\bmu_{t, S \given \bS_0},\\
    \bM_t^{-1} & = 
    \bB(\bS^*,t) 
    '\bD(\bS^*,t)^{-1}
    \bB(\bS^*,t) 
    + \bC_{t, S \given \bS_0}^{-1}. 
\end{aligned}
\end{equation}

This step is essentially the {\em Kalman update} part of our filtering setup, showing how the state-space model and observation model both inform the posterior mean and variance. For the spatial parameters $\mu_t$ and $\theta_t$, we add priors and obtain posteriors for all the unknowns using Markov chain Monte Carlo.
}

\noindent\blue{{\em Prediction:}} Since the Gaussian Process models a smooth surface of true air pollution concentrations over the entire region, we can also predict concentrations at locations that do not have any low-cost sensor (the $\times$ symbols in Figure \ref{fig:schematic}). We refer to these locations as $\Sn$ and note that they can be added into the GP part of the model to give 
\begin{align}\label{eq:interpolate}
    (\bx\sot,\bx(\bS_1,t),\dots\bx(\bS_K,t),\bx(\Sn,t))\sim GP(\blue{\bmu_t},\bC_t)
\end{align}
\blue{where $\bmu_t$ and $\bC_t$ now gives the mean vector and covariance for the GP over the set of locations $\bS_0 \cup \bS_1 \cup \ldots \cup \bS_K \cup \bS_{new}$.}

Thus, in addition to estimating the true concentrations at all the network sites, we can make predictions at any new location 
using the posterior distribution of the unknown true concentrations given the reference and low-cost data:
\begin{align*}
&p\left(\bx(\bS_1,t),\dots\bx(\bS_K,t),\bx(\Sn,t) | \bx\sot, \by(\bS_1,t),\dots\by(\bS_K,t) \right) \\
&= p\left(\bx(\Sn,t) | \bx\sot,\bx(\bS_1,t),\dots\bx(\bS_K,t) \right) \\
&\phantom{=}\times p\left(\bx(\bS_1,t),\dots\bx(\bS_K,t) | \bx\sot, \by(\bS_1,t),\dots\by(\bS_K,t) \right) 
\end{align*}
where the first term is a conditional normal distribution from a Gaussian process (i.e., the kriging predictive distribution), and the second term is the posterior distribution from the MGPF. So predictions can be obtained seamlessly after running the MCMC. These predictions, over a dense grid of locations, are used to create maps to inform spatial variability of the pollutant concentrations in the region.

\noindent\blue{{\em Marginalized estimation approach:}} \blue{The MCMC-based implementation of the MGPF algorithm described above uses the model formulation in (\ref{eq:mgpf_mat}), and hence samples the $|\bS^*|=\sum_{k=1}^K |\bS_k|$ dimensional vector $\bx(\bS^*,t)$ within each iteration. We also implemented an alternative approach, which drastically reduces the dimension of the MCMC state-space. We integrate out $\bx(\bS^*,t)$ from (\ref{eq:mgpf_mat}), and write the marginal distribution of the observed data $(\by(\bS^*,t),\bx(\bS_0,t))$ as

\begin{equation}\label{eq:mgpf_marg}
\begin{aligned}
\begin{pmatrix}
\mathbf{y}(\bS^*,t)\\[2pt]
\mathbf{x}(\bS_{0},t)
\end{pmatrix}
& \sim
N\!\left(
\begin{pmatrix}
\mathbf{a}(\bS^*,t)+\mathbf{B}(\bS^*,t)\,\mu_t \mathbbm{1}\\[2pt]
\mu_t \mathbbm{1}
\end{pmatrix}\right., \\
& \quad \qquad
\left.\begin{pmatrix}
\mathbf{B}(\bS^*,t)\,\mathbf{C}(\bS^*,\bS^*\mid \theta_t)\,\mathbf{B}(\bS^*,t)'+\mathbf{D}(\bS^*,t) & \mathbf{B}(\bS^*,t)\,\mathbf{C}(\bS^*,\bS^*_{0}\mid \theta_t)\\[2pt]
\mathbf{C}(\bS_0,\bS^*\mid \theta_t)\,\mathbf{B}(\bS^*,t)' & \mathbf{C}(\bS_{0},\bS_{0}\mid \theta_t)
\end{pmatrix}
\right).
\end{aligned}
\end{equation}


The only unknowns are now the GP parameters $\mu_t$ (scalar) and $\theta_t$ (typically 2-3 dimensional vector), drastically reducing the MCMC dimension. Subsequently, to obtain posteriors of $\mu_t$ and $\theta_t$, for each sample of these parameters, we can draw from the conditional posterior of  $\bx(\bS^*,t)$ using the Kalman update (\ref{eq:kalman update}). The resulting draws constitute samples from the marginal posterior of $\bx(\bS^*,t)$ given the observed data. Predictions at locations without sensors can then be done the same way as before. }

\subsection{Methodological benefits of the MGPF}\label{sec:goals}

\blue{Methodologically and in terms of implementation, our proposed method, MGPF is a relatively straightforward extension of the single-network GP filter. However, this extension offers} numerous methodological benefits over existing calibration methods. \blue{We first discuss how it improves over the single-network GP filter.} 

\noindent \textit{Unified predictions and maps:} 
MGPF offers a variety of qualities tailored to multiple low-cost sensor networks, that single-network calibration approaches applied individually to each network would not possess.  MGPF creates unified predictions of air quality across a region with multiple measurement sources available.  
Individual calibration 
of each network by any method will result in different predictions with different uncertainties at each new location. Rather than having to choose an ad-hoc rule to combine predictions from different networks into a final prediction, 
 MGPF will provide a single prediction with associated uncertainty for any location. 
Additionally, all the available data in the region (low-cost and reference) is used to inform predictions at all points, rather than only using a single network's data to make the predictions at those network sites. 

\noindent \textit{Robustness to preferential sampling:} MGPF improves the accuracy of the predictions and robustness to preferential sampling, compared to the predictions using just 1 network. Using data from multiple networks can help better estimate the parameters of the state-space model for the true concentration (GP parameters $\blue{\bmu_t}$ and parameters of $C_t$), since low-cost measurements are made at more total locations. Additionally, as air pollution is spatially correlated, 
it is desirable to model all correlations -- between locations of the same network, between two LCS networks, and between LCS networks and the reference data, as is done in our proposed MGPF. 
Predictions at locations that are somewhat isolated from sensors of one network but have proximal sensors from another network benefit from such joint modeling of the correlation of measurements across all networks.
Accuracy may particularly be improved if there is preferential sampling in a network and predictions are being made at new sites $\Sn$. 
For example, a network with no sensors in the most polluted areas would not measure the highest concentrations in an area, and thus predictions across that area would likely be lower than the truth, (as we will see for the PurpleAir network in Section \ref{sec:preferential}). This is mitigated when one of the networks (in our case, the SEARCH network) do not suffer from preferential sampling.  
However, if all networks are preferentially sampled in an area, 
then even when all sites are combined, there can be bias  in predictions  from any method at locations that do not have any sensor but have particularly high/low concentrations. 

\noindent \textit{Improved uncertainty quantification:} By synthesizing information from the denser combined set of locations, MGPF helps improve the uncertainty around predictions. 
However, although on average, uncertainty decreases by using multiple networks, it is possible for two networks to measure conflicting concentrations at nearby sites. In this case, it is desirable for the multi-network filter to have larger uncertainty, since there is large variation in air quality in a small area. A single network in this case may be overconfident with its predictions. 

As we will illustrate, these benefits of MGPF \blue{over the single-network GP filter} are manifested both at network sites and at locations in the region with no low-cost sensor. 

\noindent \textit{Dynamic calibration \blue{and mitigation of underestimation}:} 
\blue{MGPF has also other notable advantages relative to other calibration methods.} MGPF is a dynamic calibration approach. Both the calibration at the low-cost sites to estimate $x(\bS_1,t),\ldots,x(\bS_K,t)$ and predictions at a new location $x(\bS_{new},t)$ are informed by the concurrent available latest data $x(\bS_0,t)$, in addition to the low-cost data. The calibration is thus dynamic, informed by the latest reference data in the region, This is similar to the single network GP filter of \cite{heffernan2023dynamic} but differs from the regression-based field calibration approaches which only use the reference data during the training of the regression coefficients, and do not use concurrent reference data when calibrating at a future time point. \blue{Additionally, regression-based field calibration approaches tend to underestimate air pollutions peaks, as demonstrated empirically and proved theoretically in \cite{heffernan2023dynamic}. Filtering based methods, like single- or multi-network GP filter, mitigate this underestimation issue by using observation models for the low-cost data.}

\subsection{Heteroscedastic observation models}\label{sec:het_obs_model}

The observation models are estimated before applying the spatial filtering for calibration and mapping. In some cases, such as the SEARCH network, at least one low-cost site is collocated with reference device(s) in the region. At the collocated sites, both $x$ and $y$ are measured, so the observation model can be trained using this data over a training window. If no sites are collocated within the region, such as for the PurpleAir network in Baltimore, then training the observation model is more complicated. Several possible ways to train the observation model are to treat a reference and low-cost site that are close together as collocated, to identify days when there is little spatial variation for training using data from a remote reference site, or to train a model in a different region that has collocated sites from the same kind of device. Once the training data is selected, the observation model can be trained, and both the regression coefficients and the error model can be estimated. 

Care must be taken to ensure that the model is well specified across the full range of concentrations that it will be applied to, including high concentrations. This can be challenging when typical concentrations in many major US cities are low on most days, leading to less amount of training data with high concentrations. 
\cite{heffernan2023dynamic} proposed modeling the observation model noise as i.i.d. errors with homoscedastic variance. As we will see, at high concentrations, it is possible that the regression model remains correctly specified but that the errors $\epsilon\st$ are not i.i.d. As concentrations increase, the error variance also increases, implying heteroscedasticity which needs to be modeled. We consider a simple heteroscedastic model of the form:
\begin{align}\label{eq:het}
\begin{split}
    \epsilon\st&\sim N(0,\tau^2)\\
    &\log(\tau^2)=\alpha_0+\alpha_1\log(x\st+1)\\
    \text{or } &\tau^2=\max(0,\alpha_0+\alpha_1x\st). 
\end{split}
\end{align}
where 1 is added to $x$ because PM$_{2.5}$ is occasionally measured to be $0$, but $\log$ requires a strictly positive argument. 

This model allows for sensors to have noise dependent on the true concentration, which, as we will see in the data analysis section, is a realistic depiction of how low-cost sensors work. Other models for the error variance portion of the observation model $\tau^2$ should be considered in different applications to select the most appropriate model form. 
Ideally, the regression model and the heteroscedasticity model could be trained on a previous period of time, and then be applied to the testing period. However, if the available training data does not cover the same range of true concentrations as the testing data, which is the case when applying the model to very high concentrations that the region has not witnessed in the recent past, one has to use the period where filtering will be performed to estimate the regression and/or heteroscedasticity component of the observation model. 

\section{Simulation studies}\label{sec:gensim_main}

We conduct simulation experiments to evaluate robustness of the proposed multi–network GP filter under deliberate model misspecifications. The true pollutant field is generated using a stochastic advection–diffusion source model rather than a Gaussian process. This introduces localized plumes of varying intensity, wind–driven transport, diffusion, and background decay. This design 
represents two key forms of misspecification relative to the fitted Gaussian process filters: the data are not generated from a GP and there is strong temporal correlations on the true data that is not modeled in the filter.  

Subsequent to generating the true concentration field, low–cost observations are created through network–specific calibration and noise models. Network 1 places sensors uniformly across the domain, while Network 2 avoids a quadrant with higher pollution, leading to preferential coverage. Each network produces biased and noisy data, with colocated reference sites used for calibration. We compare single–network GP filters with the proposed multi–network GP filter (MGPF) that combines both networks.  

\begin{figure}[!h]
  \centering
    \includegraphics[trim={0 200 0 100}, clip, width=0.9\linewidth]{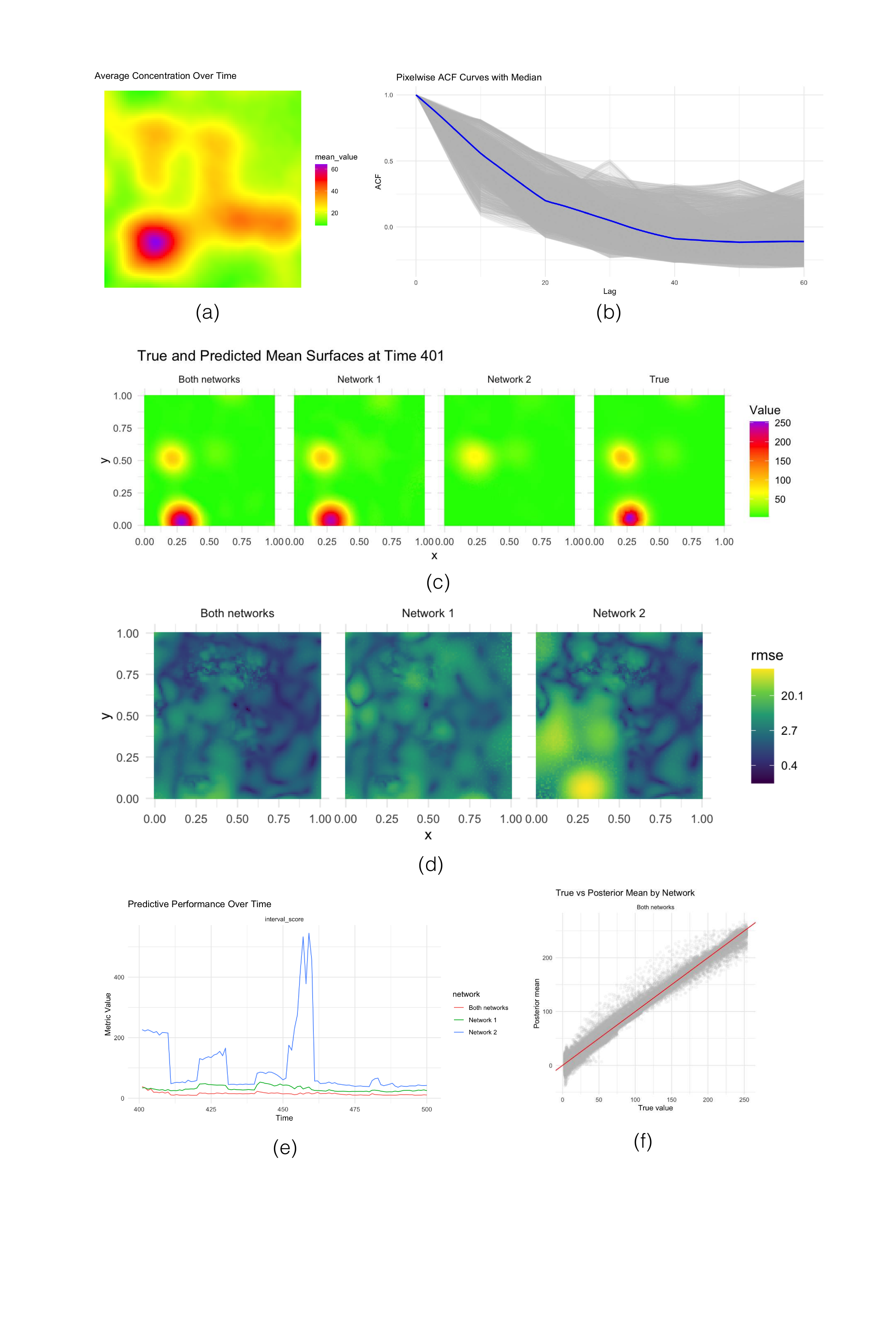}
  \caption{Summary of simulation experiments: (a) temporal mean pollutant surface, (b) median autocorrelation confirming strong temporal dependence, (c) example prediction at \(t=401\) showing improved fit by MGPF relative to single–network filters, and (d) scatterplot of true versus predicted values from MGPF.}
  \label{fig:gensim}
\end{figure}

Figure~\ref{fig:gensim} summarizes key results. Panel (a) shows the average mean pollutant surface over time, with persistent hot spots. Panel (b) presents the pixel-wise and spatially averaged autocorrelation function plot, highlighting strong temporal dependence. Panel (c) compares true and predicted mean surfaces at one example timepoint (\(t=401\)), illustrating how MGPF captures multiple peaks while the Network 2 GP filter model misses the major peak due to lack of coverage in that area. Panel (d) presents spatial maps of root mean square error (RMSE) of estimating the true pollutant field, showing that the MGPF achieves superior accuracy across the domain, with the superiority particularly evident in regions poorly covered by Network 2. Panel (e) displays interval scores over time for assessing interval estiamtion from each method. Once again, MGPF consistently achieves the lowest errors and most stable uncertainty quantification compared with single–network filters.  Finally, Panel (f) shows the overall scatterplot of true versus predicted means under MGPF across all times and locations. A tight alignment of the point cloud along the 45° line provides strong evidence on the accuracy of MGPF.  

Synthesizing all the evidence, we conclude that MGPF remains robust to both sources of misspecification, yielding accurate predictions and well–calibrated uncertainty even when the data generation process departs substantially from the model assumptions. 
 Detailed description of the simulation experiments and results are given in Supplemental Section~\ref{sec:gensim}. The technical specifications of the stochastic advection–diffusion model and pollutant field generation are in Section~\ref{sec:simdgp}, the construction of the synthetic low–cost sensor networks and their observation models are described in Supplemental Section~\ref{sec:simlcs}, and extended predictive performance results, including additional temporal and spatial evaluations, are presented in Supplemental Section~\ref{sec:simres}. We also conduct a second set of experiments more focused on issues relevant to the application of our approach in Baltimore. This includes assessment of MGPF in a design similar to our real application, as well as demonstration of the importance of correctly estimating the observation models. These experiments and the results are documented in Supplemental Section \ref{sec:sim}. 

\newpage
\section{Results: Calibrating Baltimore PM$_{2.5}$ low-cost data}\label{sec:baltimore}

\subsection{Preferential Sampling}\label{sec:preferential}

From Figure \ref{fig:locs}, we notice that the PurpleAir network seems more geographically concentrated than the SEARCH network, lacking coverage especially in the southeast corner of the city. To investigate whether there is preferential sampling at play, we use a similar approach as \cite{liang2021wildfire} to obtain location-specific estimated house prices and rents. We then look at the median and the range of prices and monthly rents across the two networks in Table \ref{tab:zillow}. We see that PurpleAir sensors have a median price of \$53,500 more than the SEARCH sensors median price, and a median rent of \$332 more than SEARCH. Additionally, the ranges of both house prices and rents in the PurpleAir network are smaller than the respective SEARCH network ranges, indicating that the PurpleAir network does not capture the entire spectrum of house prices, and by extension all income levels, in Baltimore. 
The histograms of the prices and rents, as well as more details about how they are obtained, are shown in Supplement \ref{sec:supp_zillow} and Figure \ref{fig:zillow_hist}. Figure \ref{fig:acs_hist} shows a histogram of the house prices in Baltimore according to the American Community Survey \citep{ACS}. The median house price is \$210,300, which is less than both the SEARCH and PurpleAir medians, but closer to the SEARCH median. \cite{liang2021wildfire} also found that PurpleAir sensors tended to be in higher income areas. 
Therefore, preferential sampling may impact the predictions from the PurpleAir network. 

\begin{table}[h]
\centering
\caption{Median and range of house prices and monthly rents estimates at addresses near the network sensors.}
\begin{tabular}{>{\centering\arraybackslash}p{3em}ccccc}
\toprule
\multicolumn{2}{c}{ } & \multicolumn{2}{c}{Network} & \multicolumn{1}{c}{Difference } \\
\cmidrule(l{3pt}r{3pt}){3-4}
 &  & PurpleAir & SEARCH & (PurpleAir - SEARCH)\\
\midrule
 & median & 344,500 & 291,000 & 53,500\\
\cmidrule{2-5}
\multirow{-2}{3em}{\centering\arraybackslash \textbf{price (\$)}} & max - min & 691,300 & 767,400 & -76,100
\\
\bottomrule
 & median & 2,500 & 2,168 & 332\\
\cmidrule{2-5}
\multirow{-2}{3em}{\centering\arraybackslash \textbf{rent (\$)}} & max - min & 3,538 & 4,530 & -992\\
\bottomrule
\end{tabular}
\label{tab:zillow}
\end{table}

\subsection{Training observation models}

There is only one reference device in Baltimore, at Lake Montebello. The SEARCH network has permanently collocated two low-cost sensors  with the reference device. Thus we have ample collocated data to estimate the observation model parameters for the SEARCH network. We use 2 years of data, 2020-2021, to train the following observation model: 
\begin{align*}
    y\st=\beta_0+\beta_1x\st+\bbeta_2\bz\st+\bbeta_3x\st\bz\st+\epsilon\st
\end{align*}
where $\bz$ is a vector of the covariates --- relative humidity (RH), temperature (T) and a binary indicator for weekend (W). The PM$_{2.5}$ sensor used in this network has a lab-correction equation \citep{levy2018field}, so we use the lab-corrected measurement as $y\st$. 

\begin{figure}[t]
\centering
\includegraphics[width=3in]{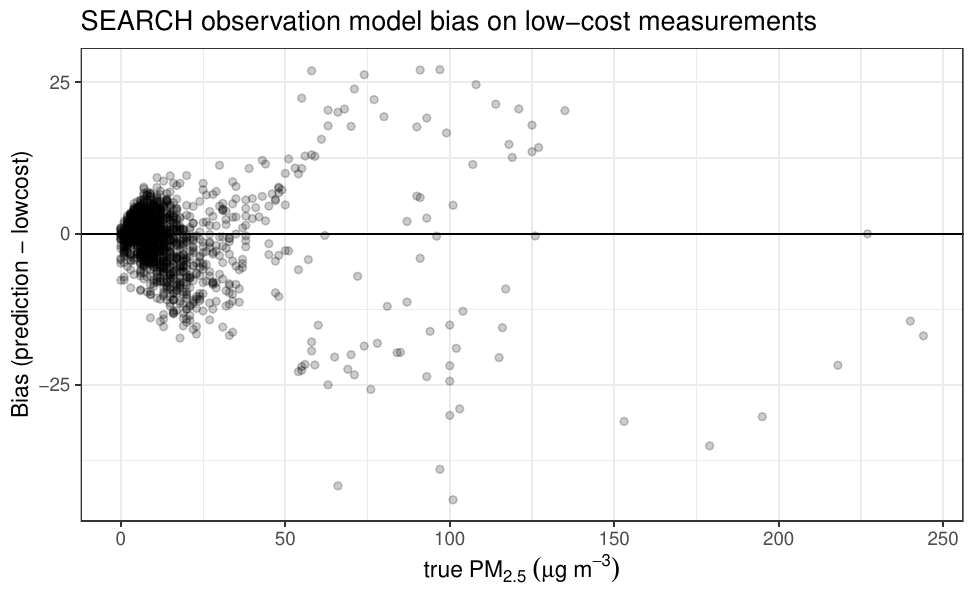}
\includegraphics[width=3in]{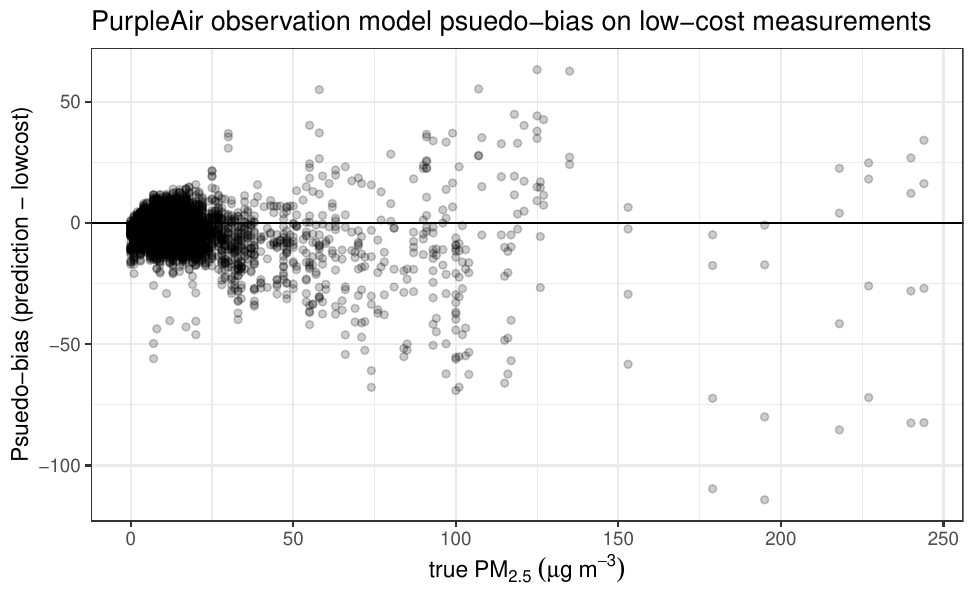}
\caption{(Pseudo)-Bias of the (left) SEARCH or (right) PurpleAir observation model in June and July 2023 at the collocated Lake Montebello site (SEARCH) or the closest sites to the Lake Montebello site (PurpleAir)}\label{fig:obs_resid} 
\end{figure}

During this two year training period of 2020-2021, the highest observed $x$ is only 77 $\mu g/m^3$, while in the testing period of June and July 2023, the highest $x$ is 244 $\mu g/m^3$. Therefore, care must be taken to ensure that the observation model is appropriate for the testing data. Figure \ref{fig:obs_resid} (left) shows the bias when this fitted observation model assuming homoscedastic errors is applied to predict the true concentrations at Lake Montebello during the out-of-sample time period of interest, June and July 2023. The bias is centered at around 0, indicating that the estimated parameters for the regression part (mean) of the observation model do seem to be good fit for the higher concentrations. However, the error variance increases with the concentrations, thereby indicating the need for a heteroscedastic model. 

We fit a heteroscedastic model for the observation model variance. We use the 
log model from Equation (\ref{eq:het}), i.e., $\log(\tau^2)=\alpha_0+\alpha_1\log(x\st+1)$. 
We train the variance model on the period of June and July 2023 because the two-year training period used for estimating the regression coefficients do not have concentrations greater than 77 $\mu g/m^3$, and extrapolating the variance model to higher concentrations may not be appropriate (see Section \ref{sec:sim} and Figure \ref{fig:het_comparison} of Supplement \ref{sec:supp_het}).
Once parameters of the the error variance model are estimated from June-July 2023, we re-estimate the regression model coefficients by fitting a generalized least squares with these heteroscedastic errors $\epsilon$ using the 2020-2021 training data. This does not change the regression coefficients of the observation model very much, indicating again that the regression coefficients for the mean were well estimated, despite the variance being misspecified. We also considered fitting a model to the log concentrations but this model did not fit the data well (see Supplement \ref{sec:supp_het}, Figure \ref{fig:log_obs}).


For the PurpleAir network, we have no collocated devices in the entire region, and since the sensors used are different from the SEARCH sensors, we cannot assume that the same observation model holds for both networks. 
Several attempts to train an observation model using Baltimore data proved unsuccessful (see Supplement \ref{sec:supp_het}, Figure \ref{fig:retain_PA}). 
Therefore, we instead use the US nationwide calibration equation for PurpleAir sensors \citep{barkjohn2021development}, which can be solved for $y\st$ to give
\begin{align}\label{eq:barkjohn}
    y\st&=-10.97+1.91 x\st+0.16 RH\st+\epsilon\st.
\end{align}
This is the observation model we use for the PurpleAir network. Note that the SEARCH network observation model uses three covariates, RH, T, and W, while the nationwide calibration equation for PurpleAir found that only RH was needed as a covariate, showing how different sensors have different biases. Additionally, there is no interaction between RH and $x$ in the PurpleAir observation model, while an interaction was found to be beneficial to the SEARCH observation model. 

We see in Figure \ref{fig:obs_resid} (right) that the pseudo-bias (as the reference is not exactly collocated) for the PurpleAir sensors on the test data is roughly centered around 0, so the mean model in (\ref{eq:barkjohn}) is reasonable for the June and July 2023 data, but as with the SEARCH network, there is evidence of heteroscedasticity. We use the June and July 2023 data to train the model for $\tau^2$, again using the log model from Equation (\ref{eq:het}), $\log(\tau^2)=\alpha_0+\alpha_1\log(x\st+1)$. A comparison of training the variance models using data from June and July 2023 as opposed to June 2022 - May 2023 is shown in  Supplement \ref{sec:supp_het} Figure \ref{fig:het_comparison}, revealing the need to train the variance model during the window with high concentrations (June and July 2023). 

The fitted observation model variances, as well as a plot of the log squared bias which is the outcome in the fitted model we use, are shown in Figure \ref{fig:obs_var}. 
The fitted observation models (the regression coefficients for both the mean and the variance models for both networks) are shown in Table \ref{tab:coefs}. 

\begin{table}[h]
\centering
\caption{Observation model parameters for SEARCH and PurpleAir networks}
\begin{tabular}{|p{2in}|p{1in}|p{1in}|}
\bottomrule
\textbf{Term} & \textbf{SEARCH} & \textbf{PurpleAir}\\
\hline
Intercept & -0.9756 & -10.9733\\
\hline
true PM$_{2.5}$ & 1.0789 & 1.9084 \\
\hline
RH & 0.0422 & 0.1645\\
\hline
Temp & -0.0357 & \\
\hline
weekend & 0.4086 &\\
\hline
true PM$_{2.5}$ * RH & -0.0030 &\\
\hline
true PM$_{2.5}$ * Temp & 0.0058 &\\
\hline
true PM$_{2.5}$ * weekend & -0.0736 &\\
\hline
variance model intercept & -1.2136 & 0.4973\\
\hline
variance model slope & 1.1774 & 0.8802 \\
\toprule
\end{tabular}
\label{tab:coefs}
\end{table}

\subsection{Prior specification}

Subsequent to training the observation models, we apply the MGPF to filter hourly data. To stabilize the predictions and avoid very local fluctuations greatly impacting the GP covariance structure, we use the following specification for our GP:
\begin{align*}
    (\bx\sot,\bx\sat,\bx\sbt)&\sim GP(\mu_t\mathbbm{1},\bC_t)\\
    \bC_t(d)&=\sigma_t^2\exp\{-\phi_t d\}+\sigma_{nugget,t}^2\mathbb{I}_{d=0};\\
    \mu_t&\sim \blue{HN(0,V)};\quad\sigma_t^2\sim U(0,s_{max});\\
    \phi_t&\sim U(\phi_{min},\phi_{max});\quad\sigma_{nugget,t}^2\sim U(0,s_{nugget,max})
\end{align*}

\begin{figure}[t]
\centering
\includegraphics[width=6.5in]{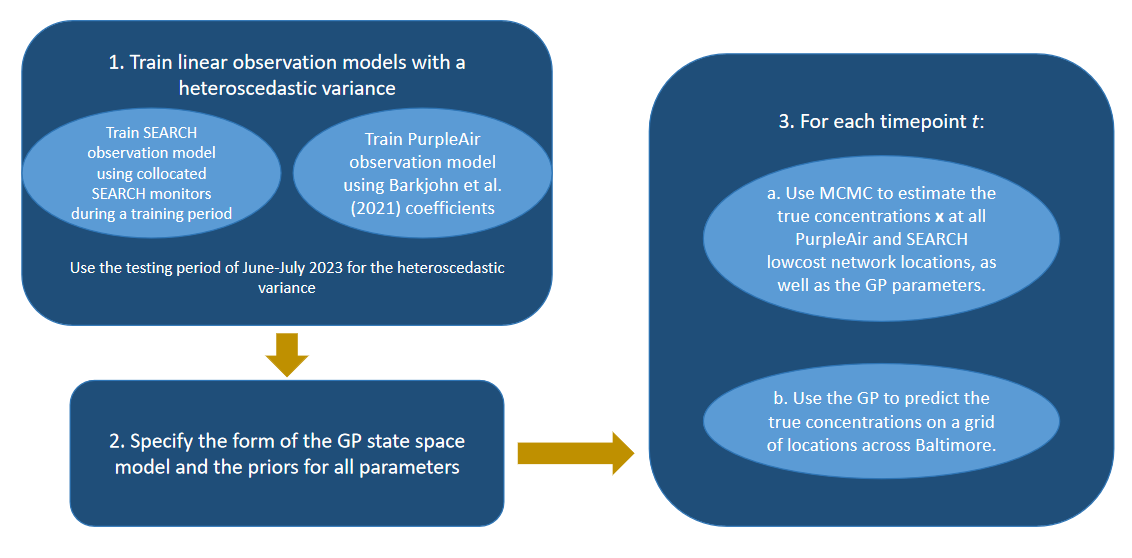}
\caption{Schematic of the MGPF filtering process in Baltimore}\label{fig:schematic_process}
\end{figure}

\blue{where $HN$ indicates the half normal distribution and $V$ is a large variance.}
We use an exponential covariance structure with spatial variance $\sigma_{t}^2$ and spatial decay $\phi_t$ to model the spatial structure in the true pollutant concentrations. The nugget $\sigma_{nugget,t}^2$ is added to model micro-scale variations in PM$_{2.5}$ levels. The upper bound on the nugget, $s_{n,max}$ is chosen to be the variance of the SEARCH lab-corrected data for that time-point. This data has already gone through one round of correction, unlike the PurpleAir data, so it is on a more similar scale to the true concentrations. This bound essentially allows for all the data variance to be in the nugget, which is fairly generous, to allow for time-points where the spatial variability in concentrations will be very high. The bound on the spatial variance $\sigma_t^2$ is twice the variance of the predictions from the inverse model, again a generous bound. We allow for a larger bound of $\sigma_t^2$ than $\sigma_{n,t}^2$ because we believe that the spatial variance should generally be larger than the nugget variance at most time-points when there is limited variability in air quality in the area. 
Finally, the range of the uniform prior for $\phi_t$ is selected so that the correlation at the farthest points in the network is between 2\% and 98\%. These bounds prevent the Bayesian sampler from going into extreme and unlikely parameter values, and ensure that there is at least some smoothness in air pollution concentrations. \blue{A schematic summarizing the model training and filtering process in Baltimore is shown in Figure \ref{fig:schematic_process}.}

\subsection{Performance in June and July 2023}

\begin{figure}[t]
\centering
\includegraphics[width=4.5in]{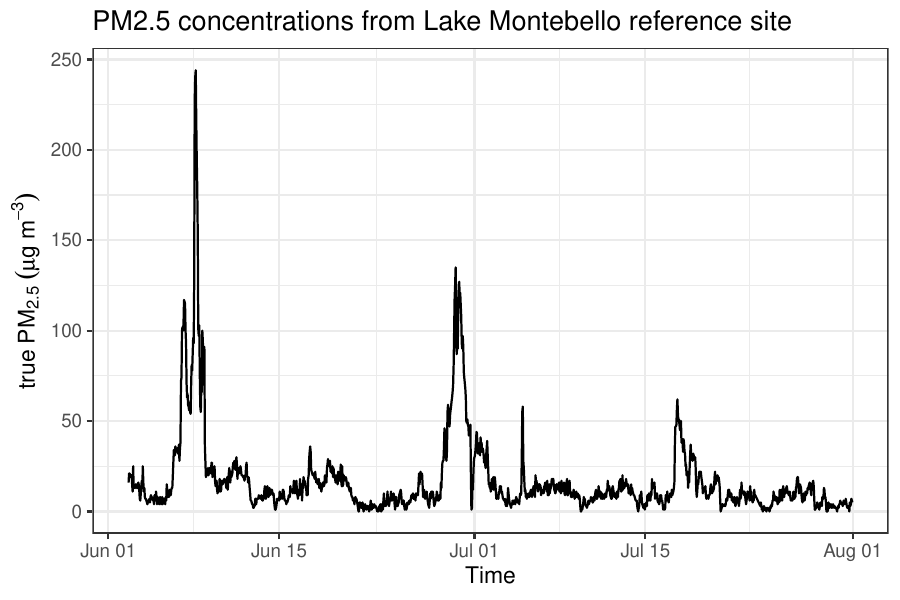}
\caption{\blue{Time-series of PM$_{2.5}$ concentrations in $\mu g m^{-3}$ measured by the reference device at Lake Montebello in June and July 2023}}\label{fig:ts} 
\end{figure}

\begin{figure}[t]
\centering
\includegraphics[trim={1 120 2 3}, clip, width=5.5in]{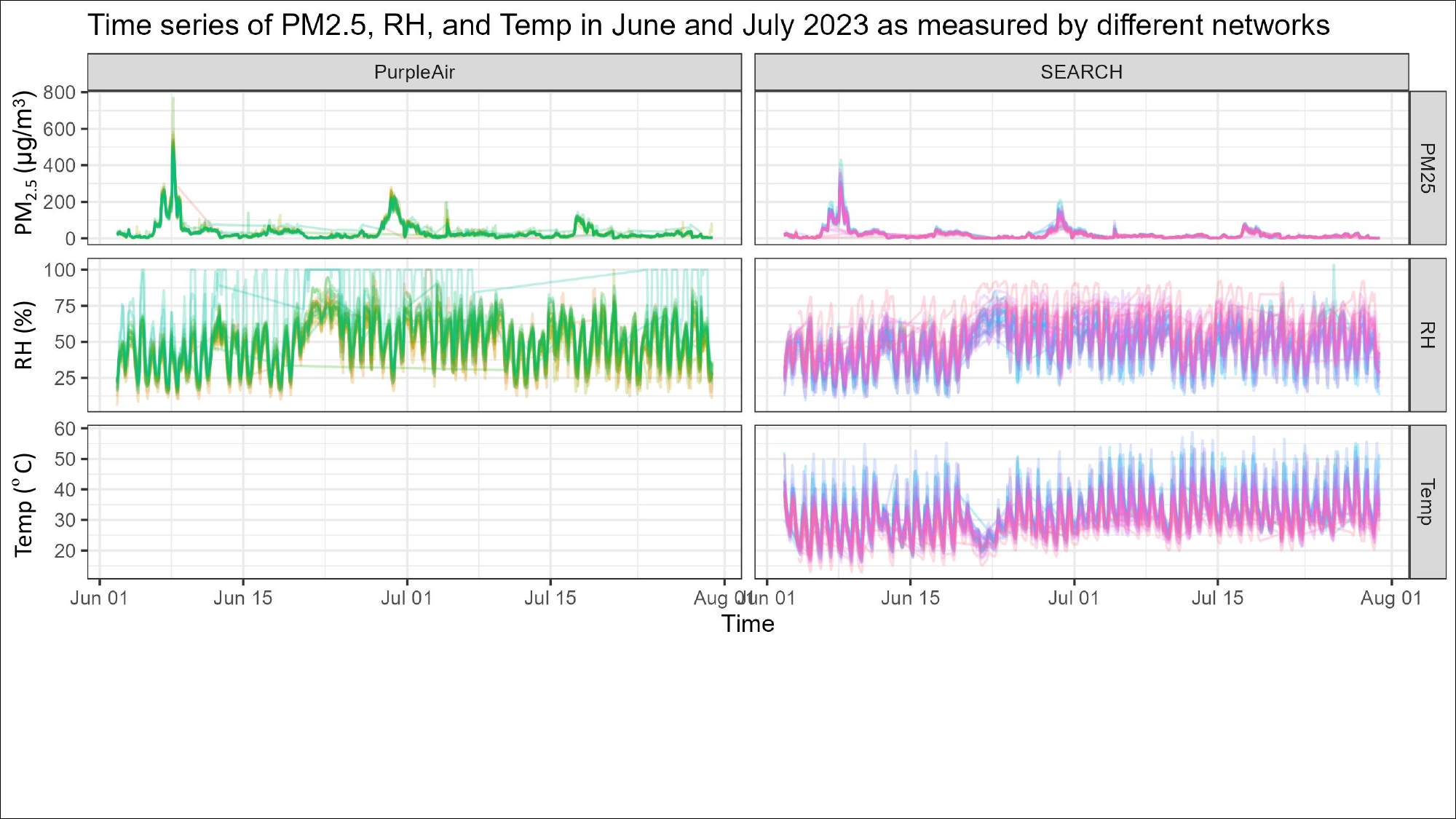}
\caption{\blue{Time series of PM$_{2.5}$, RH, and Temp at the low-cost sites by network. Each color represents a different low-cost sensor. Temperature for PurpleAir sites is omitted since it is not used in the observation model.}}\label{fig:ts_lc}
\end{figure}

We now present the results of applying the MGPF to the SEARCH and PurpleAir networks. There are three main periods during June and July 2023 where concentrations were elevated: June 7-8, June 28-30, and July 17-18, with concentrations 
going up to almost 250 $\mu g/m^3$. The distribution and time series of true concentrations \blue{at Lake Montebello are given in Table \ref{tab:true_pm} and Figure \ref{fig:ts}. It clearly shows the large spikes during the wildfire smoke days.  
The time series of the low-cost PM$_{2.5}$ measurements, RH, and Temp are given in Figure \ref{fig:ts_lc}. Note that we do not see any drastic changes in the RH or Temp time-series during the wildfire smoke days when we clearly see the spikes in the PM$_{2.5}$ time-series. This presents evidence that the spikes were not due to local meteorological conditions but due to a regional source (wildfire smoke). A scatterplot of the ratio between low-cost and true PM$_{2.5}$ measurements and the meteorological variables used in observation models is shown in Figure \ref{fig:scatterplot_ratio}.}  

At every hour, we perform filtering using only the PurpleAir network, only the SEARCH network, and using both networks. One important metric to understand the impact of multi-network filtering is to assess the difference in uncertainty. We will calculate a percent difference in the length of the \blue{credible} interval (CI) using two networks ($l_2$) compared to one network ($l_1$):
\begin{align}\label{eq:percent_diff}
    \%diff&=\frac{l_2-l_1}{l_1}*100
\end{align}

\begin{figure}[t]
\centering
    \includegraphics[height=3.75in,trim={284 0 273 0},clip]{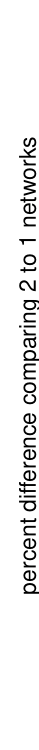}
    \includegraphics[height=3.75in,trim={20 0 0 0},clip]{figures/percent_diff.pdf}
    \caption{(Top left) Percent difference in CI lengths, for each monitor from the PurpleAir and SEARCH networks, averaged across time. (Bottom left) Percent difference in CI lengths, for each time-point, averaged across the PurpleAir and SEARCH networks. (Right) Percent difference in CI lengths averaged across all interpolated sites in Baltimore, using both networks compared to using only one network, capped at 100\%. The median of the percent differences, and the proportion $p$ of percent differences that are negative, are listed. The purple lines are the true concentrations at the reference site at Lake Montebello. }\label{fig:CIlengths} 
\end{figure}

We compare the predicted concentrations and the length of CIs at network sites when both networks are used to when only that network is used in the filtering (Figure \ref{fig:CIlengths}, top left and bottom left). We see the 
percent differences in CI lengths averaged by monitor (top left) and by time-point (bottom left) at the network sites. We see decreases in CI length when two networks are used compared to using a single network for every monitor.  The decrease is bigger at the SEARCH network sites when PurpleAir is also used for filtering, than the other way around. There are $83\%$ of hours with a decrease in CI length using two networks, and the median change in CI length across the network is $-9.49\%$. Many, but not all, of the time-points where uncertainty increases correspond to the periods with higher concentrations (purple time series), showing that an individual network may be overconfident when concentrations are high. 
Some diagnostics of applying the method on this dataset, such as a looking at the spatial parameter estimates and a comparison of the MGPF predictions and predictions from the observation model alone, are shown in Supplement \ref{sec:supp_data_analysis}. \blue{We also assessed patterns of missingness in the low-cost data in terms of spatial distribution, temporal trends, and distributions of variables. All of this provided no evidence of informative missingness (see Supplemental Section \ref{sec:supp_missing}). Hence, we do not model the missingness pattern as reasoned  in the last part of Section \ref{sec:mgpf}.}

We also make predictions of PM$_{2.5}$ across Baltimore, on a fine grid of locations, and plot in Figures \ref{fig:CIlengths} (top right and bottom right), for each network, the time-series of average change in uncertainty when adding the second network. We see that the CI length when using two networks is smaller than when using one network $81-83\%$ of the time. The median percent change in CI length across the city using the multi-network GP filter is $-17\%$ compared to filtering using the SEARCH network, and $-11\%$ compared to filtering using the PurpleAir network. We also see that for a few time-points, the CIs from MGPF can increase in length by over $100\%$. This is  usually because one network (often the PurpleAir network) is overconfident and has very narrow intervals when used alone. This tends to happen when the PurpleAir predicted surface is very flat (Supplement \ref{sec:supp_data_analysis}, Figure \ref{fig:increase_uncertainty}). The percent decreases at interpolated sites are larger than at the network sites, showing that the biggest advantage of using two networks is in making predictions at new locations. The absolute difference in CI lengths at network sites and interpolated locations is shown in Figures \ref{fig:supp_CI_network} and \ref{fig:supp_CI_interpolated}. Additional model validation is shown in Figures \ref{fig:pars} and \ref{fig:inv_filter}. 

\begin{figure}[t]
\centering
\includegraphics[width=6.25in]{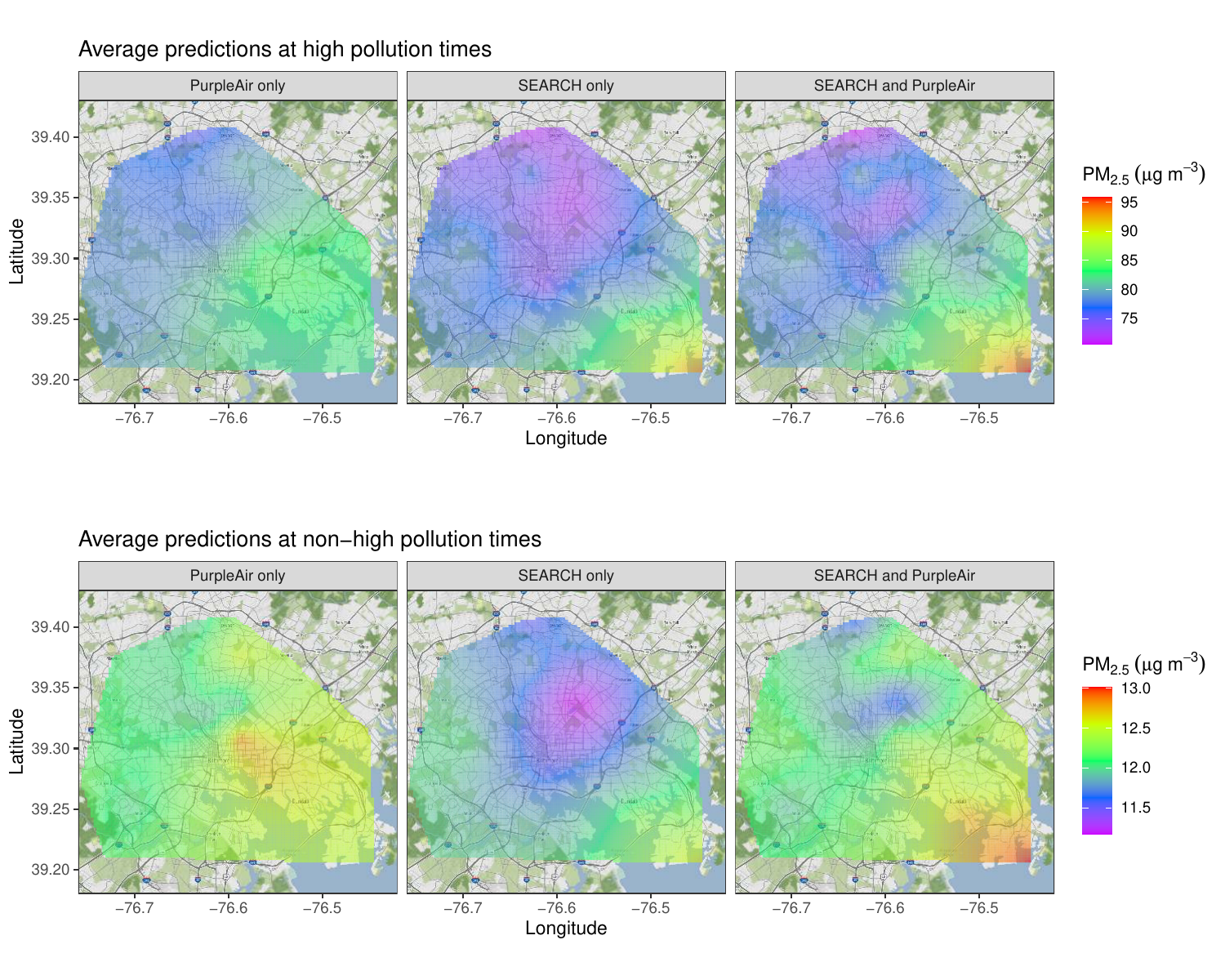}
\caption{Mean of predicted PM$_{2.5}$ in Baltimore (Top) over the days with high pollution (Bottom) over all remaining days. 
}\label{fig:avg_pred} 
\end{figure}

Additionally, we look at maps of concentrations overall during this two month period. The mean of the predictions at the interpolated sites is shown in Figure \ref{fig:avg_pred} for high pollution days (top), i.e., the days with the top 10\% of concentrations measured by the reference sensor at Lake Montebello, and on the remaining days which are more representative of typical concentrations in Baltimore (bottom). 
There are clear differences between the predictions from different methods. In particular, in the southeast portion of the city, the PurpleAir network has lower predicted concentrations. This is due to the preferential sampling in the PurpleAir network, having no sensors to measure the higher concentrations in that part of the city. 
PurpleAir estimates a flatter surface on average, while SEARCH and the two networks combined have more variability across the city. 

\begin{figure}[t]
\centering
\includegraphics[width=6.25in]{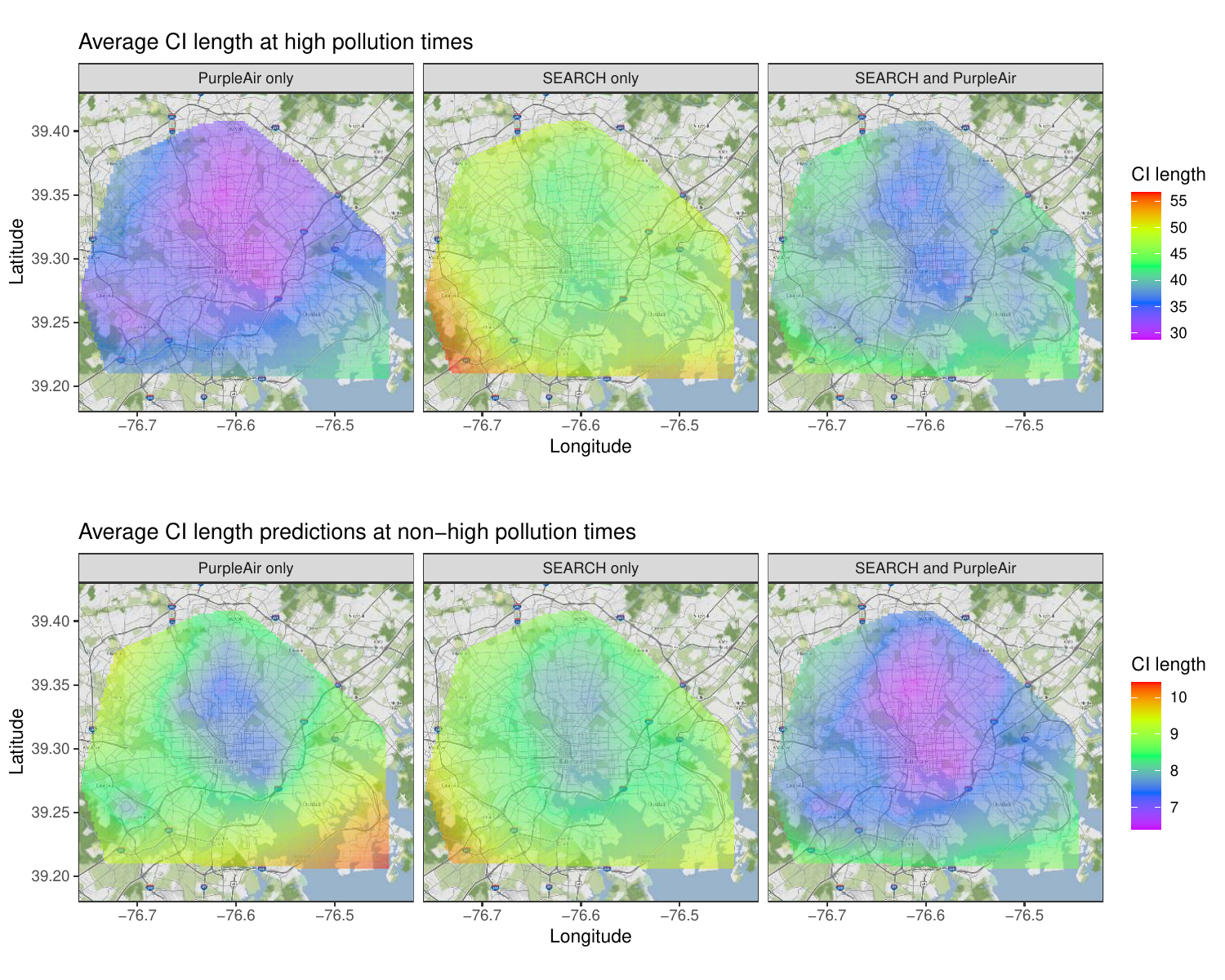}
\caption{Average \blue{credible} interval length of predicted PM$_{2.5}$ in Baltimore (Top) over the days with high pollution (Bottom) over all remaining days. 
}\label{fig:avg_CI} 
\end{figure}


Figure \ref{fig:avg_CI} shows the average length of CIs across the two months. We see different behaviors for high and non-high time-points. For high time-points, SEARCH benefits most from uncertainty reduction by adding PurpleAir, while filtering only with the PurpleAir leads to overly anti-conservative intervals which are widened by the multi-network filtering. For non-high time-points, filtering only with SEARCH or PurpleAir produces comparable interval lengths, and both networks benefit greatly from adding the other network. From this, we see that there are benefits in terms of prediction accuracy (especially in the PurpleAir network) and in uncertainty (especially in the SEARCH network) from using a multi-network approach to filtering. Figures \ref{fig:avg_pred}-\ref{fig:avg_CI} don't show the network sites for clarity of presentation. The figures including the network sites are in Figures \ref{fig:map_avg_supplement}-\ref{fig:map_avg_CI_supplement}. We also present maps of the average concentration and CI length over the entire two-month period, as well as an example of a map for a single time-point and a comparison of CI lengths in certain periods, in Supplement \ref{sec:supp_data_analysis} (Figures \ref{fig:extreme_CI}-\ref{fig:preds_CI_124}). 



Another approach to combine data from multiple networks is to individually calibrate each network using a linear regression calibration equation, and then to interpolate using inverse distance weighting (IDW) \citep{mueller2004map}. This approach does not model spatial correlation across different locations, which leads to an interpolated surface with much more fluctuation on average, especially around network sites (Figures \ref{fig:idw_avg}, \ref{fig:idw_avg_high_low}), and it suffers from underestimation (Figure \ref{fig:idw_one}). The pseudo-RMSE of the IDW method is also higher than the MGPF (Figure \ref{fig:idw_prmse}). More details are given in Supplement \ref{sec:supp_data_analysis}.  

\section{Conclusions}

\blue{In this manuscript we developed 
multi-network Gaussian process filter (MGPF), an extension of the single-network GP filter of \cite{heffernan2023dynamic}} to calibrate data from multiple low-cost sensor (LCS) networks, each with different biases and noise levels. We apply this method to combine PM$_{2.5}$ data from the SEARCH and the PurpleAir LCS networks, in Baltimore, Maryland \blue{and produce unified, uncertainty quantified maps of PM$_{2.5}$ in the city.} 
Regression equations have been published that calibrate \blue{LCS data from individual networks, e.g.,} the PurpleAir \citep{barkjohn2021development} and SEARCH \citep{datta2020statistical} networks, but using these equations in isolation means that spatial patterns are ignored. \blue{Additionally, as \cite{heffernan2023dynamic} showed, such direct regression-based calibration equations can underestimate high concentrations. The GP filter approach of 
\cite{heffernan2023dynamic} was developed to mitigate this issue -- adopting a filtering approach to calibration instead of regression. Filtering acknowledges the inferior quality of the LCS data by modeling it as a biased and noisy version of the true concentrations, and modeling the spatial correlation among all observations. MGPF enjoys all of these advantages of the single-network GP filter. At the same time it considerably expands the scope of the filtering-based approach, by accommodating data from multiple networks with network-specific biases, and producing unified calibrated maps. }   


Other potential approaches to combine data from two networks exist. One is to use spatial filtering to calibrate the low-cost measurements from each site individually to make predictions 
and then take an average 
of these predictions at each location as the final prediction. This approach 
could be biased by the preferentially sampled network which is part of the average. Alternatively, estimates at sensor sites can be made for each network individually, and then inverse distance weighting (IDW) \citep{mueller2004map} can be used to interpolate the predictions across the network (Supplement \ref{sec:supp_data_analysis}). \blue{This approach can create large and unrealistic local fluctuations in predicted concentrations where the two networks have sites nearby, as we saw empirically (see Figures \ref{fig:idw_avg} and \ref{fig:idw_avg_high_low} for comparison of maps produced by MGPF and IDW).} 
Additionally, IDW does not have associated uncertainty quantification. \blue{Finally, this alternate approach of calibrating the networks individually and then combining involves making many decisions,} such as whether to interpolate using each network and then combine or combine estimates first and then interpolate, whether and how to weight observations, and what weight to use in IDW. The MGPF circumvents having to make these decisions by using a principled hierarchical model, incorporating the calibration and interpolation steps into one, and using all data to do both of these steps. 

Our approach is similar to 
\blue{\cite{zimmerman2005complementary} and \cite{fuentes2005model}, who also study two networks with different measurement errorss, focusing on predicting observations when the bias is not identifiable. However, we accommodate three networks, where the reference network measurements has no bias or measurement error, and our goal is to predict outcomes for this reference network using data from all networks.} 
Also, we allow for non-constant bias. We further tailored our method to the problem of low-cost sensors in Baltimore, by focusing on the issues of heteroscedastic measurement error, preferential sampling, and calibration of high concentrations in a setting with a few high-quality sensors and some low-cost sites that are collocated with these high-quality devices. 

\blue{Air pollution concentrations are correlated over time and it would be important to model this correlation if the goal is forecasting future concentrations or downscaling to finer time-resolution than the hourly maps we produce. Neither is the objective of this paper which focuses on producing hourly spatial maps for each hour combining hourly data from all three sources. Hence, we follow the guidance given in Section 3.7 of \cite{heffernan2023dynamic} and do not model temporal correlations, filtering at each time point separately. 
In simulation studies we consider data generation processes with temporal correlation (Figure \ref{fig:restime}) and the MGPF performs very well. 
In principle, the spatial filtering of MGPF can be extended to a spatio-temporal filtering that models both spatial and temporal correlations. Supplemental Section S4 of \cite{heffernan2023dynamic} outlines ideas for this extension for the single-network filtering and most of those can be adapted to multi-network setting. There is also a very rich literature on spatio-temporal models for forecasting air pollutants, \citep[see. e.g., ][]{sahu2015bayesian,nicolis2019bayesian}. It would also be important to explore if these approaches can be adapted to the setting of joint filtering data of varying quality from multiple networks to produce unified forecasted maps.} 

The uncertainty estimates in the maps produced by MGPF can be used to identify areas with higher uncertainty, where adding a network site can be most beneficial. 
Also, while we focus on PM$_{2.5}$ in this manuscript, the MGPF could be extended to other pollutants. Assuming the same observation model for an entire network, as we did in this work, is not possible for gas sensors, so alternative approaches to create sensor-specific observation models would be needed. 
\blue{Another future extension 
would be to include a third network in our MGPF for Baltimore operating in Curtis Bay, a neighborhood in south Baltimore with many sources of pollution, 
\citep{deanes2025relation}. 

Although the computational burden of Gaussian process models scale cubically in term of the number of locations, we have not focussed on scalability of the algorithm in this manuscript. This is because low-cost sensor air pollutants networks in cities or some local region
are not too large. Possible exception to this will be a large statewide or nationwide analysis. In such settings, we can easily adapt MGPF for large number of locations by switching from a full GP prior to a Nearest Neighbor Gaussian Process \citep[NNGP, ][]{nngp,finley2019efficient,Datta2022SparseCholesky} prior. NNGP has linear computational complexity and can be used as a prior in any hierarchical setup and can thus be seamlessly integrated into the MGPF hierarchical model.}

The two-network MGPF filtering approach can be used as the new standard for calibrating low-cost data in Baltimore and interpolating predictions throughout the city. 
The gains in prediction accuracy and uncertainty quantification were clearly demonstrated in June and July 2023, thus the approach has proved itself able to filter high concentrations. 
The observation model regression coefficients and heteroscedastic model variance that we trained should be usable for the foreseeable future in Baltimore. Thus, the MGPF is a better choice to predict air quality in Baltimore compared to filtering either network individually. 

\textbf{Supplemental information:} Additional information and supporting material for this article is available in the attached supplemental material \blue{pdf file. A gif file showing a video of the concentration surfaces over time for the simulations in Section \ref{sec:gensim_main} is also included as Supplement. Code for all the analysis is provided in a GitHub repository \url{https://github.com/cmheffernan/MGPF}.}

\noindent \blue{\textbf{Acknowledgments and Disclosures:} 
This work was supported by the U.S. Environmental Protection Agency (Assistance Agreement no. RD835871 to Yale University); the Bloomberg American Health Initiative at the Johns Hopkins Bloomberg School of Public Health in collaboration with Alliance for a Healthier World at Johns Hopkins University; the Johns Hopkins University Data Science and AI Institute Demonstration Project Award (AD and KK); National Science Foundation Graduate Research Fellowship Program (grant No. DGE2139757 to C.H.); National Institute of Environmental Health Sciences (NIEHS) (grant R01 ES033739 to A.D., R.P. K.K., and C.H.). The authors declare they have nothing to disclose. 
The authors would like to thank Dr. Misti Levy Zamora and Dr. Colby Buehler for their contributions in developing the SEARCH network, processing and quality control of the data, and suggestions on improving the manuscript. We also acknowledge and thank all individuals who made their Purple Air Sensor data freely available for scientific analysis. The authors also acknowledge use of chatgpt.com for helping devise simulation scenarios, improve the writing in some places, and write and debug code to conduct some of the analysis.}

\bibliographystyle{biom}
\bibliography{mybib}

\renewcommand\thesection{S\arabic{section}}
\renewcommand\theequation{S\arabic{equation}}
\renewcommand\thefigure{S\arabic{figure}}
\renewcommand\thetable{S\arabic{table}}
\setcounter{figure}{0}
\setcounter{section}{0}
\setcounter{equation}{0}
\setcounter{table}{0}

\newpage

\section{Zillow house prices and rents}\label{sec:supp_zillow}

\begin{figure}[ht]
\centering
\includegraphics[width=5in]{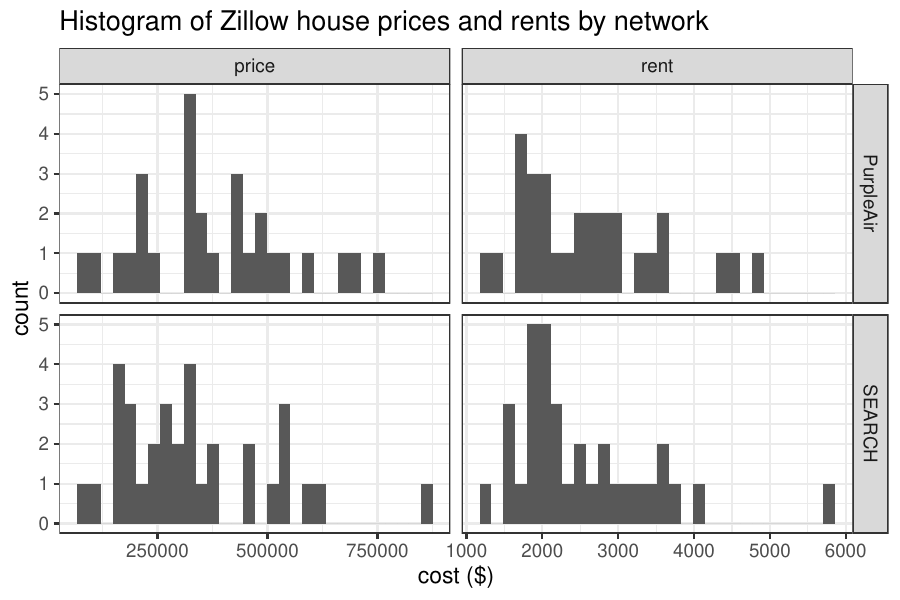}
\caption{Histogram of the house prices and rents in Baltimore at the locations of the PurpleAir and SEARCH networks.}\label{fig:zillow_hist} 
\end{figure}

We follow the approach of \cite{liang2021wildfire} in understanding the extent of preferential sampling in the two low-cost PM$_{2.5}$ networks in Baltimore. We input the co-ordinates of the sensors into Google Maps, and identify the closest house to each address. Then, we use Zillow.com, a housing website that provides estimates of the prices of many homes, to obtain an estimated price of that nearest house and the cost to rent the house, which are shown in Figure \ref{fig:zillow_hist}. 

\begin{figure}[ht]
\centering
\includegraphics[width=5in]{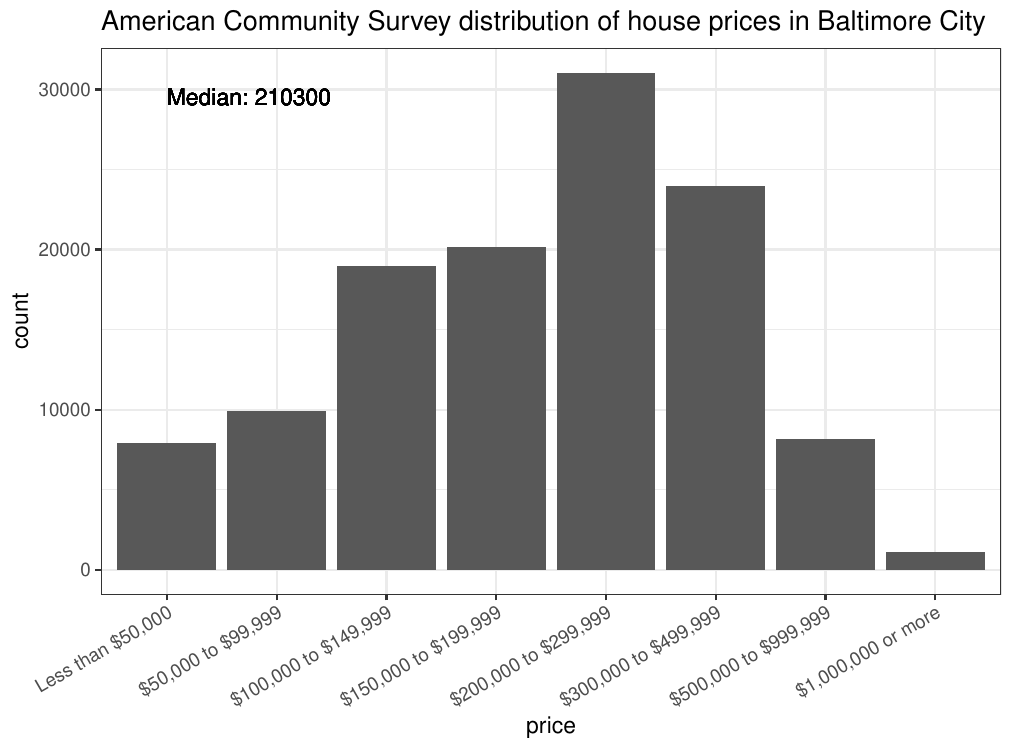}
\caption{Histogram of the house prices in Baltimore City according to the American Community Survey.}\label{fig:acs_hist} 
\end{figure}

Figure \ref{fig:acs_hist} shows house prices according to the American Community Survey. Most of the houses fall in the \$200,000-\$300,000 range. Compared to SEARCH, the PurpleAir network has fewer sensors next to homes worth less than the median of \$210,300 (Figure \ref{fig:zillow_hist}), showing that SEARCH better matches the demographics of the city, though still with a tendency towards locations where house prices are higher. 

\FloatBarrier
\section{Training heteroscedastic models}\label{sec:supp_het}

Table \ref{tab:true_pm} shows the distribution of the true concentrations across the training period for the SEARCH network (2020-2021) and the testing period of June and July 2023. We see that the concentrations are overall considerably higher in the testing period, and that the concentrations are more variable. 

\begin{table}
\caption{Summary statistics of true concentrations measured by the reference site in the training period for the mean part of the observation model (2020-2021) and the testing period (June and July 2023).}
\centering
\begin{tabular}{l|r|r|r|r|r|r|r}
\hline
dataset & mean & sd & min & Q1 & median & Q3 & max\\
\hline
test & 17.39 & 24.08 & 0 & 6 & 10 & 18 & 244\\
\hline
train & 7.01 & 5.67 & 0 & 3 & 6 & 9 & 77\\
\hline
\end{tabular}
\label{tab:true_pm}
\end{table}

\begin{figure}[ht]
\centering
\includegraphics[width=5in]{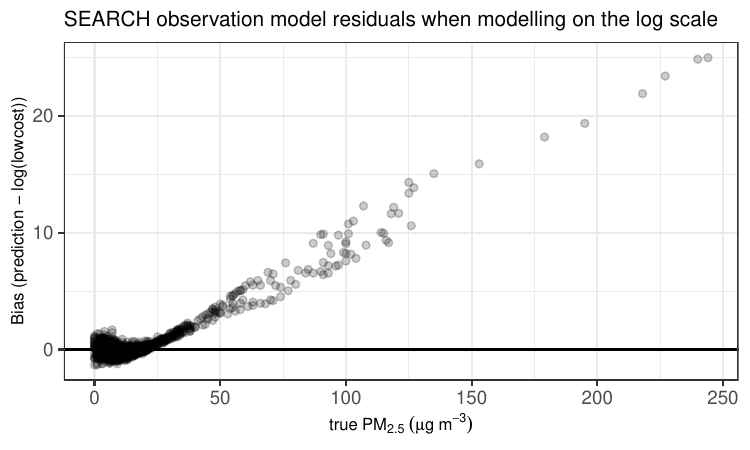}
\caption{Bias from training the SEARCH observation model on the log scale. }\label{fig:log_obs} 
\end{figure}

Since the bias in Figure \ref{fig:obs_resid} show evidence of heteroscedasticity. One possible approach to mitigate this would be to fit a homoscedastic observation model in the log-scale, i.e., $\log(y\st+1)=\beta_0+\beta_1x\st+\bbeta_2\bz\st+\bbeta_3x\st\bz\st+\epsilon\st$ (where the $+1$ is needed since the measurements can be 0). We fit this model on the SEARCH data in 2020-2021, but we see in Figure \ref{fig:log_obs} that the bias in the residuals when this model is applied to June and July 2023 is very large as concentrations increase. 
Thus, a log model is completely inadequate for this data. Additionally, for the PurpleAir network, retraining a model on the log scale would be difficult due to the lack of collocation, as we discuss next, so we do not attempt to fit this log-scale model. 

We present an attempt to retrain the PurpleAir observation model regression coefficients, using one year of past data. Since we do not have collocated sites, we must assume that the Lake Montebello true concentrations are equal to the true concentrations for some other time/site combinations. From this, a model can be fit. We present two ways of selecting the sites/times to use for training 
\begin{enumerate}
    \item using the timepoints where there was the least variation in the raw measurements from the PurpleAir sites. At these times, air quality can be assumed to be roughly similar across the city, and the reference data at Lake Montebello can be taken as a crude proxy for collocated reference. The 20\% of timepoints with the smallest relative range are considered here, and all sites at these timepoints are used. 
    \item using the sites that are closest to Lake Montebello, and assuming the true concentration there is the same as Lake Montebello at all times. Thus, only four sites are selected, but all timepoints from the year are used. 
\end{enumerate}

\begin{figure}[ht]
\centering
\includegraphics[width=5in]{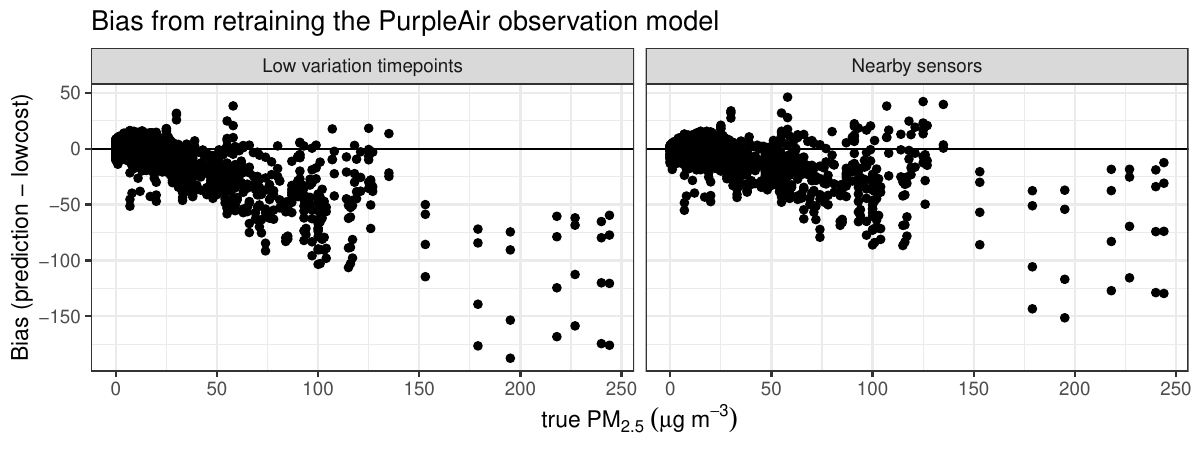}
\caption{Bias from retraining the PurpleAir observation model using one year of past data. }\label{fig:retain_PA} 
\end{figure}

The bias of the predictions from fitting the observation model using these approaches are shown in Figure \ref{fig:retain_PA}. We see that there is a strong negative bias when training on the low variance timepoints, and a slightly smaller bias when using the nearby sensors, that is most visible for concentrations above $\sim 50 \mu g/m^3$. Other approaches to select a training period or low-variance timepoints also resulted in similar types of negative biases. Since these biases are more extreme than what is observed in Figure \ref{fig:obs_resid} (right), which uses the \cite{barkjohn2021development} PurpleAir calibration equation (Equation (\ref{eq:barkjohn})), we choose to use Equation (\ref{eq:barkjohn}) as our final observation model regression coefficients. 

\begin{figure}[ht]
\centering
\includegraphics[width=5in]{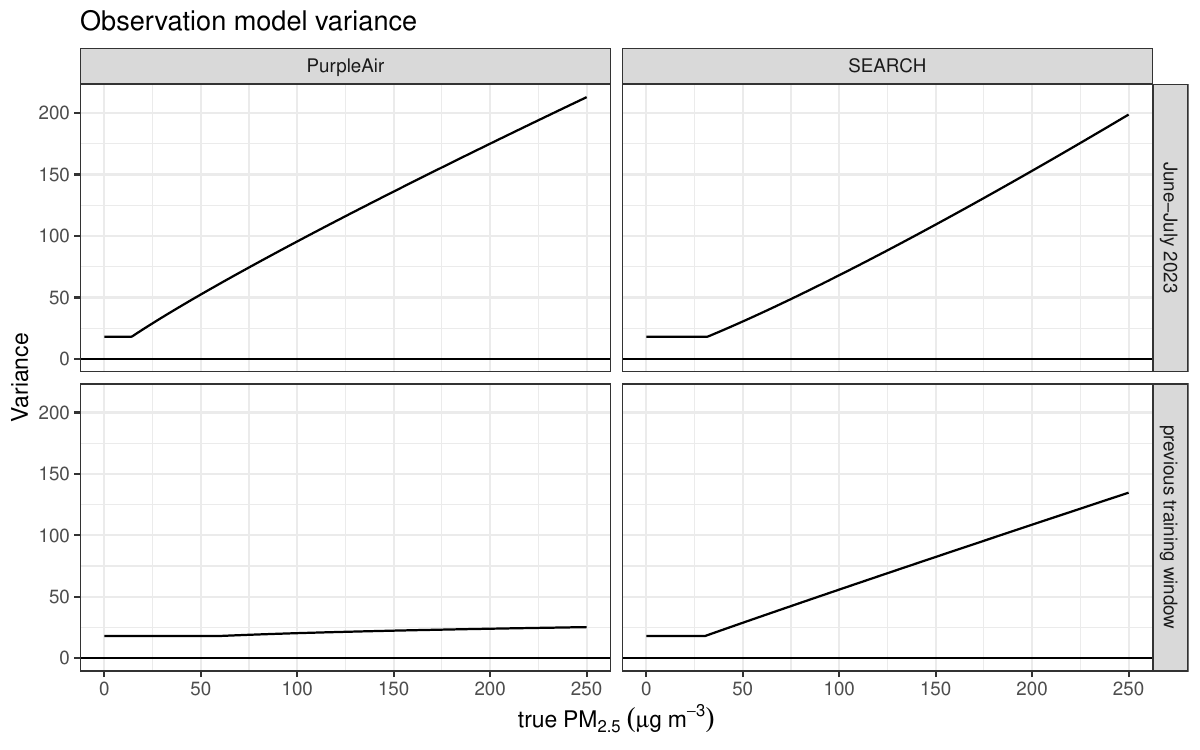}
\caption{Estimated variance $\tau^2$ as a function of the true concentration $x$, using different windows to train the variance model. For PurpleAir, the previous training window is June 2022-May 2023. For SEARCH, the previous training window is 2020-2021.}\label{fig:het_comparison} 
\end{figure}

We also present a comparison of training the heteroscedastic part of the observation model using different time periods. For the SEARCH network, since we train the  observation model regression coefficients using 2020-2021 data, it is natural to train the heteroscedastic variance model using the same time period. However, in Figure \ref{fig:het_comparison} we see that the estimated observation model variance is very different using 2020-2021 (labelled ``previous training window"), and June-July 2023 to train this part of the model. In both cases, the heteroscedastic model is trained on the square of the pseudo-bias obtained from the SEARCH observation model coefficients trained on 2020-2021 (from Figure \ref{fig:obs_resid}). In particular, using 2020-2021, where the highest concentration was 77 $\mu g/m^3$, requires extrapolation to estimate the variance at higher concentrations, and the big difference in variance from extrapolation and from training on a wider range illustrates that the extrapolated variances are likely less reliable.

PurpleAir data, using Equation (\ref{eq:barkjohn}) as the regression model, also has evidence of heteroscedasticity. Since no collocated data is available, we use the four closest sites to Lake Montebello as approximately collocated. We compare using June and July 2023 to train the heteroscedastic model to using the previous year of data, from June 2022 - May 2023. Figure \ref{fig:het_comparison} shows that when the previous year of data is used, the estimated heteroscedastic variance only goes up to about 25. This is completely outside of the range of other model fits, and a look at the out-of-sample bias in Figure \ref{fig:obs_resid} (right) shows that this estimate does not match what is observed in 2023. This example illustrates the dramatic error in extrapolation that can occur when fitting the observation model (in particular, the variance model) on data that does not cover the same range as the testing data. We also demonstrated via a simulation study in Section \ref{sec:sim_obs}. 

\begin{figure}[t]
\centering
\includegraphics[width=4.5in]{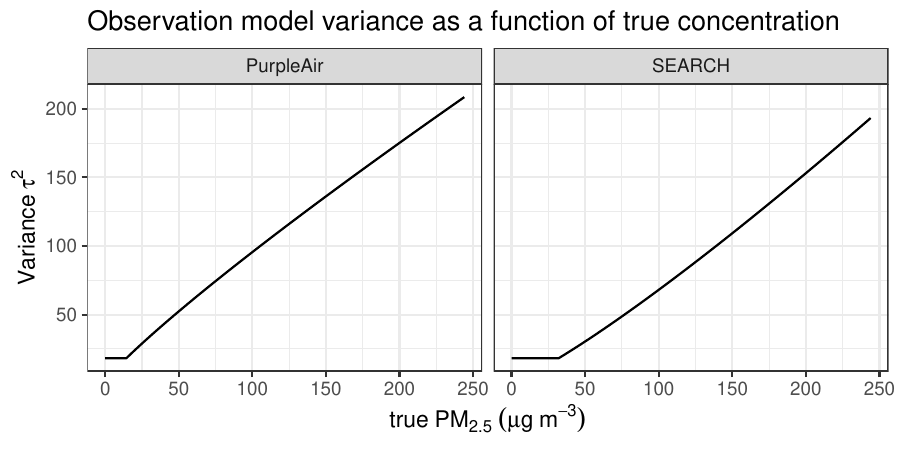}
\includegraphics[width=4.5in]{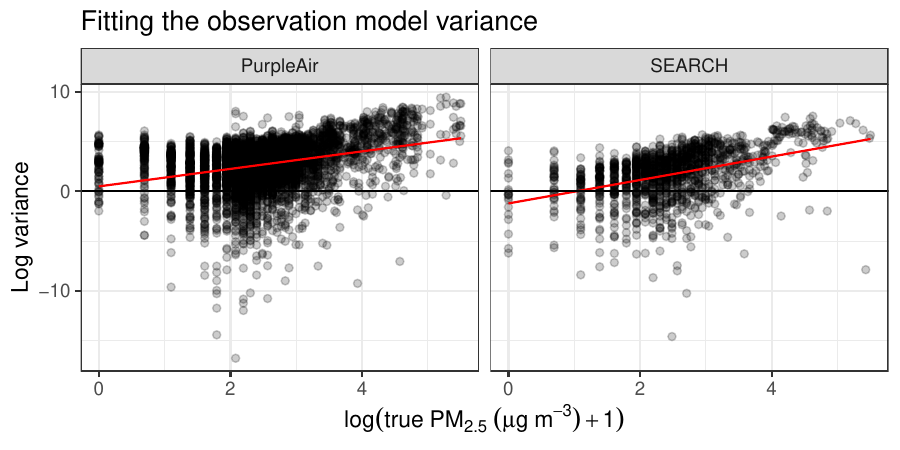}
\caption{ (Top) Variance of the observation model as a function of the true PM$_{2.5}$, for the two low-cost networks. (Bottom) Log variance of the observation model as a function of the log true PM$_{2.5}$, for the two low-cost networks, with points denoting the log squared bias of each observation during June and July 2023. }\label{fig:obs_var} 
\end{figure}

From these results, we conclude that using training data with considerably reduced variability in air pollution levels to estimate the observation model variances seems to lead to faulty extrapolation, especially in the PurpleAir data. Therefore, we use the June and July 2023 data to estimate this variance. Figure \ref{fig:obs_var} (top) shows these variances once again, which are the variances from the top row of Figure \ref{fig:het_comparison}. In Figure \ref{fig:obs_var} (bottom), we see the log squared bias plotted against the log-true PM$_{2.5}$, which illustrates that the selected variance model is a reasonable fit to the error. To ensure that the variances do not become unrealistically small, which would result in the filtering putting considerable weight on the low-cost measurement and little weight on spatial information, we impose a minimum value for the variance, which is the naive $\tau^2$ estimate from fitting the ordinary linear regression model on the 2020-2021 SEARCH data. 

\FloatBarrier
\section{Additional analysis of SEARCH and PurpleAir data}\label{sec:supp_data_analysis}

\begin{figure}[H]
\centering
\includegraphics[width=5.5in]{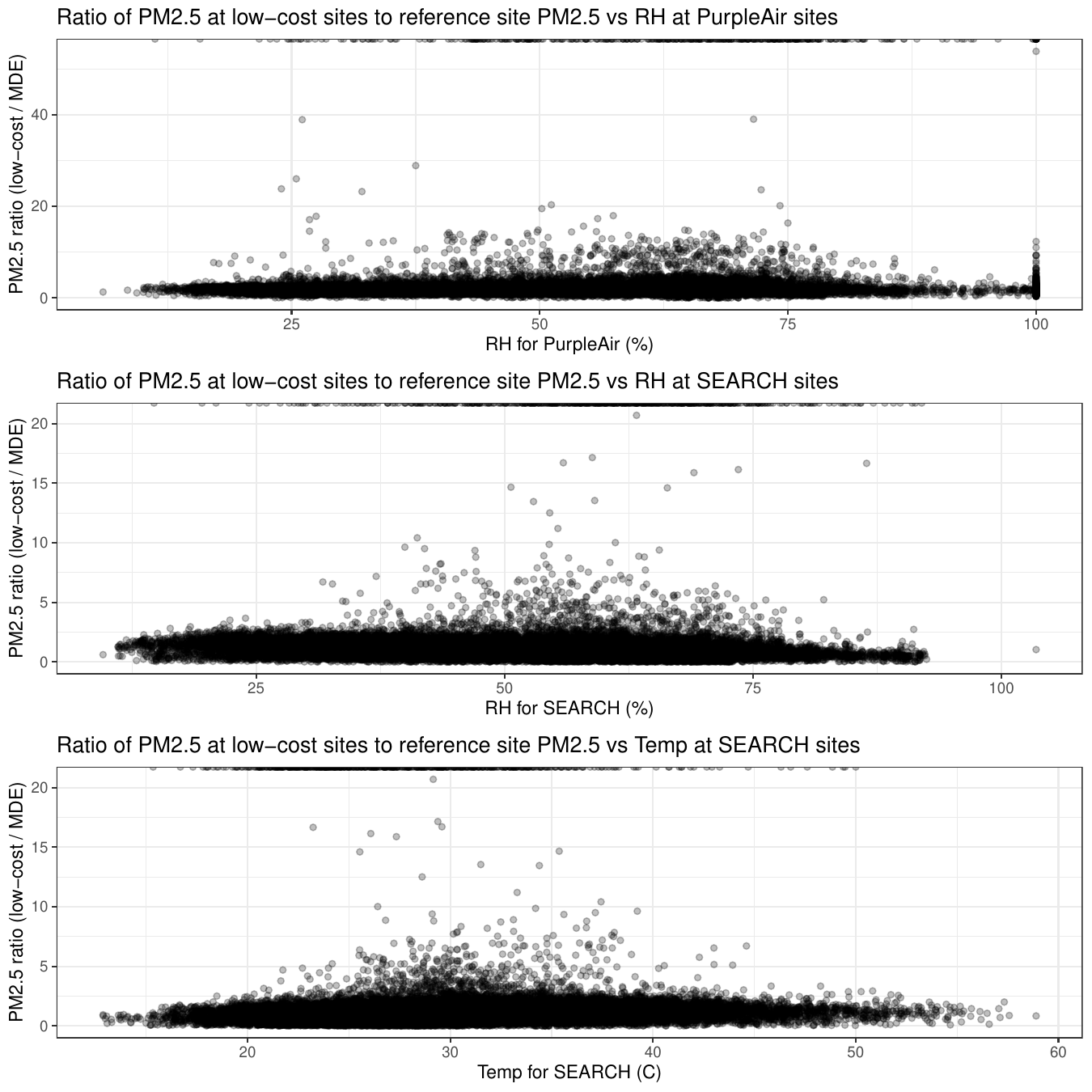}
\caption{\blue{Ratio of PM$_{2.5}$ from low-cost and reference sites vs meteorological variables. Temperature for PurpleAir sites is omitted since it is not used in the observation model.}}\label{fig:scatterplot_ratio}
\end{figure}

\blue{Figure \ref{fig:scatterplot_ratio} shows the ratio of PM$_{2.5}$ as measured by the low-cost networks to the reference device, versus the meteorological variables (RH and Temp) that are used in the observation models. We see that there is a relationship between the meteorological variables and the multiplicative bias, with bias being generally lower at low RH values in the PurpleAir network and higher at low temperatures in the SEARCH network. }

\begin{figure}[h]
\centering
\includegraphics[width=5.4in]{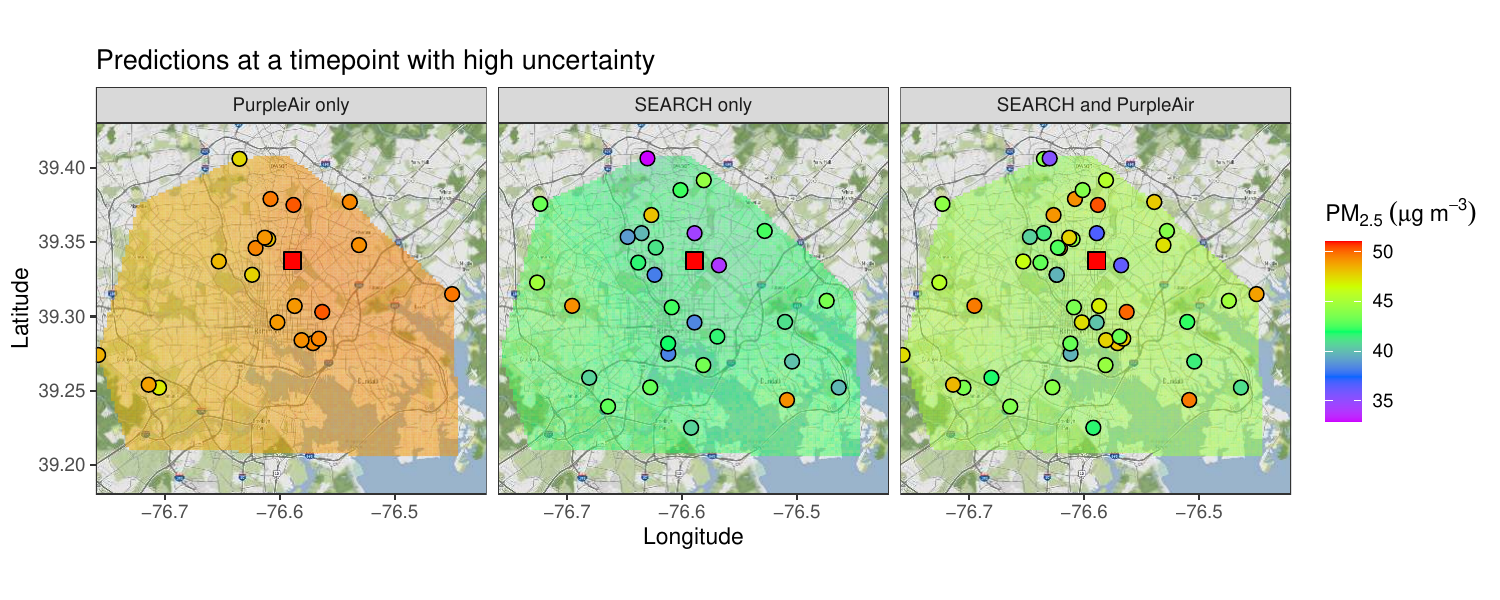}
\includegraphics[width=5.4in]{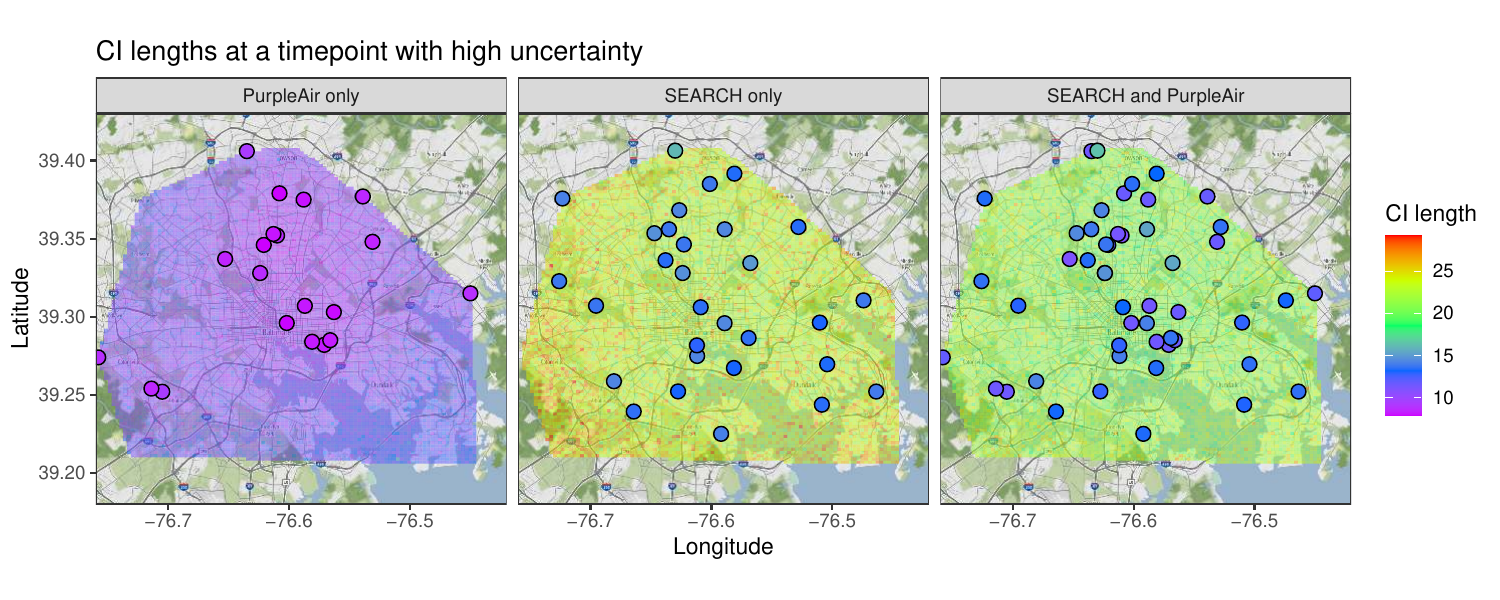}
\caption{Map of the (top) predictions and (bottom) length of the \blue{credible} intervals, on June 30, 2023 at 10am. Circles represent low-cost network sites. The square represents the Lake Montebello site.}\label{fig:increase_uncertainty} 
\end{figure}

Figure \ref{fig:increase_uncertainty} presents one of the timepoints where the length of the \blue{credible} interval increases when two networks are used instead of one (see Figure \ref{fig:CIlengths}). As we see in Figure \ref{fig:increase_uncertainty} (top left), at many of these timepoints, the PurpleAir network predicts a very flat surface, with low uncertainty (Figure \ref{fig:increase_uncertainty} bottom left) at all locations in the city. The SEARCH network better captures variability, as seen from the wider \blue{credible} interval lengths in Figure \ref{fig:increase_uncertainty} bottom middle. 
The variability of the maps produced from the two-network MGPF is also higher than the uncertainty from just using the PurpleAir network, as it incorporates the information from the SEARCH network. 

\begin{figure}[ht]
\centering
\includegraphics[width=4.5in]{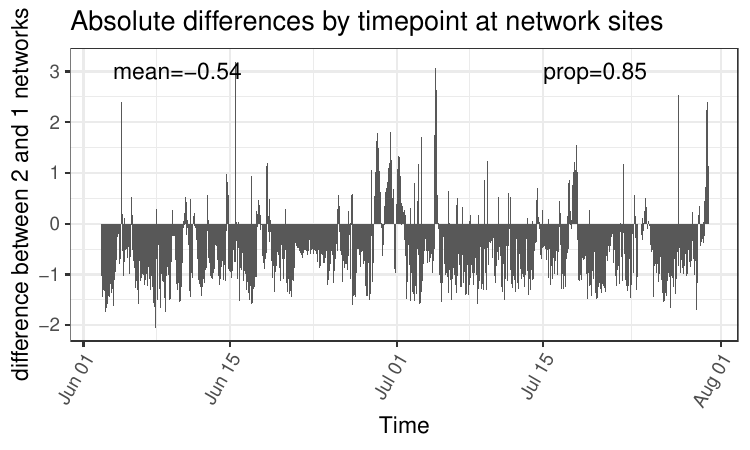}
\caption{Absolute difference in CI lengths, for each timepoint, averaged across the PurpleAir and SEARCH networks. The median of the differences, and the proportion $p$ of differences that are negative, are listed.}\label{fig:supp_CI_network}
\end{figure}

\begin{figure}[ht]
\centering
\includegraphics[width=4in]{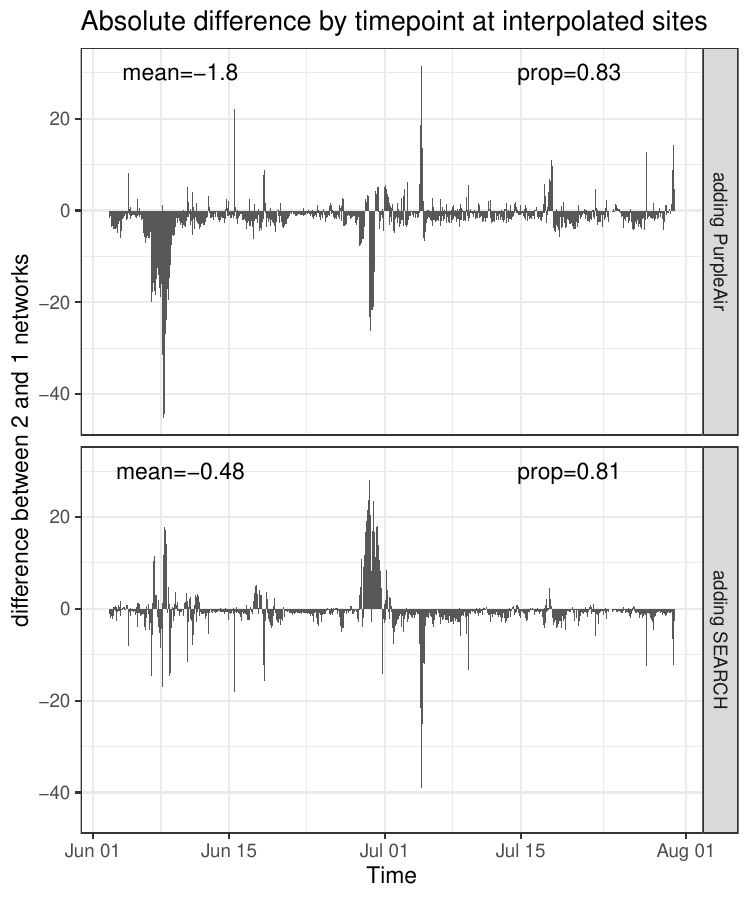}
\caption{Difference in CI lengths averaged across all interpolated sites in Baltimore, using both networks compared to using only one network. The median of the differences, and the proportion $p$ of differences that are negative, are listed.}\label{fig:supp_CI_interpolated} 
\end{figure}

\begin{figure}[t]
\centering
\includegraphics[width=5.5in]{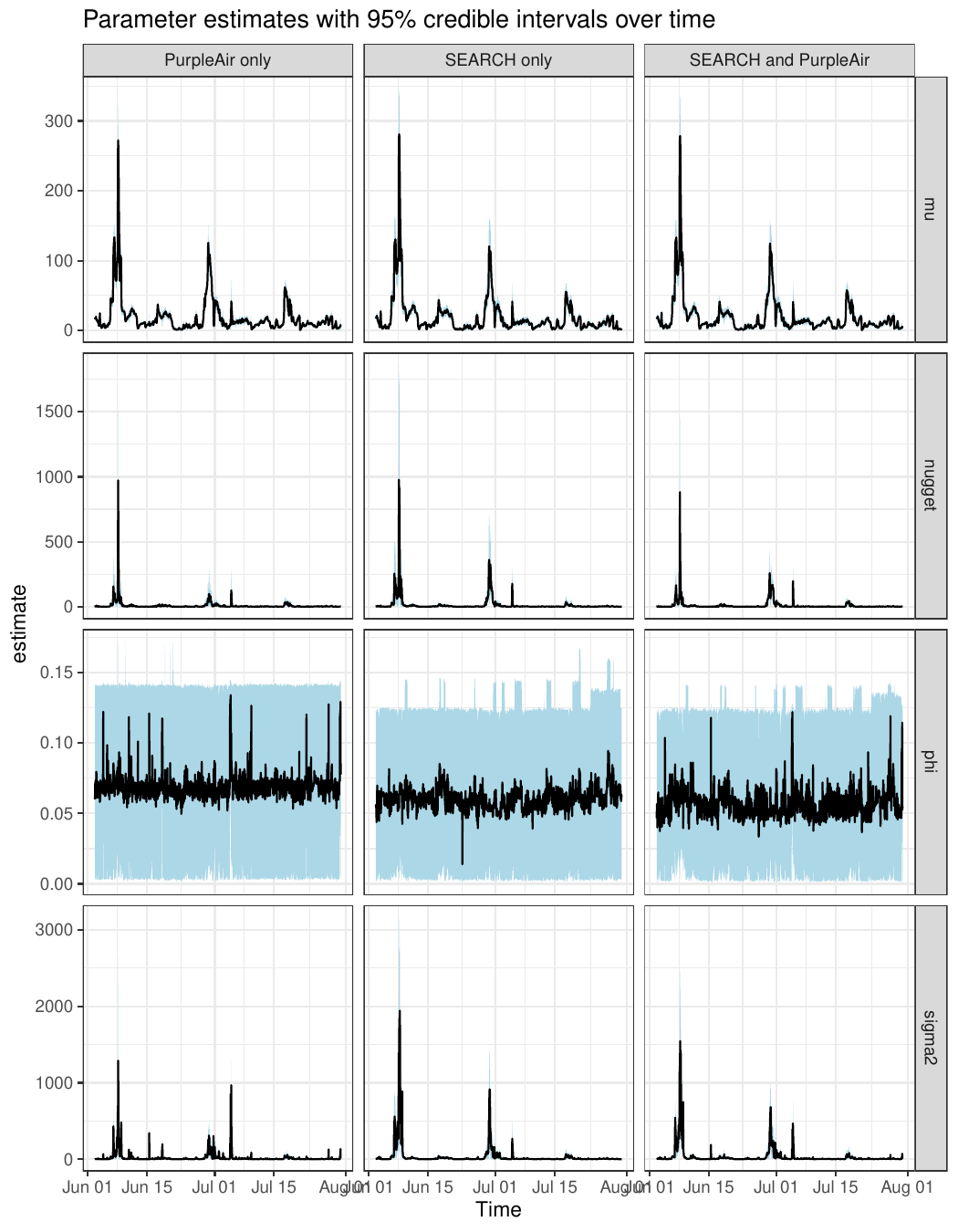}
\caption{Hourly parameter estimates \blue{ (black line) and 95\% credible intervals (blue bands)} from the Gaussian Process model, when applying the MGPF to Baltimore data. }\label{fig:pars} 
\end{figure}

We present some further diagnostics of the analysis of low-cost data in Baltimore. First, the absolute differences in CI lengths at the network sites and the interpolated locations are shown in Figures \ref{fig:supp_CI_network}-\ref{fig:supp_CI_interpolated}. \blue{Also, the parameters of the Gaussian Process part of the MGPF, along with their 95\% credible intervals are shown in Figure \ref{fig:pars}. The spatial variance and nugget terms are both larger when the GP mean is higher, which is during the peak periods. Outside of these periods, there are few timepoints where these variances are very large. For the spatial decay parameter $\phi$, the posterior intervals are wide. This is often noted in Bayesian analysis with Gaussian process as these parameters are not individually identifiable \citep{zhang2004inconsistent}. For all the other parameters, the estimates look as expected and the 95\% intervals are not overly wide, indicating that the selection of bounds and the model fit are adequate. The parameters $\mu_t$, $\sigma^2_t$, and $\sigma_{n,t}^2$ have wider intervals at the timepoints where the true concentrations are higher.}

\begin{figure}[ht]
\centering
\includegraphics[width=6in]{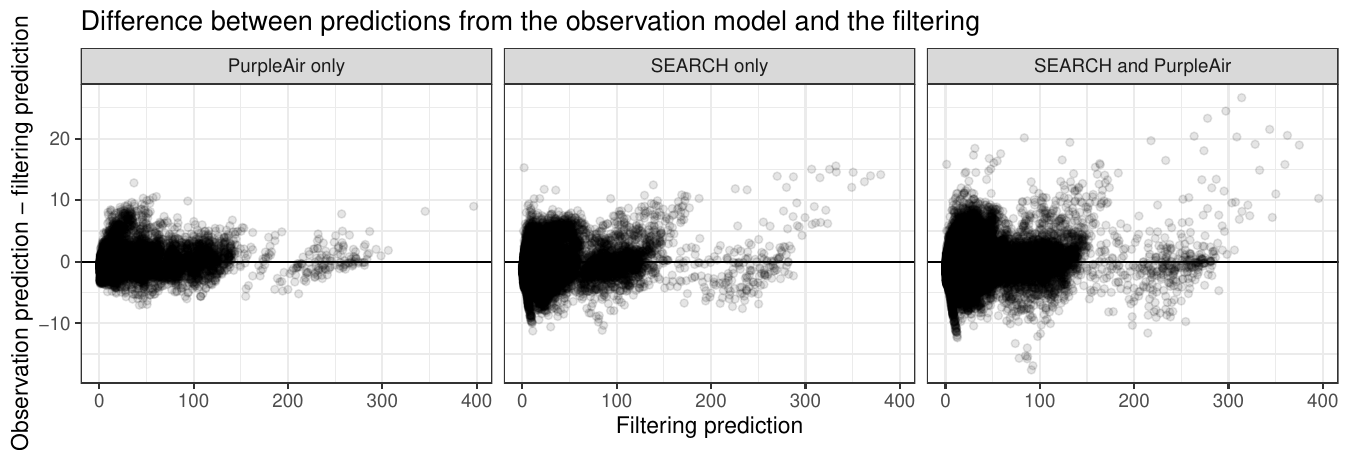}
\caption{Comparison of the predictions from the observation model and from the GPF (left and center panels) and MGPF (right panel) spatial filtering. }\label{fig:inv_filter} 
\end{figure}

We also compare the predictions from the observation model and the spatial filtering methods in Figure \ref{fig:inv_filter}. We see that there is considerable fluctuation between the predictions, which means that there are many timepoints where the spatial model changes the naïve predictions from the observation model considerably, due to the added spatial information. Therefore, we do not see evidence of overly relying on the observation model for predictions. Also, the differences center around 0, indicating that the filtering does not systematically increase or decrease the prediction compared to the observation model. 

\begin{figure}[ht]
\centering
\includegraphics[width=6.25in]{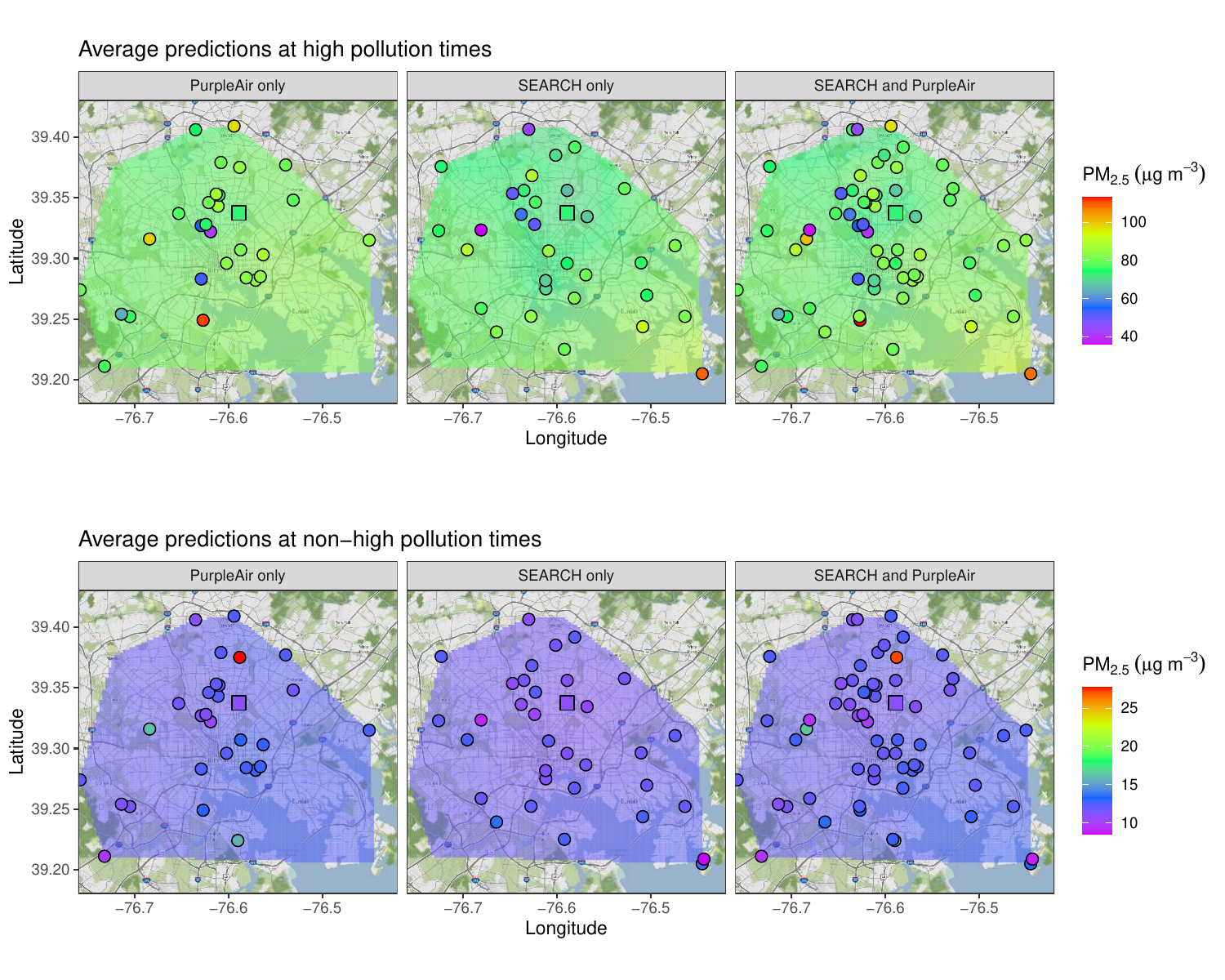}
\caption{Mean of predicted PM$_{2.5}$ in Baltimore (Top) over the days with high pollution (Bottom) over all remaining days. Circles represent low-cost network sites. The square represents the Lake Montebello site. 
}\label{fig:map_avg_supplement} 
\end{figure}

\begin{figure}[ht]
\centering
\includegraphics[width=6.25in]{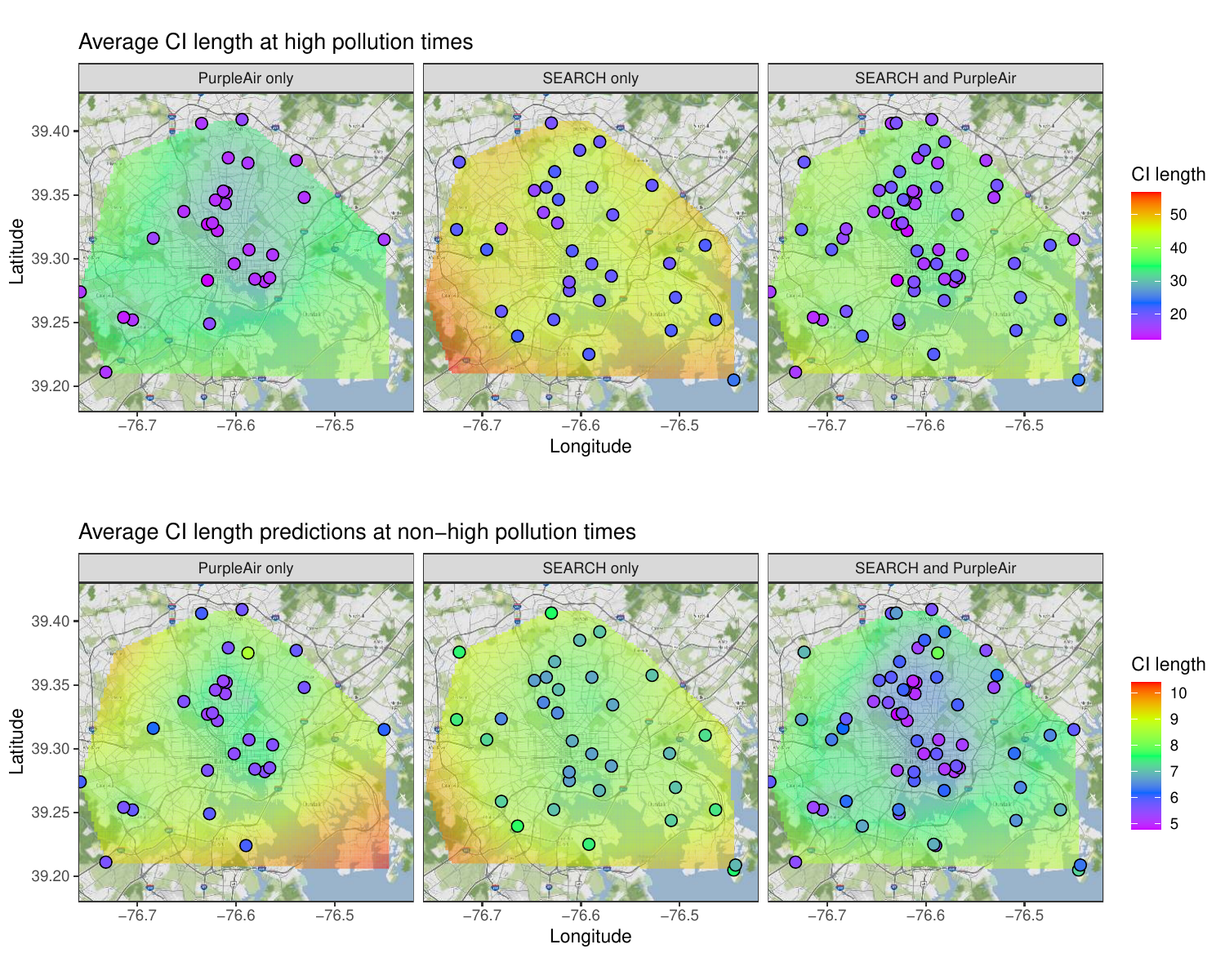}
\caption{Average \blue{credible} interval length of predicted PM$_{2.5}$ in Baltimore (Top) over the days with high pollution (Bottom) over all remaining days. Circles represent low-cost network sites. 
}\label{fig:map_avg_CI_supplement} 
\end{figure}

In Figures \ref{fig:map_avg_supplement}-\ref{fig:map_avg_CI_supplement}, we also present versions of Figures \ref{fig:avg_pred}-\ref{fig:avg_CI} that include the sensor sites as well as the interpolated sites. A few sensors have much higher or lower average 
than the majority of the network, so it is more difficult to see the city-wide pattern when including the sensor sites. Additionally, the CI lengths at the network sites are much smaller than at the interpolated sites because of the presence of the low-cost sensors and the inclusion of a nugget in the model.  

\begin{figure}[ht]
\centering
\includegraphics[width=6in]{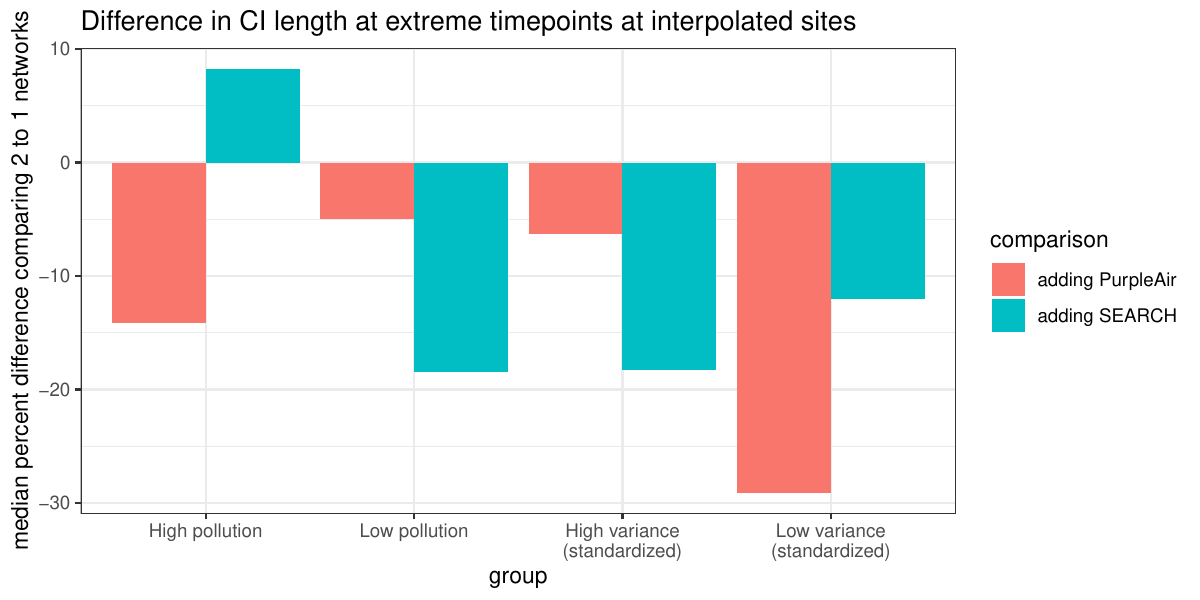}
\caption{Median percent difference in CI lengths averaged across all interpolated sites in Baltimore, focusing on timepoints in the top/bottom 10\% of concentrations at Lake Montebello or network-wide variability in predictions}\label{fig:extreme_CI} 
\end{figure}

We now focus on certain time periods: those with high/low PM$_{2.5}$ at Lake Montebello, and those where PM$_{2.5}$ is most/least variable across the city (Figure \ref{fig:extreme_CI}). Rather than looking at the raw variance, we consider the coefficient of variation, scaling the standard deviation of the predictions at network sites by the mean of the predictions, so that the high variance timepoints don't overlap too much with the high concentration timepoints. We look at the median percent difference in \blue{credible} interval lengths so that the results are not overly influenced by a few very large percent increases. For high concentration timepoints, adding the PurpleAir network to the SEARCH network decreases the CI length across the city by a median of 15\%, while adding the SEARCH network to the PurpleAir network increases CI lengths, since the PurpleAir network tends to be overconfident at high concentrations, since it tends to estimate flatter pollution surfaces. For low pollution timepoints, there is a decrease in uncertainty from adding both networks, but the greatest improvement is obtained from adding the SEARCH network, because PurpleAir alone tends to have wider intervals for these timepoints. For high or low variability timepoints, both networks contribute to a decrease in CI length, with adding the SEARCH network causing greater improvement for high variability timepoints, and adding PurpleAir a greater improvement for low variability timepoints, with up to a 30\% reduction in uncertainty. 

\begin{figure}[ht]
\centering
\includegraphics[width=6.25in]{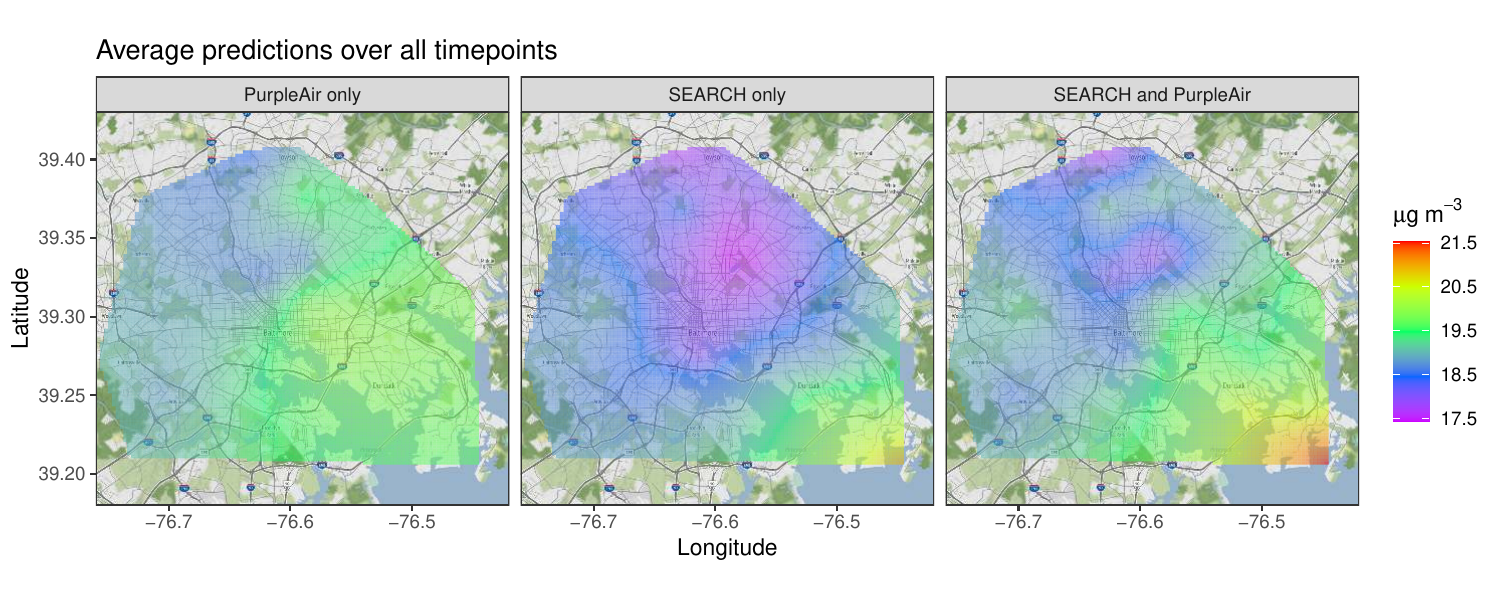}
\caption{Mean of predicted PM$_{2.5}$ in Baltimore over all times in June and July 2023.
}\label{fig:map_avg_alltime_supplement} 
\end{figure}

\begin{figure}[ht]
\centering
\includegraphics[width=6.25in]{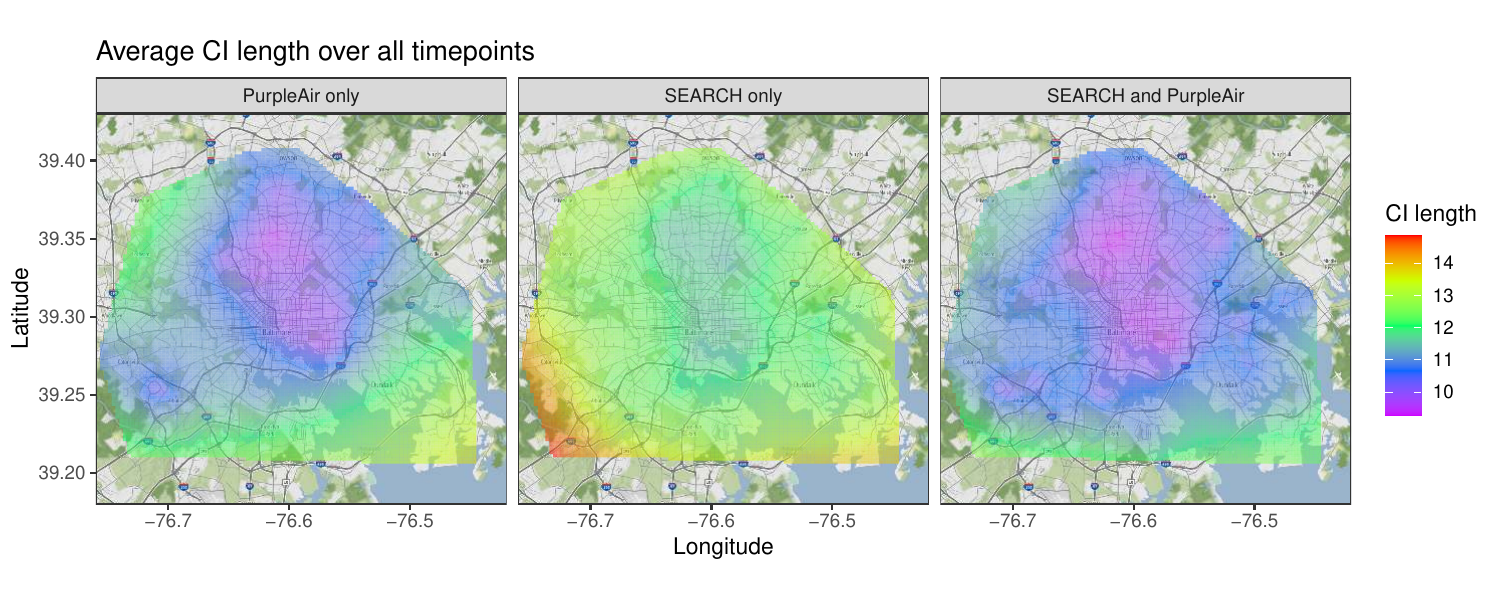}
\caption{Average \blue{credible} interval length of predicted PM$_{2.5}$ in Baltimore over all times in June and July 2023.
}\label{fig:map_length_alltime_supplement} 
\end{figure}

\begin{figure}[t]
\centering
\includegraphics[width=6.25in]{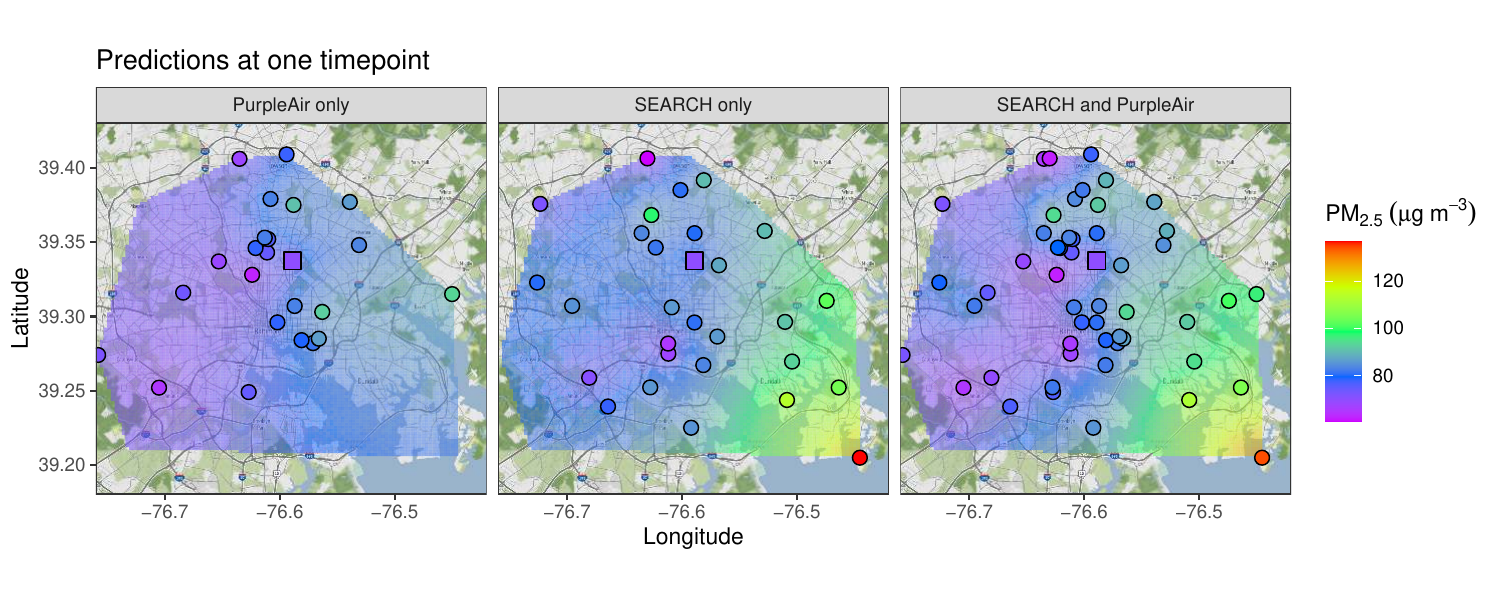}
\includegraphics[width=6.25in]{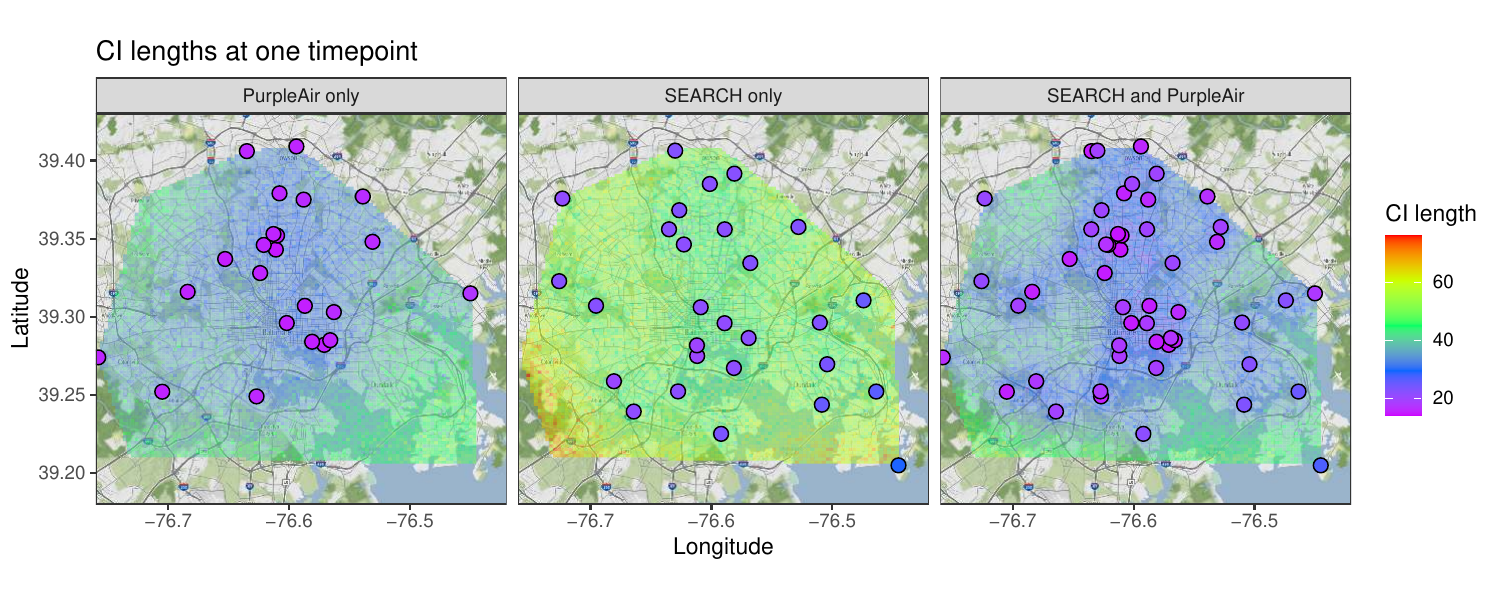}
\caption{(Top) Predictions or (bottom) CI lengths, at one hour using one network or both. Circles represent low-cost network sites. The square represents the Lake Montebello site. The timepoint shown in these maps is June 7, 2023 at 7pm. }\label{fig:preds_CI_124} 
\end{figure}

We present the maps of the average (Figure \ref{fig:map_avg_alltime_supplement}), 
and CI length (Figure \ref{fig:map_length_alltime_supplement}) overall all timepoints. The previous maps separated out high and non-high timepoints. The same areas of the city, namely the east and southeast, have the highest concentrations and CI lengths averaging over all timepoints as when looking at high or non-high timepoints separately. We also present a map of the point estimates and uncertainty at one timepoint, June 7, 2023 at 7pm, in Figure \ref{fig:preds_CI_124}. This timepoint was selected because it is one of the timepoints with high concentrations on average, but not the highest concentrations observed during this period. From these maps, we see that the issues of preferential sampling in the PurpleAir network and high uncertainty in the SEARCH network that were evident when averaging over the whole two month period are also present at individual timepoints.  

\begin{figure}[ht]
\centering
\includegraphics[width=6.25in]{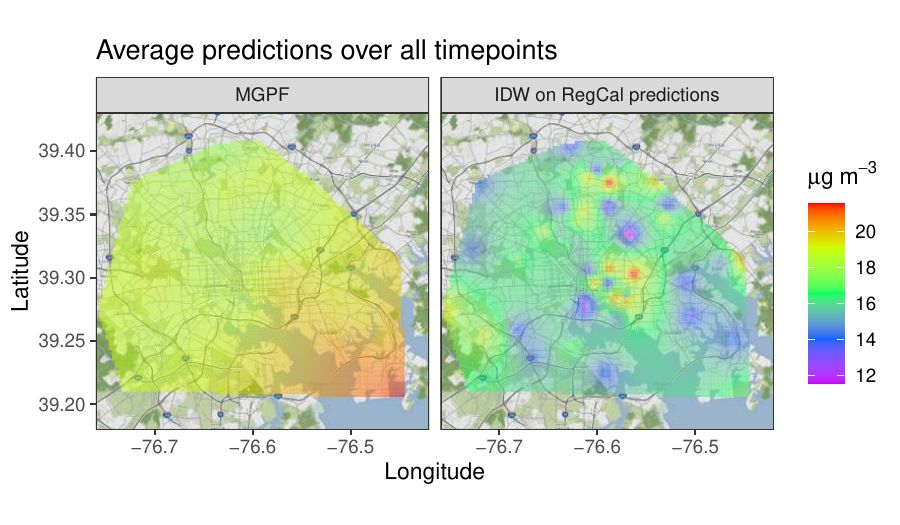}
\caption{Average predicted PM$_{2.5}$ in Baltimore over all times in June and July 2023, including using IDW on individual network calibrations
}\label{fig:idw_avg} 
\end{figure}

\begin{figure}[ht]
\centering
\includegraphics[width=6.25in]{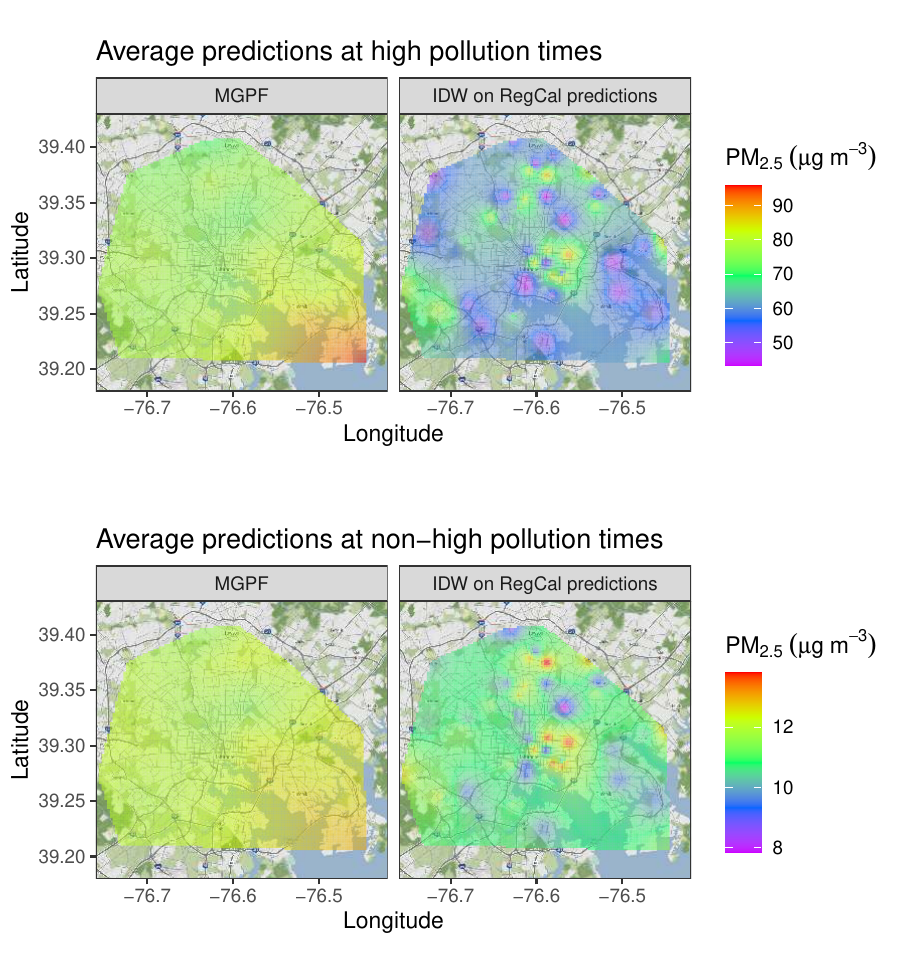}
\caption{Average predicted PM$_{2.5}$ in Baltimore (Top) over the days with high pollution (Bottom) over all remaining days in June and July 2023, including using IDW on individual network calibrations
}\label{fig:idw_avg_high_low} 
\end{figure}

We also show a comparison of the MGPF to individually calibrating each network and then interpolating using inverse distance weighting (IDW) \citep{mueller2004map}. Each network can be calibrated using a regression calibration model (RegCal) \citep{heffernan2023dynamic}, which models the true concentration as a linear function of the low-cost measurement. Each network can have its own calibration equation, and then IDW can interpolate these values to all other locations in the city. The average predictions across the city using this method are shown in Figures \ref{fig:idw_avg} and \ref{fig:idw_avg_high_low}), compared to the MGPF method. Since information is not shared across sites in the regression-calibration-followed-by-IDW method, there is much more local fluctuation across the city, and the locations of many of the sites are visible as local minimums or maximums in PM$_{2.5}$. It is unlikely that where the sensors were placed exactly corresponds to locations which has lowest or highest concentrations, and this is an artifact of the IDW interpolation. 

\begin{figure}[ht]
\centering
\includegraphics[width=6.25in]{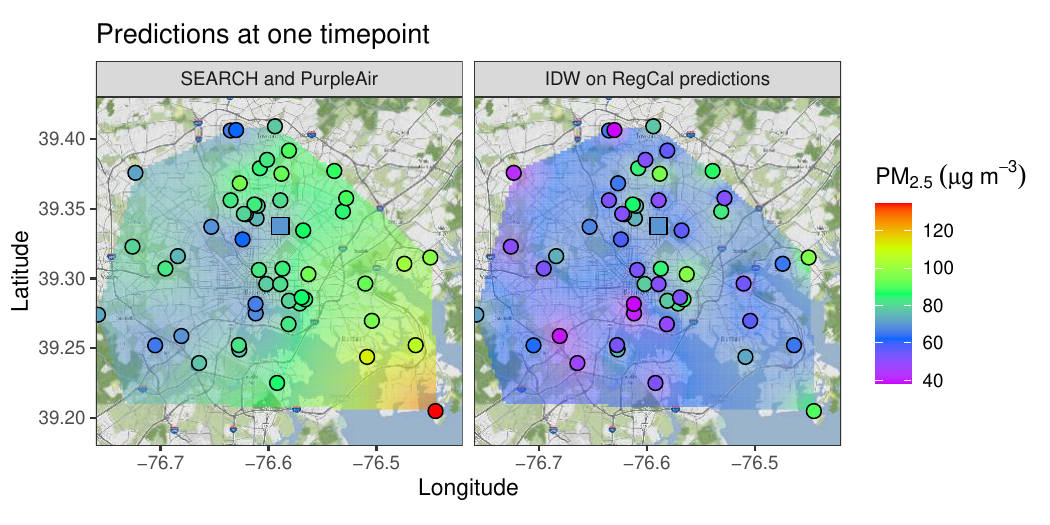}
\caption{Average predicted PM$_{2.5}$ in Baltimore during one timepoint, including using IDW on individual network calibrations
}\label{fig:idw_one} 
\end{figure}

Additionally, \cite{heffernan2023dynamic} showed that these linear models tend to underestimate high concentrations, while the MGPF guards against this. Thus, at high concentrations, the predicted concentrations from doing IDW on RegCal predictions are consistently and considerably lower than using the MGPF (Figure \ref{fig:idw_one}). This results in lower concentrations on average as well (Figure \ref{fig:idw_avg}, Figure \ref{fig:idw_avg_high_low} (top)). 

\begin{figure}[ht]
\centering
\includegraphics[width=5in]{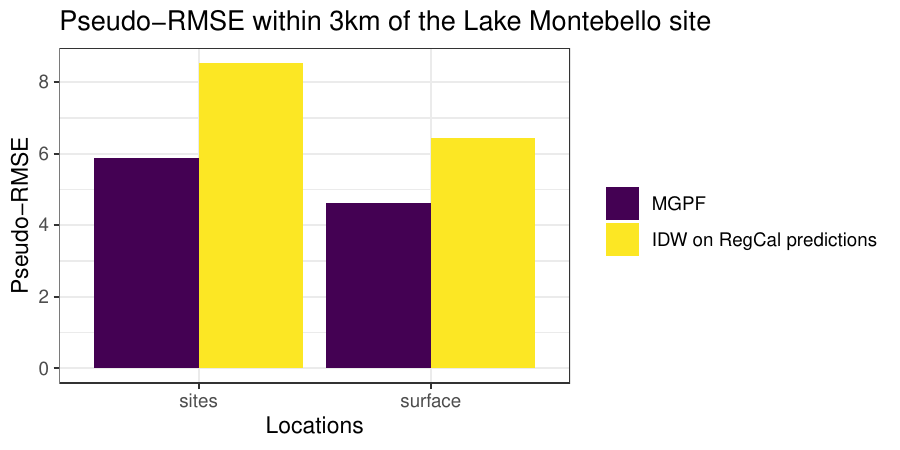}
\caption{Pseudo-RMSE when using filtering on the SEARCH network, PurpleAir network, or both networks together, as well as using IDW on the predictions from regression calibration models (RegCal). Only sites or locations along the interpolated surface within 3km of Lake Montebello are considered. 
}\label{fig:idw_prmse} 
\end{figure}

Finally, to quantitatively assess the accuracy of doing IDW on RegCal predictions on two networks compared to our MGPF method, we calculate a pseudo-RMSE. Since we only know the true PM$_{2.5}$ concentrations at Lake Montebello, we consider locations within a 3km radius of this site, and use the true concentrations measured at Lake Montebello as approximate proxy reference data at these nearby locations. The pseudo-RMSE across the two month period is shown in Figure \ref{fig:idw_prmse}. The figure shows both the RMSE at the network sites, and on the interpolated surface. We see that doing IDW on RegCal has much higher pseudo-RMSE than using the MGPF.

\FloatBarrier
\section{\blue{Missingness in lowcost networks}}\label{sec:supp_missing}

\begin{figure}[ht]
\centering
\includegraphics[width=4.5in]{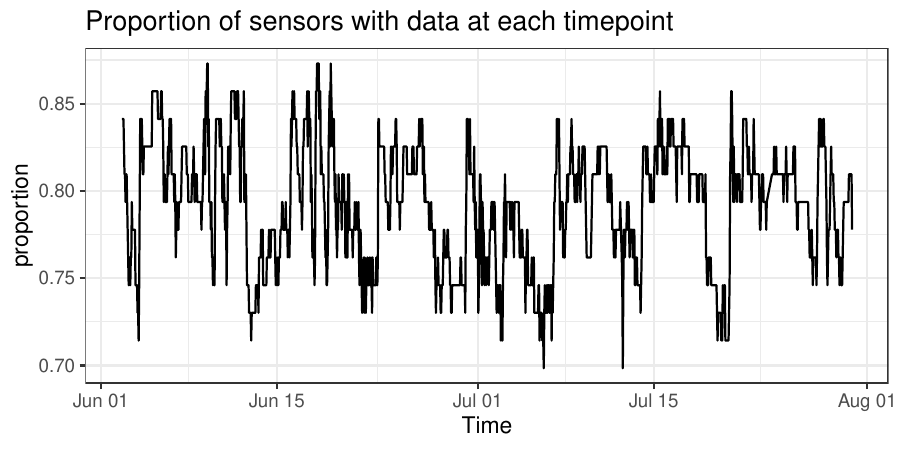}
\caption{
\blue{Proportion of data available from low-cost sensors with data by timepoint.}}\label{fig:missingness_time} 
\end{figure}

\begin{figure}[ht]
\centering
\includegraphics[width=4.5in]{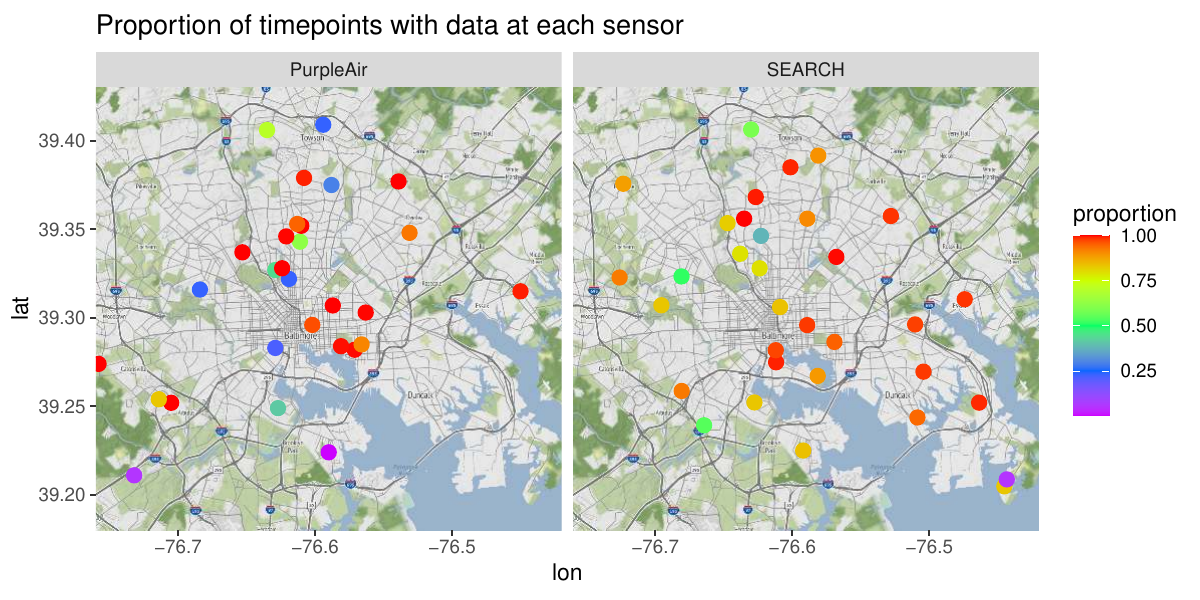}
\caption{\blue{Proportion of timepoints with data by low-cost sensors.}}\label{fig:missingness_space} 
\end{figure}

\blue{MGPF is designed to work with different sets of low-cost sensor sites being active at different timepoints which is often the case. However, we also check the pattern of missingness to be sure that there is no informative missingness. Figure \ref{fig:missingness_time} shows the proportion of lowcost sensors with data at each timepoint and Figure \ref{fig:missingness_space} shows the proportion of timepoints with data at each lowcost sensor. We see that the average amount of data available from each sensor is around $80\%$. There is also no notable large pattern in time or space for the variation in the missingness percentage. }

\begin{figure}[H]
\centering
\includegraphics[trim={0 0 411 20},clip,width=3.5in]{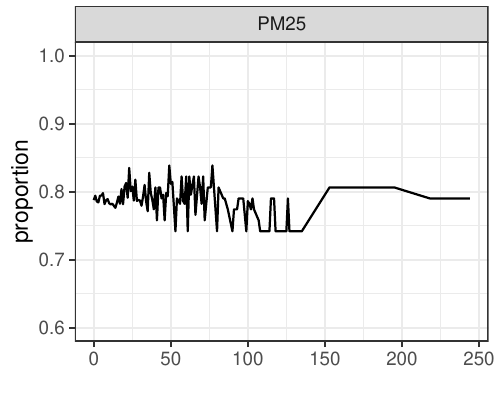}
\caption{\blue{Proportion of sensors with data by PM$_{2.5}$ 
at the reference site}}\label{fig:missingness_by_vars} 
\end{figure}

\blue{Finally, the proportion of sensors with data by the values of the PM$_{2.5}$ concentration 
is shown in Figure \ref{fig:missingness_by_vars}. There is minimal fluctuation in the proportion of sensors with data by values of PM$_{2.5}$, 
(entire range of variation is only $75\%-85\%$). This indicates that the true PM$_{2.5}$ concentration do not  inform of the missingness, which aligns with our understanding of the primary reasons for missingness in these low-cost sensors.}

\newpage
\blue{
\section{General simulations}\label{sec:gensim}

In this Section, we present more details of the main set of simulation experiments and results summarized in Section \ref{sec:gensim_main}. The experiments are in a general setup, detached from the real data application and with multiple model misspecifications with respect to the fitted filtering models. These experiments are used to assess robustness of the proposed multi-network filtering to such misspecifications. 

\subsection{True pollutant surface data generation using stochastic advection diffusion source model}\label{sec:simdgp}

We generate the true concentration surface, not from a Gaussian process model, but using a stochastic advection diffusion source model that broadly mimics air pollution dynamics. This is the first aspect of misspecification, as the filtering methods fit a Gaussian process to the true concentration surface, and does not use any information about the actual advection-diffusion process that generates the data. 

The simulated data on true concentrations describe the evolution of a pollutant field over space and time in a
square region. It starts with a few  sources in the area initially and every few timepoints new sources are introduced. All sources resemble  localized elliptical plumes with
irregular shapes and heavy–tailed strengths, meaning most are weak but a few can be very
large. The centers of these sources are chosen randomly across the region, but with a
greater probability of occurring inside a small subsquare that represents a high
concentration zone. This is a purposeful design choice, as we want to see the impact of preferential sampling on the filtering methods if one of the monitoring networks is underrepresented in this high concentration region. 

Once introduced, each source gradually fades over its lifetime while its
emissions are spread out through diffusion. The entire pollutant field is transported by a
time–varying wind that pushes material mainly eastward with smaller oscillations in the
north–south direction. In addition, a global decay continuously reduces concentrations
throughout the region, mimicking background removal processes. Together, these components
create realistic spatio–temporal patterns with clusters of intense activity in the hotspot
area, long–range transport from the wind, smoothing from diffusion, and steady decline from
decay. A gif file included as a supplement provides video illustration of the concentration surfaces across time. The technical details of the data generation follows.

We set the spatial domain be the square \([L, H]^2\) with \(L=-0.2\) and \(H=1.2\).
Set the lattice size to be
\(
100\,(H-L)+1=141
\)
and define equally spaced grid points
\(
x_i=L+i\,\Delta x,\ i=0,\dots,140
\)
and
\(
y_j=L+j\,\Delta y,\ j=0,\dots,140
\),
with \(\Delta x=\Delta y=(H-L)/140=0.01\).
Time evolves in steps of size \(\Delta t=0.01\) for \(t=1,\dots,T\) with \(T=500\).
Let \(X_{i,j}^t\) denote the concentration at location \((x_i,y_j)\) and time step \(t\). 
To define the discrete operators, note that the five point discrete Laplacian acting on a matrix \(X^t=\{X_{i,j}^t\}\) is
\[
\bigl(\Delta_h X^t\bigr)_{i,j}
=\frac{X_{i+1,j}^t+X_{i-1,j}^t+X_{i,j+1}^t+X_{i,j-1}^t-4X_{i,j}^t}{\Delta x^2}.
\]
The advection field is the time varying wind with components 
\[
v_x(t)=0.2+0.4\,\sin\!\Bigl(\frac{2\pi t}{40}\Bigr), 
\qquad
v_y(t)=0.09+0.2\,\cos\!\Bigl(\frac{2\pi t}{60}\Bigr).
\]
Here $v_x(t)$ is the dominant eastward wind and $v_y(t)$ is the weaker north-south wind. 
Upwind differences are used for the first derivatives
\[
\bigl(D_x X^t\bigr)_{i,j}=
\begin{cases}
\displaystyle \frac{X_{i,j}^t-X_{i-1,j}^t}{\Delta x}, & v_x(t)\ge 0,\\[1.2ex]
\displaystyle \frac{X_{i+1,j}^t-X_{i,j}^t}{\Delta x}, & v_x(t)< 0,
\end{cases}
\qquad
\bigl(D_y X^t\bigr)_{i,j}=
\begin{cases}
\displaystyle \frac{X_{i,j}^t-X_{i,j-1}^t}{\Delta y}, & v_y(t)\ge 0,\\[1.2ex]
\displaystyle \frac{X_{i,j+1}^t-X_{i,j}^t}{\Delta y}, & v_y(t)< 0.
\end{cases}
\]
A reflective boundary is used by copying edge values when computing spatial differences, 
which enforces a zero gradient at the boundary (no flux across the edges).

A uniform removal term acts everywhere in the domain at rate $\lambda=10$.
In other words, if no advection, diffusion, or sources were present, the concentration would decay with a global decay, evolving as
\[
\frac{dX}{dt} = -\lambda X,
\qquad
X(t) = X(0)\,e^{-\lambda t},
\]
so that pollutant levels decay exponentially to zero. This term represents background loss
processes such as chemical breakdown or deposition.

Let $\mathcal S(t)$ denote the collection of all sources that are active at time $t$.  
For each active source $s \in \mathcal S(t)$, let $S_{s,i,j}(t)$ be the contribution of
that source at grid location $(x_i,y_j)$ and time $t$.  
The total source field at $(i,j,t)$ is then
\[
S_{i,j}(t) = \sum_{s \in \mathcal S(t)} S_{s,i,j}(t).
\]
New sources are introduced at \(t=1\) for five initial components and then one additional component every ten steps. A preferential placement rule selects the source center \((x_s,y_s)\) in the cluster window \([0.1,0.3]^2\) with probability \(\rho=0.2\) and otherwise uniformly on \([L,H]^2\). The spatial footprint of a source is an oriented elliptical Gaussian with small smooth irregularity,
\[
B_{s}(x,y)=\exp\!\left(
-\frac{x_\theta^2}{2\sigma_{x,s}^2}-\frac{y_\theta^2}{2\sigma_{y,s}^2}
\right)\,J(x,y),
\]
where
\[
\begin{pmatrix} x_\theta \\ y_\theta \end{pmatrix}
=
\begin{pmatrix} \cos\theta_s & \sin\theta_s \\ -\sin\theta_s & \cos\theta_s \end{pmatrix}
\begin{pmatrix} x-x_s \\ y-y_s \end{pmatrix},
\]
\(\theta_s\) is the orientation angle in radians, drawn uniformly between $-\pi/4$ and $\pi/4$, and
\[
J(x,y)=1+a\sin(kx)\sin(ky)+0.1\,\xi(x,y).
\]
Here \(\xi\) is a smoothed mean zero field, \(a=0.3\), and \(k=3\).
Ellipse scales \(\sigma_{x,s},\sigma_{y,s}\) are drawn from
\(\text{Uniform}(0.06,0.1)\) if the center is inside \([L+\delta,H-\delta]^2\) with $\delta=0.02$, and from \(\text{Uniform}(1.2,2.4)\) if the center is outside the crop window to emulate external (regional) sources, influencing concentrations over larger areas.

Each source has a lifetime \(L_s=\text{round}\!\bigl(1+\sqrt{\sigma_{x,s}^2+\sigma_{y,s}^2}\bigr)\) and fades linearly in time. Its instantaneous field on the grid is
\[
S_{s,i,j}(t)=\begin{cases}
\bigl(1-\frac{t-t_{0,s}}{L_s}\bigr)\,A_s\,B_s(x_i,y_j), & 0\le t-t_{0,s}\le L_s,\\
0, & \text{otherwise,}
\end{cases}
\]
where \(t_{0,s}\) is the start time and \(A_s\) is a heavy tailed random strength. 
Independently for each source,
\[
A_s \sim 
\begin{cases}
1+9\,U_{\text{Beta}(2,5)}, & \text{with prob } 0.95,\\
\min\!\bigl(10\,U^{-1/2},\,100\bigr), & \text{with prob } 0.05\times 0.9,\\
\text{Uniform}(100,300), & \text{with prob } 0.05\times 0.1,
\end{cases}
\]
with \(U\sim \text{Uniform}(0,1)\) and \(U_{\text{Beta}(2,5)}\sim \text{Beta}(2,5)\).

Each source also carries a diffusion coefficient \(D_s\sim \text{Uniform}(0.005,0.01)\).
Let \(\widetilde S_{s,i,j}(t)=S_{s,i,j}(t) / \max_{i,j} S_{s,i,j}(t)\) be the source mask normalized to the unit scale. The diffusion contribution is a masked Laplacian
\[
\bigl(\mathcal D X^t\bigr)_{i,j}
=\sum_{s\in\mathcal S(t)} D_s\,\widetilde S_{s,i,j}(t)\,\bigl(\Delta_h X^t\bigr)_{i,j}.
\]

With explicit Euler integration, the concentration evolves as
\[
X_{i,j}^{t+1}
= X_{i,j}^{t}
+\Delta t\Bigl(
-\,v_x(t)\,\bigl(D_x X^t\bigr)_{i,j}
-\,v_y(t)\,\bigl(D_y X^t\bigr)_{i,j}
+ \bigl(\mathcal D X^t\bigr)_{i,j}
- \lambda\,X_{i,j}^{t}
+ S_{i,j}(t)
\Bigr).
\]

At each step the field is cropped to the unit square \([0,1]^2\) by retaining indices with \(x_i\in[0,1]\) and \(y_j\in[0,1]\).
Let \(v_{\min}(t)\) and \(v_{\max}(t)\) be the minimum and maximum of the cropped field at time \(t\). Values are linearly rescaled to the range \([3,253]\) (a reasonably broad range for PM$_{2.5}$ concentrations) via
\[
\text{value}_{i,j}^t
= 3+\frac{X_{i,j}^t-v_{\min}(t)}{v_{\max}(t)-v_{\min}(t)}\times 250.
\]
The output is the table with columns \(x\), \(y\), \(\text{value}\), and \(t\) obtained by stacking the cropped grids over time.

\begin{figure}[!h]
  \centering
  \begin{subfigure}{0.49\textwidth}
    \centering
    \includegraphics[width=\linewidth, trim={200 0 200 0},clip]{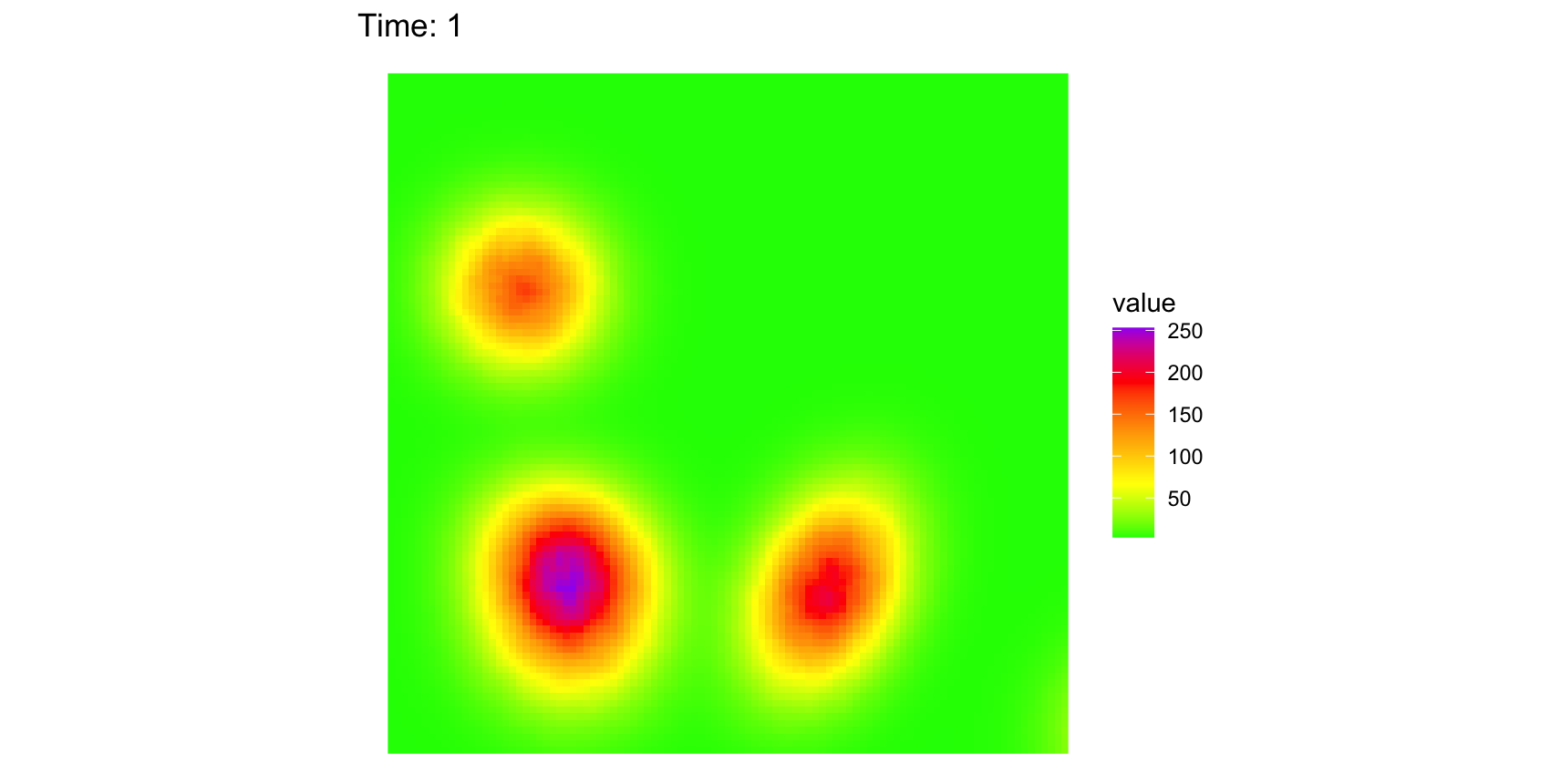}
    \caption{t \(= 1\)}
  \end{subfigure}
  \begin{subfigure}{0.49\textwidth}
    \centering
    \includegraphics[width=\linewidth, trim={200 0 200 0},clip]{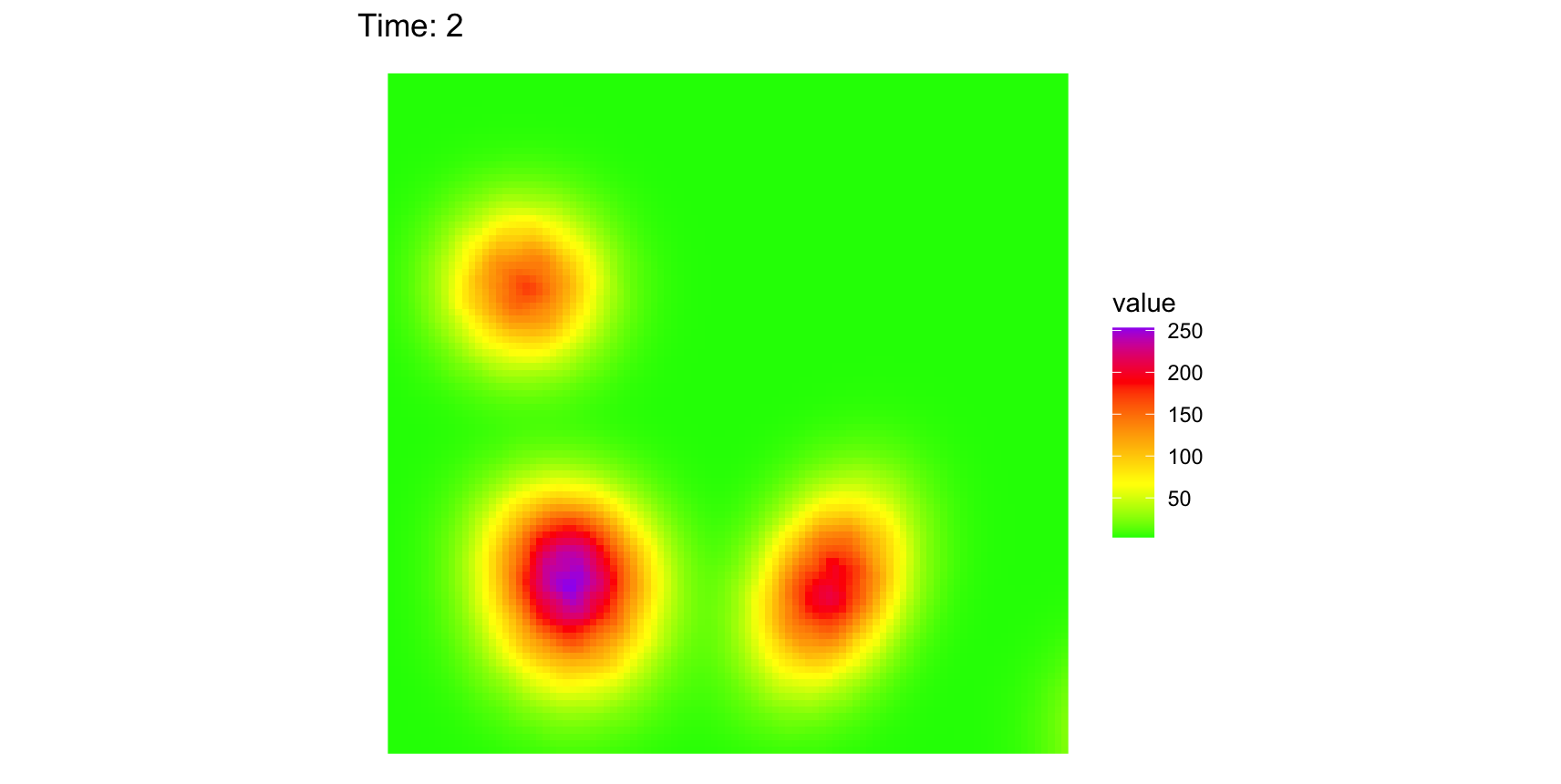}
    \caption{t \(= 2\)}
  \end{subfigure}
  \vspace{0.75em}
  
  \begin{subfigure}{0.49\textwidth}
    \centering
    \includegraphics[width=\linewidth, trim={200 0 200 0},clip]{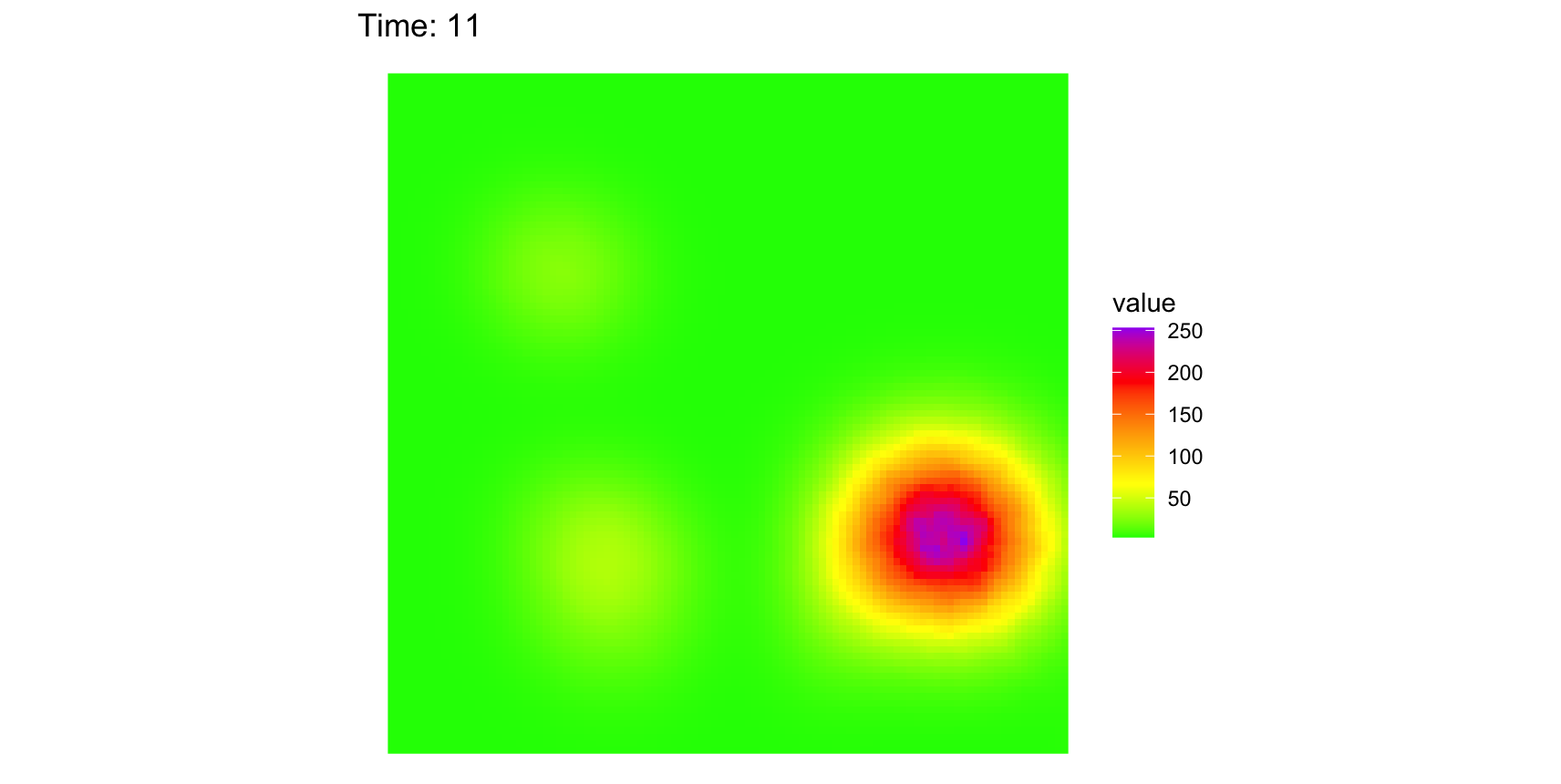}
    \caption{t \(= 11\)}
  \end{subfigure}
  \begin{subfigure}{0.49\textwidth}
    \centering
    \includegraphics[width=\linewidth, trim={200 0 200 0},clip]{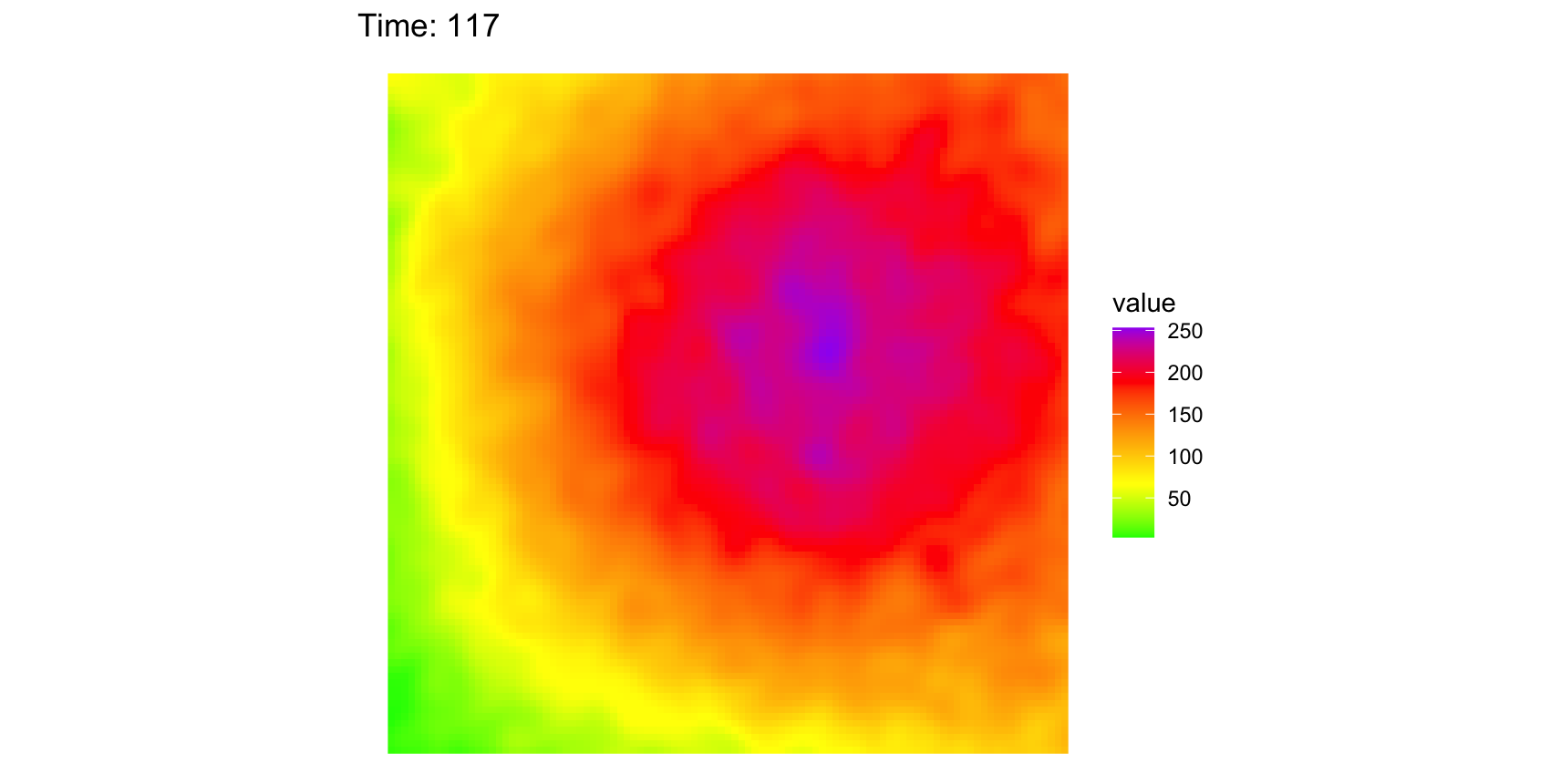}
    \caption{t \(= 117\)}
  \end{subfigure}
  \vspace{0.75em}
  \begin{subfigure}{0.49\textwidth}
    \centering
    \includegraphics[width=\linewidth, trim={200 0 200 0},clip]{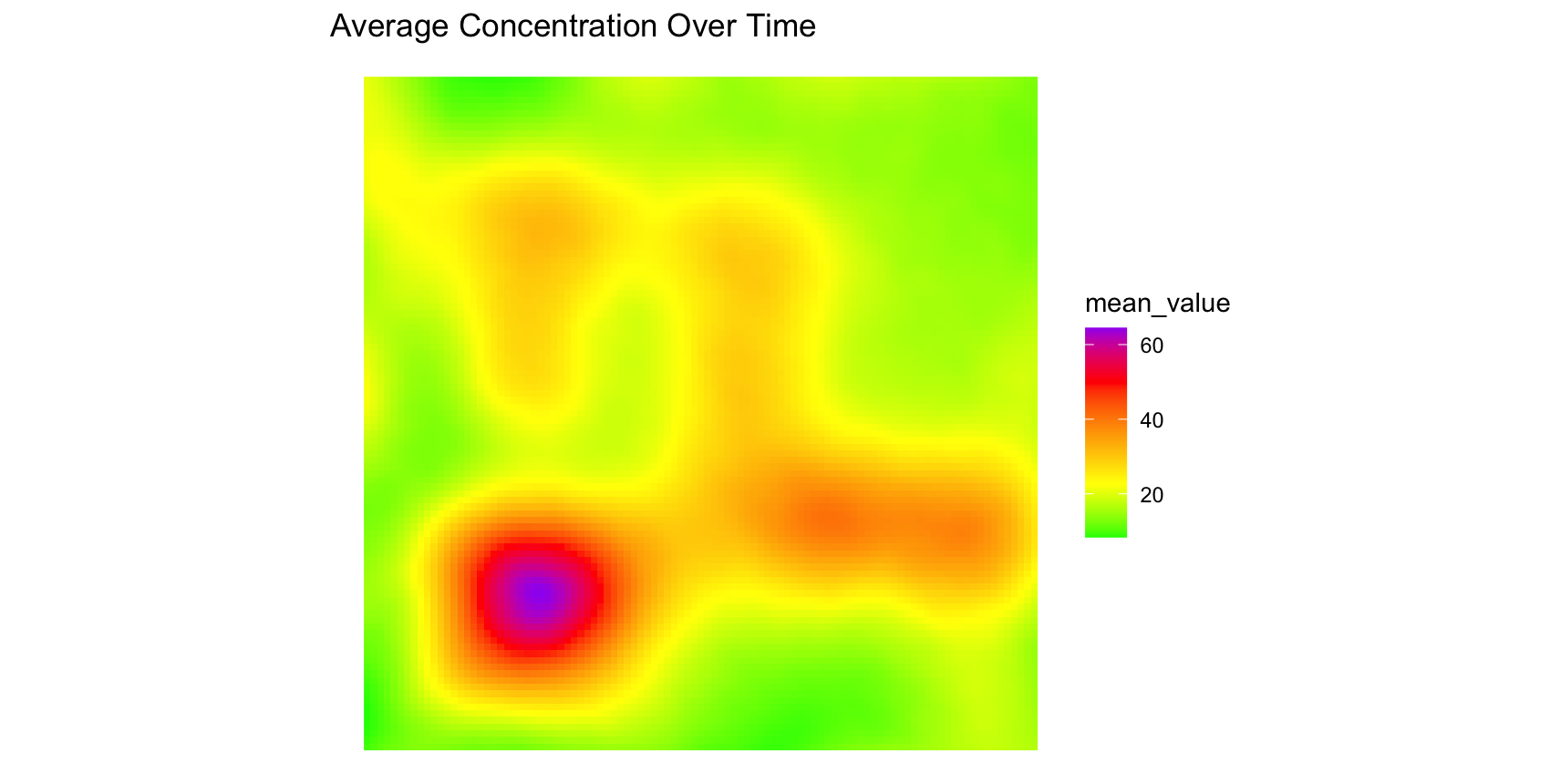}
    \caption{Average over time}
  \end{subfigure}
  \begin{subfigure}{0.49\textwidth}
    \centering
    \includegraphics[width=\linewidth]{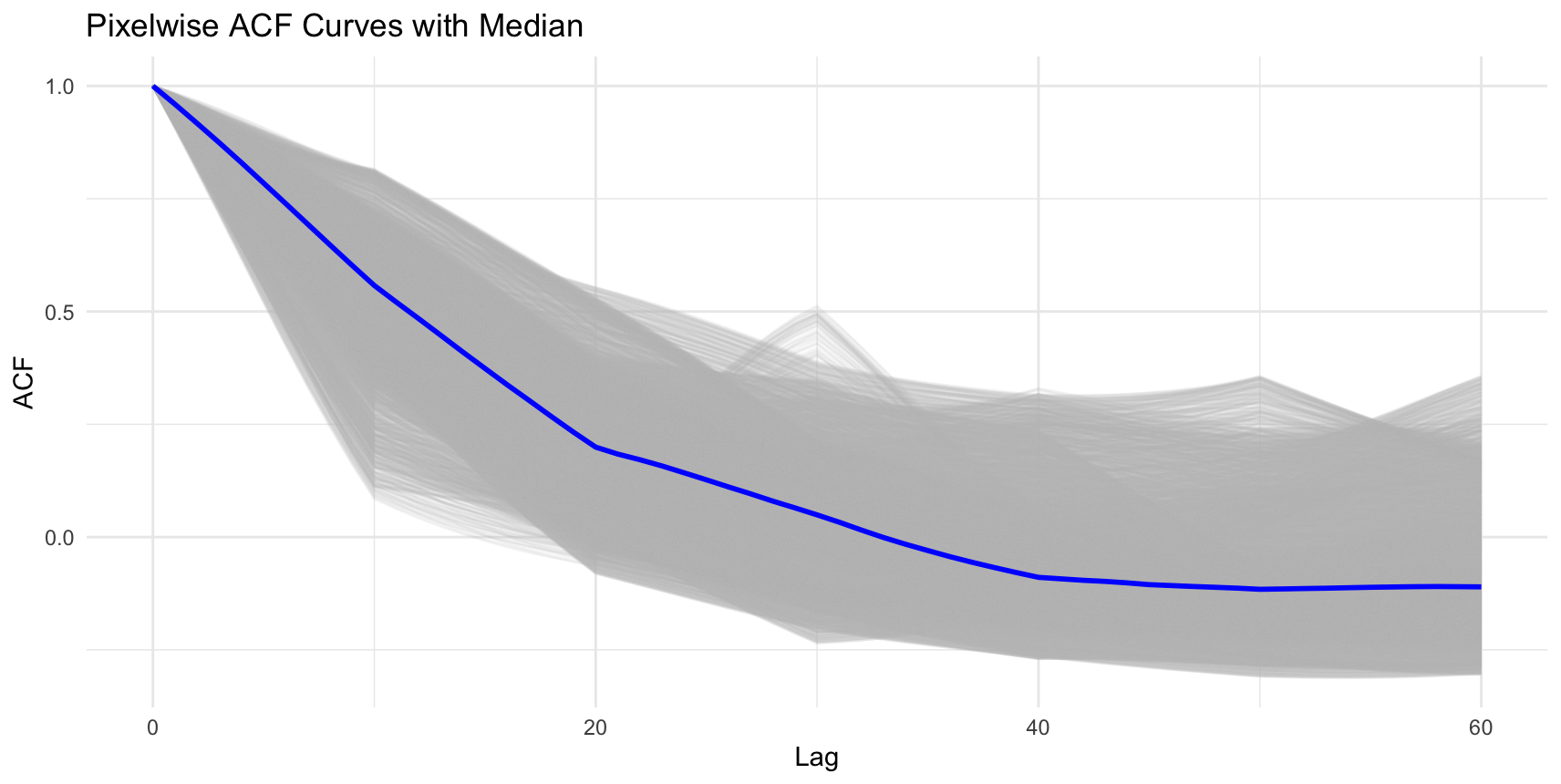}
    \caption{Pixel-wise (grey) and mdian (blue) auto-correlation plots}
  \end{subfigure}

  \caption{\blue{Spatio-temporal dynamics of the simulated true air pollutant surface.}}
  \label{fig:simtrue}
\end{figure}

Figure \ref{fig:simtrue} summarizes key aspects of the simulation of the true surface. 
Panels (a)--(d) display snapshots of the surface at selected time points. At 
\(t=1\) and \(t=2\) the field is characterized by three localized regions of high 
concentration. The snapshots of these two time points are very similar, reflecting the temporal correlation. By \(t=11\) these sources have weakened and a new source has appeared. At \(t=117\) the field 
has diffused into a broad, spatially extensive plume. Panel (e) shows the temporal 
average of concentrations across the simulation, highlighting persistent hot spots 
where emissions remain elevated. Panel (f) presents pixel-wise autocorrelation 
functions (grey) along with their median (blue), confirming strong short-term 
dependence and gradual decay in temporal autocorrelation with increasing lag. 

As we do not explicitly model temporal correlation in our GP filter, the strong temporal correlation in the true data represents another aspect of misspecification between the data generation process and the fitted model. 

\subsection{Generation of low-cost observations}\label{sec:simlcs} 

For each timepoint, we construct a synthetic dataset to mimic two low cost air quality sensor networks operating in the area, observing the same underlying spatio-temporal pollutant field but with some bias and noise, specific to each network. 
Network 1 consists of \(n_1=100\) sensors placed uniformly over the full unit square \([0,1]^2\), representing a spatially balanced design. Network 2 consists of \(n_2=100\) sensors placed uniformly over \([0,1]^2\) but restricted to the complement of the lower-left quadrant \([0,0.5]^2\), representing a preferential design that avoids part of the region which tends to have higher concentration. 

At each discrete time \(t\) the latent true pollutant surface is \(X(\cdot,t)\), defined on \([0,1]^2\). Recall from the previous section that this field is available on a grid and is interpolated to the sensor coordinates using smooth bivariate interpolation on a \(100\times100\) lattice. This gives the true concentration value \(X(s_i,t)\) at each sensor location \(s_i\).

Observed low cost measurements are generated from a calibration and noise model. For a sensor in network \(k\in\{1,2\}\), the measurement at time \(t\) is
\[
Y_i(t) = a_k + b_k\,X(s_i,t) + \varepsilon_i(t), \qquad \varepsilon_i(t)\sim N(0,\sigma_k^2),
\]
with independent errors across sensors and times. The calibration parameters are set to \(a=(1,2)\), \(b=(1.2,1.5)\), and \(\sigma=(2,1)\), so that the two networks differ in offset, gain, and noise level.

To provide colocated reference sites, we identify for each network the sensor closest to the network’s centroid. Specifically, for network \(k\) we compute the centroid \((\bar x_k,\bar y_k)\) of its coordinates and select the sensor minimizing squared Euclidean distance to this centroid. Figure \ref{fig:networks} presents the network locations along with the colocated sites.  

\begin{figure}
    \centering
    \includegraphics[width=0.99\linewidth]{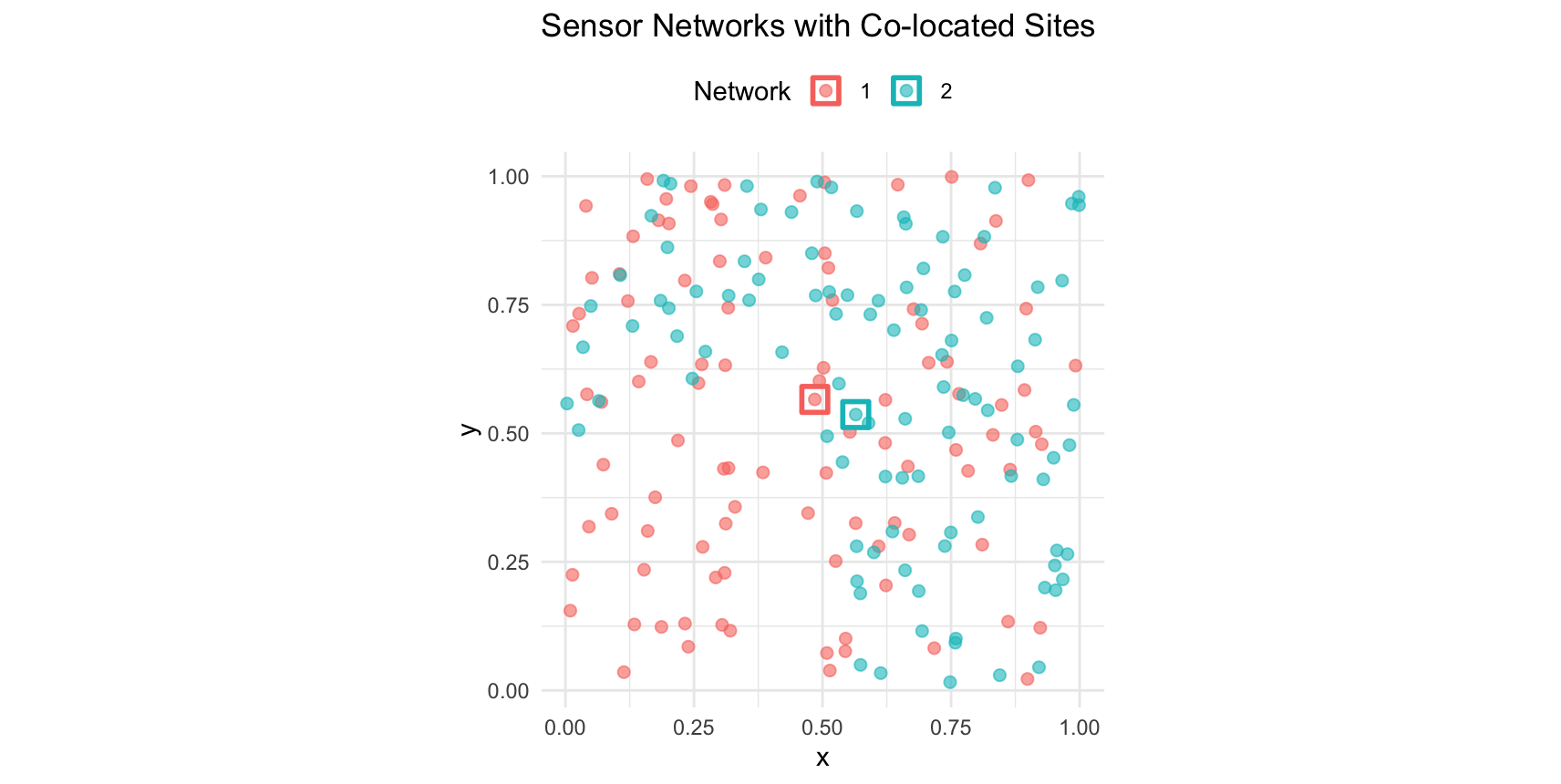}
    \caption{\blue{Two low-cost networks along with their co-located sites (outlined by boxes).}}
    \label{fig:networks}
\end{figure}

\subsection{Results}\label{sec:simres}
We use the first 400 timepoints to estimate the observation model for each network based on their respective collocated sites, and then use this estimated observation model in the filtering part to predict the true pollutant surface for time-points 401 to 500. We consider three filtering methods -- two single network GP filter (each using one of the two low-cost network data), and the multi-network GP filter that uses data from both networks. We compare the methods in terms of root mean squared error (RMSE), mean absolute error (MAE), continuously ranked probability score (CRPS), 95\% interval coverage, 95\% interval width, and 95\% interval score. Among these metrics, RMSE and MAE evaluate the point estimation, the remaining evaluate distributional/interval estimation with CRPS and interval score being proper scoring rules \citep{gneiting2007strictly}.

\begin{figure}
    \centering
    \includegraphics[width=0.99\linewidth]{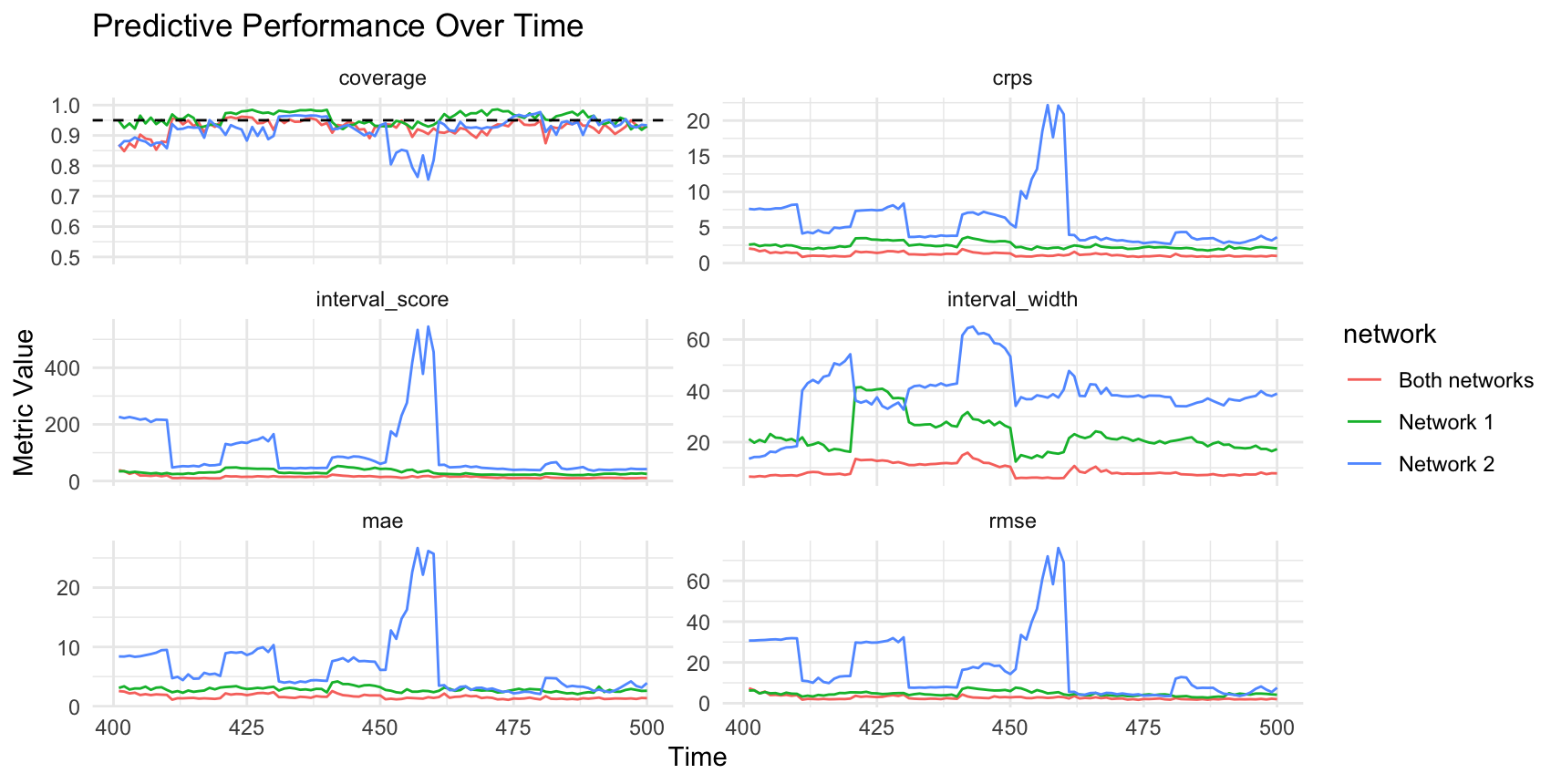}
    \caption{\blue{Predictive performance metrics over time for GP filter models based on Network 1 (green), 
Network 2 (blue), and both networks jointly, i.e., the proposed multi-network GP filter (red). Subfigures display (top left) coverage, 
(top right) continuous ranked probability score (CRPS), (middle left) interval score, 
(middle right) interval width, (bottom left) mean absolute error (MAE), and 
(bottom right) root mean squared error (RMSE).}}
    \label{fig:restime}
\end{figure}

Figure~\ref{fig:restime} presents the time trends in these metrics for the three modeling strategies. The coverage plots show that all approaches generally achieve close to nominal 95\% coverage, though Network 2 alone occasionally drops well below target. The CRPS, interval score, MAE, and RMSE panels reveal that predictions based solely on Network 2 exhibit substantially larger errors and instability, with spikes indicating periods of poor predictive accuracy and inflated uncertainty quantification. In contrast, models based on Network 1 or both networks maintain consistently low error values and stable uncertainty, with the joint model typically performing best. The interval width plots further emphasize that GP filter with Network 2 produces much wider predictive intervals, reflecting its lower information quality due to restricted spatial coverage. However, this plot also shows that GP filter using Network 1 also has considerably higher uncertainty due to using less information than available. The multi-network GP filter yield narrower, more precise intervals. 
Overall, the results highlight the benefits of combining both networks: the joint model leverages complementary spatial information to improve predictive accuracy by mitigating bias from a network (Network 2) having preferential spatial distribution of the sensors, and improve uncertainty quantification over single-network GP filter using either of the networks.

As seen from Figure \ref{fig:restime}, the performance of GP filter using network 2 only is particularly bad at certain stretches of time, characterized by drastic drop in coverage and increase in the accuracy metrics. Examples includes the timeperiod between timepoints 401 to 410 and again between timepoints 450-460. 
\begin{figure}[h!]
    \centering
    \includegraphics[width=0.99\linewidth,trim={0 150 0 150},clip]{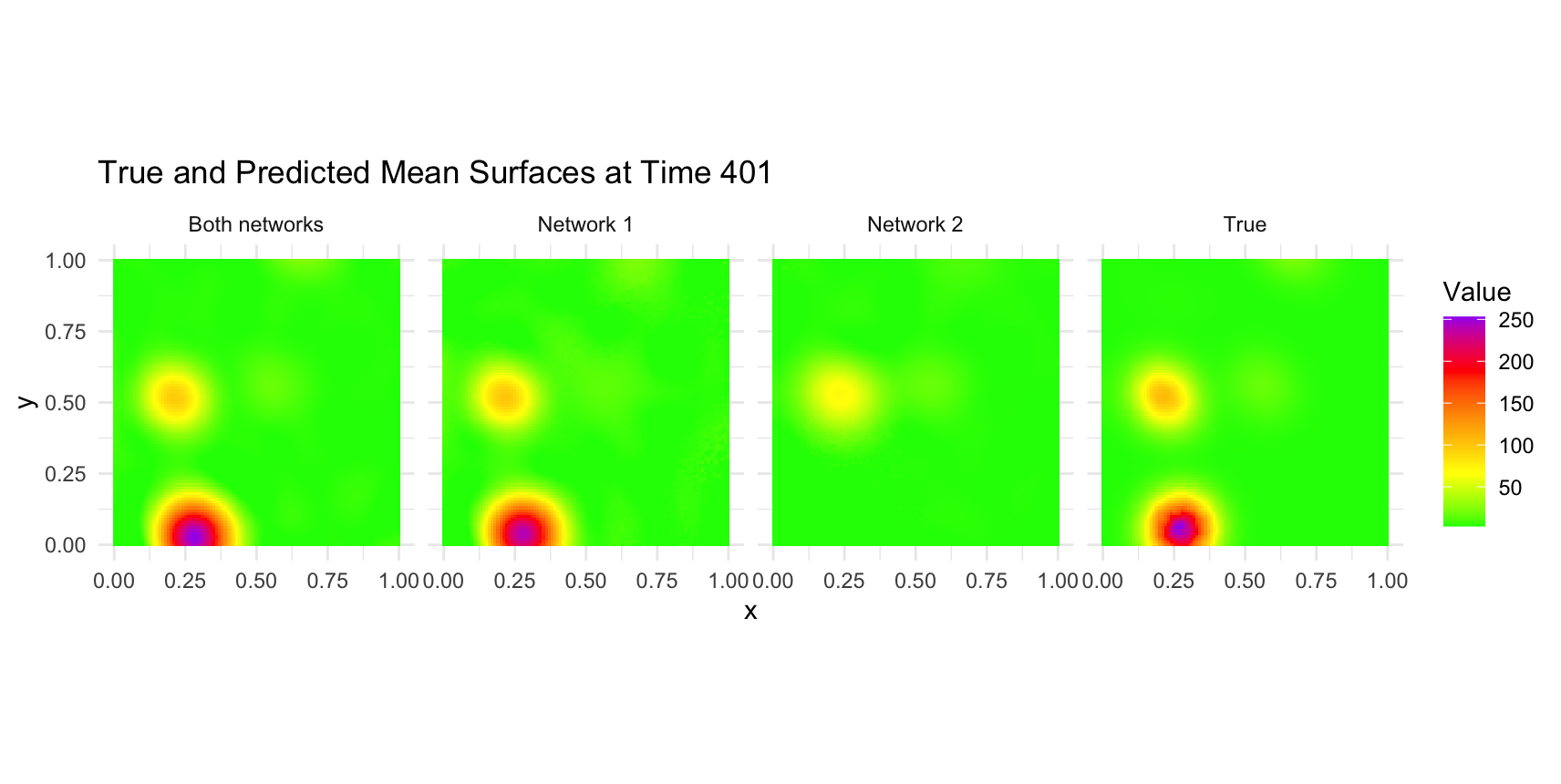}
    \includegraphics[width=0.99\linewidth,trim={0 150 0 150},clip]{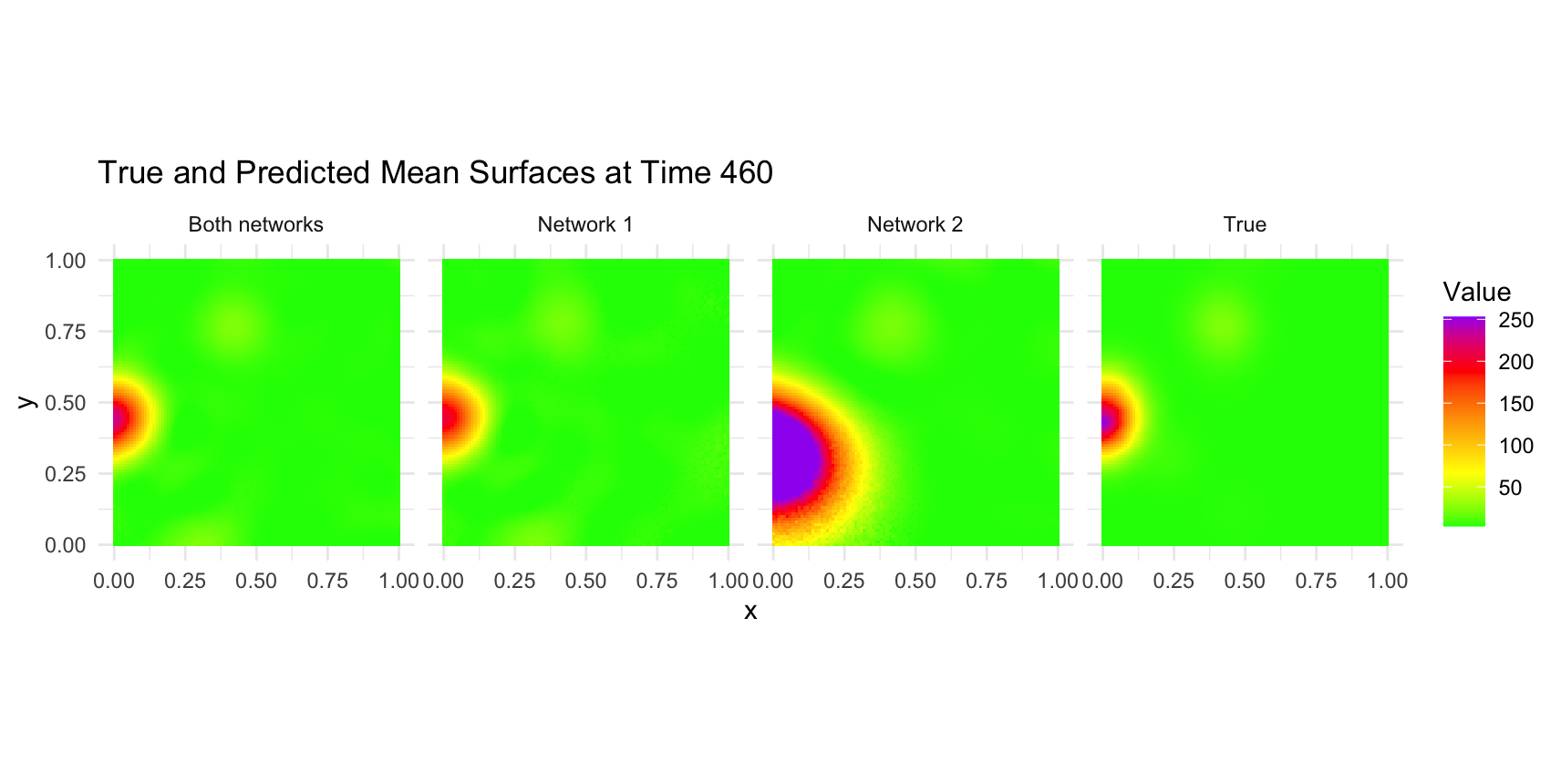}
    \caption{\blue{Predicted mean concentration surfaces compared with the true surface at 
two time points. Top row: results at time \(t=401\), showing predictions from the 
joint model using both networks, the Network 1 model, the Network 2 model, and the 
true surface. Bottom row: results at time \(t=460\), with the same layout.}}
    \label{fig:resexamples}
\end{figure}
Figure~\ref{fig:resexamples} presents true pollutant surfaces and predicted mean concentration surfaces for one day from each period. 
The top row shows results at time \(t=401\). The true surface contains two clear concentration peaks, one near the bottom boundary and another toward the left interior. The multi-network GP filter and the GP filter using Network 1  both recover these features reasonably well, while the Network 2 model completely misses the bottom peak due to lack of spatial coverage in that area. 

\begin{figure}[h!]
    \centering
    \includegraphics[width=0.99\linewidth]{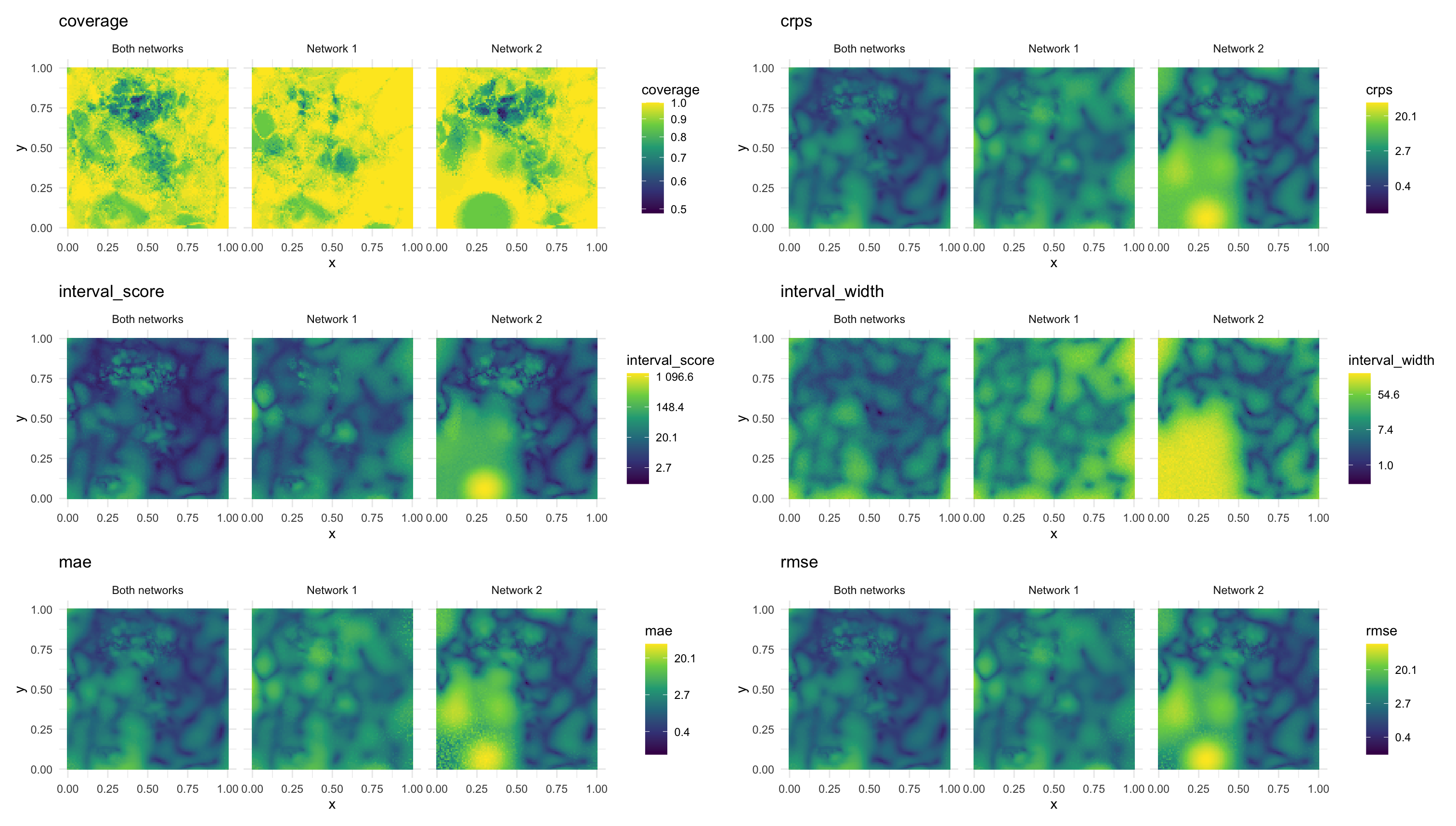}
    \caption{\blue{Spatial maps of predictive performance metrics for GP filter using both networks, Network 1, and Network 2. Shown are (top left) coverage, (top right) CRPS, (middle left) interval score, (middle right) interval width, (bottom left) MAE, and (bottom right) RMSE.}}
    \label{fig:resspatial}
\end{figure}

The bottom row shows results at time \(t=460\). At this time the true field is characterized by a single strong source along the left boundary. The multi-network GP filter and the GP filter using Network 1 again capture this structure, although with some differences in intensity with MGPF capturing the truth better. GP filter with Network 2 produces a distorted pattern with exaggerated spread. This is because due to lack of spatial coverage of Network 2 in the bottom left square, GP filter using Network 2 only observes a small part of the peak and has to rely  on extrapolation to predict the rest of the peak in the area where it does not have coverage. This results in substantial overestimation. 
Overall, these two snapshots nicely illustrate how preferential spatial coverage of a network can lead to both under- and over-estimation of peaks, and how the multi-network approach mitigates this. 

We also look at the geographical patterns in the performance of the methods over space, by averaging the metrics over time. Figure~\ref{fig:resspatial} displays spatial maps of the predictive performance metrics. 
These maps provide a spatially resolved view of where prediction quality differs across the domain. We see that all the  metrics for GP filter with Network 2 are poor in the bottom left of the area where the network has no coverage. We also see that the Multi-network GP filter is consistently the best across all metrics, generally producing lowest errors and best uncertainty quantification throughout the maps.  

\begin{figure}[h]
    \centering
    \includegraphics[trim={0 0 0 90}, clip, width=0.4\linewidth]{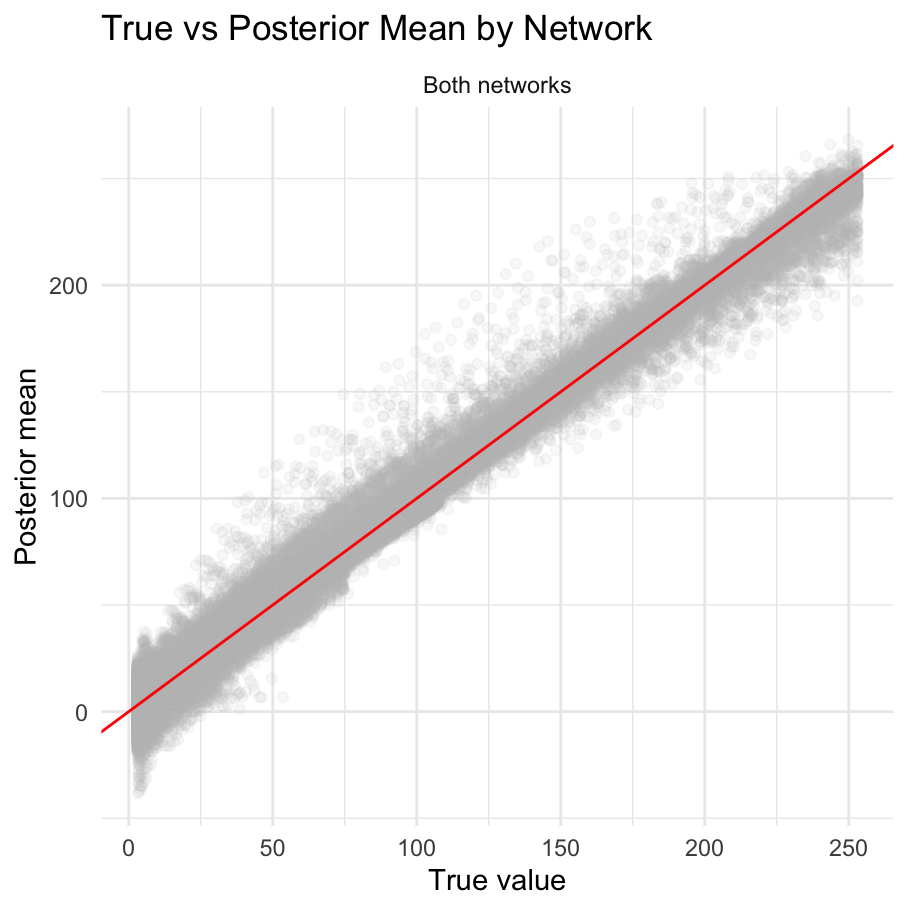}
    \caption{\blue{Scatterplot of the true concentration values and the mean predicted concentrations from MGPF for all time-points and all spatial locations.}}
    \label{fig:scatter}
\end{figure}

Finally, we look at the overall fit quality of the MGPF model, which is clearlt superior to the two single-network GP filters. Figure \ref{fig:scatter} presents the scatterplot the true concentration values and the mean predicted concentrations from MGPF for all time-points and all spatial locations. We see very strong alignment of the point cloud around the 45 degree line indicating a high quality fit. Overall, this provides strong evidence towards robust performance of the method despite the two misspecifications: the data not being generated as a Gaussian process and the true data having strong temporal correlations (Figure \ref{fig:simtrue} (f)) which is not modeled in MGPF.
}

\newpage
\section{Simulation studies based on the data analysis}\label{sec:sim}

We also performed another set of simulation experiments to validate the \blue{performance of MGPF in settings similar to the main application of MGPF for producing maps of fine particulate matter in Baltimore.} 

\subsection{Filtering results}

We aim to illustrate the better accuracy, lower bias, and lower uncertainty of the MGPF with two networks compared to using either network individually. 
We simulate data as follows: we sample 30 sites for each network and 1 reference site from a unit square. The reference site is sampled near the center of the square, and the network sites are sampled either randomly or with preferential sampling. If there is preferential sampling, we randomly sample 20\% of the sites from the full unit square, and the remaining 80\% of the sites from a smaller square within the unit square. To simulate true concentrations, we sample from a Gaussian Process with a small mean $\mu_t$ to represent a low level of ambient concentrations. We also create two point sources, which are located in the area where the preferentially sampled network will have fewer sensors. These point sources each have a randomly sampled concentration they emit, with locations closer to the source having more pollution from that source. This means that concentrations in the area that is not preferentially sampled are higher on average than in the area that has preferential sampling. We sample 10 datasets, each with 100 time-points. More details about the simulation setup and data generation are included in Supplement \ref{sec:supp_sim}. 

Given the true concentrations, we simulate relative humidity, and generate a low-cost measurement from an observation model. The first low-cost network observation model regression coefficients are the PurpleAir coefficients. For the second observation model, we use an observation model whose regression coefficients and variance are both proportional to the first observation model. This way, the amount of uncertainty in each network scales with the magnitudes of the observations from that network. The observation models equations are included in Supplement \ref{sec:supp_sim}. 

In Figure \ref{fig:sim}, we show the percent difference in CI lengths using two networks compared to one network, where we use Equation (\ref{eq:percent_diff}) to calculate the percent difference. On the top left, we see a benefit of using both networks when it comes to uncertainty reductions at network sites, with uncertainty going down by 6-7\% on average. 
There is little difference in the CI reductions between networks, or comparing random and preferential site sampling. 
Looking at CI length across the whole sampling region (top right), we see much larger reductions everywhere, just over 30\%. Again there is little difference between networks.  


We also look at the bias and RMSE of the method, at all interpolated locations in the network (bottom of Figure \ref{fig:sim}). We see that when the sites are randomly sampled, bias is fairly low. However, under preferential sampling, filtering only using Network B, which was preferentially sampled, results in a large negative bias. Filtering using the two networks together results in substantially less bias than filtering only using Network B. For RMSE, under preferential sampling, Network B also has much larger RMSE than Network A. Overall, using both networks results in smaller RMSE than either network individually, both under random and preferential sampling. 

This set of simulations demonstrates that the MGPF can indeed accomplish the goals that were outlined in Section \ref{sec:goals}. A unified set of predictions across the region was made, and these predictions had better accuracy (lower RMSE) and better uncertainty (lower CI length) than using either network individually. Additionally, we demonstrated that in the case of preferential sampling, as we see in the PurpleAir network, there can be systematic bias when areas with more extreme concentrations are more/less sampled, which is mitigated in the multi-network filtering. 

\subsection{Details of simulation setup}\label{sec:supp_sim}

For each dataset, we simulate locations from a unit square with point sources at $(0.2,0.1)$ and $(0.9,0.2)$. These point sources will be used to generate a local pollution surface. A single reference location is sampled from a smaller central square of length 1/3. Then, for each network, 30 low-cost site locations are sampled. If there is no preferential sampling, these locations are all drawn from the full unit square. If there is preferential sampling, these locations are then transformed to be preferentially sampled as follows: the first 6 locations are left as is. The remaining 24 locations are restricted to being within the top-left quadrant of the unit square. Any location that was already in this quadrant is left unchanged, while for the remaining locations, the coordinate(s) that fall outside this quadrant are resampled. 

Then, a Gaussian Process is used to simulate a true ambient (background) concentration surface with means $\mu_t$ generated from a shifted Beta distribution with a minimum of 2 and most of the data below 12. The spatial scale parameter $\phi_t$ is generated so that the correlation at the farthest points across the network is high. The spatial variance $\sigma_t^2$ is also generated from a Beta distribution, scaled according to the value of $\mu_t$, and the nugget variance $\sigma_{n,t}^2$ is restricted to being small relative to $\sigma_t^2$. Once the ambient concentrations are generated from the GP, emissions from the point sources are generated. The emission at each source is generated from a shifted Beta distribution with minimum of 20, and most of the data under 180. The effect of the source decreases exponentially with respect to the squared distance from the source. 

The sum of the ambient concentration (GP) and the local concentration (driven by the two point source emissions) is the final concentration surface, which gives a final surface of: 
\begin{align*}
    x\st&=x_{ambient}\st+x_{local}\st\\
    x_{local}\st&=z_{1,t}\exp\{-d_{1,\bs}^2\psi_{1,t}\}+z_{2,t}\exp\{-d_{2,\bs}^2\psi_{2,t}\}\\
    x_{ambient}(\cdot,t)&\sim GP(\mu_t,\sigma_t^2\exp\{-\phi_t d\}+\sigma_{n,t}^2\mathbb{I}_{d=0})\\
    z_{1,t}&\sim a_1 Beta(b_1,c_1)\\
    z_{2,t}&\sim a_2 Beta(b_2,c_2)
\end{align*}
where $d_{i,\bs}$ is the distance from location $\bs$ to point source $i$. 

We note that the true data generation process is thus misspecified with respect to MGPF. In the true DGP only part of the true concentration is a GP (the ambient part), whereas in MGPF the entire true concentration is modeled as a GP.  

Relative humidity is then generated from a uniform distribution. One low-cost network uses the PurpleAir observation model coefficients, while the other one uses a multiple (1.5) of those coefficients. Then the low-cost sensor measurements are generated using a heteroscedastic observation model variance model with the standard deviation of network B equal to 1.5 times the standard deviation of network A.This gives the observation models:
\begin{align*}
    &\by\sat=-10.97+1.91 x\sat+0.16 RH\sat+\epsilon_A\sat,\\
    &\epsilon_A\sat=10.0+0.5x\sat,\\
    &\by\sbt=-16.46+2.86\bx\sbt+0.25\bz\sbt+\epsilon_B\sbt,\\
    &\epsilon_B\sbt=22.5+1.13x\sbt.
\end{align*}

\subsection{Observation model considerations}\label{sec:sim_obs}

In the previous simulations, we assumed that the training data had the same distribution as the test data. However, in practice, the training dataset may not cover the full range of the testing data, so we investigate the impact of training on a smaller range in this simulation. 
Since each observation model gets trained independently, it is enough to consider one network to illustrate the potential issues with training an observation model. We generate data using the PurpleAir observation model regression coefficients, with the observation model variance being a linear function of the true concentrations $x$. We generate 2,000 training time-points and 100 testing time-points. We consider two cases: first, the range of the training data is the same as the range of the testing data. Second, the range of the training data is 1/3 the range of the testing data, with no high concentrations. This is similar to what we see in the case study. We fit the regression part of the observation model in both cases, and make predictions of the true concentrations from this observation model to see whether the point estimates from the model are correct. We also train the heteroscedastic variance model, and 
compare the resulting fit to the true heteroscedastic model. 

\begin{figure}[t]
\centering
\includegraphics[width=5in]{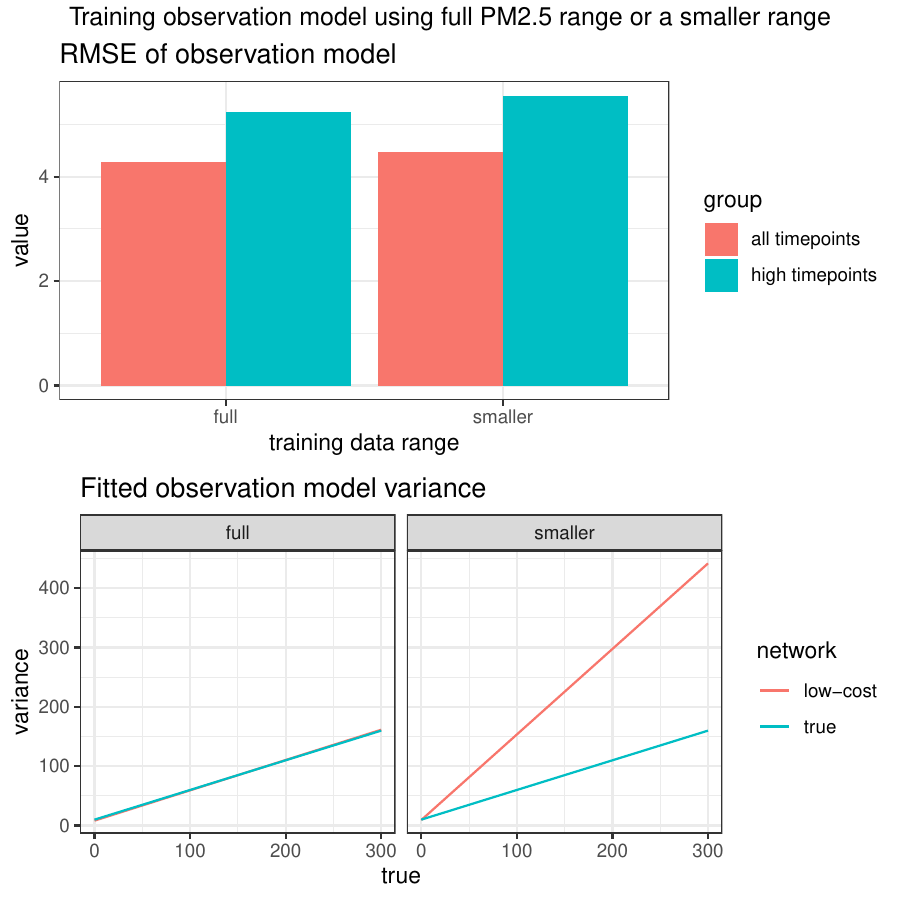}
\caption{RMSE of predictions from inverting the observation model regression equation and predicting on test data, and estimate of the low-cost data observation model error variance ($\tau^2$) model. The x-axis of the top graph and facets of the bottom panel show whether the training data had the full range of the testing data, or a smaller range. In the bottom left panel, the two lines are almost overlapping and thus the red line is not visible. 
}\label{fig:sim_train} 
\end{figure}

Figure \ref{fig:sim_train} (top) shows that the test RMSE of the observation model does not change much when training on the full range or a smaller range. This is still the case when only calculating the RMSE of higher concentration time-points, which the smaller training data range does not cover. Thus, the regression coefficients are well estimated in both cases. However, we see from the bottom figure that the $\tau^2$ estimates at a high concentration are considerably closer to the truth (the blue line) when the training data has the full range. Using a smaller range, the $\tau^2$ estimates are considerably larger than the true variance. This indicates that the heteroscedastic model training suffers when the range of the training data is too small, informing our decision to train the SEARCH observation model variance based on June and July 2023 data. 

\begin{figure}[h]
\centering
\includegraphics[width=2.9in]{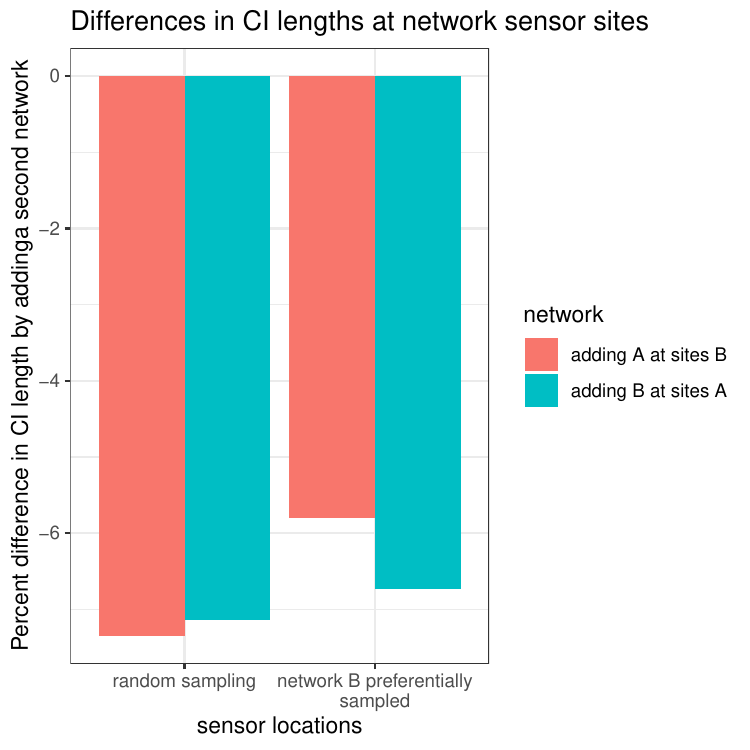}
\includegraphics[width=2.9in]{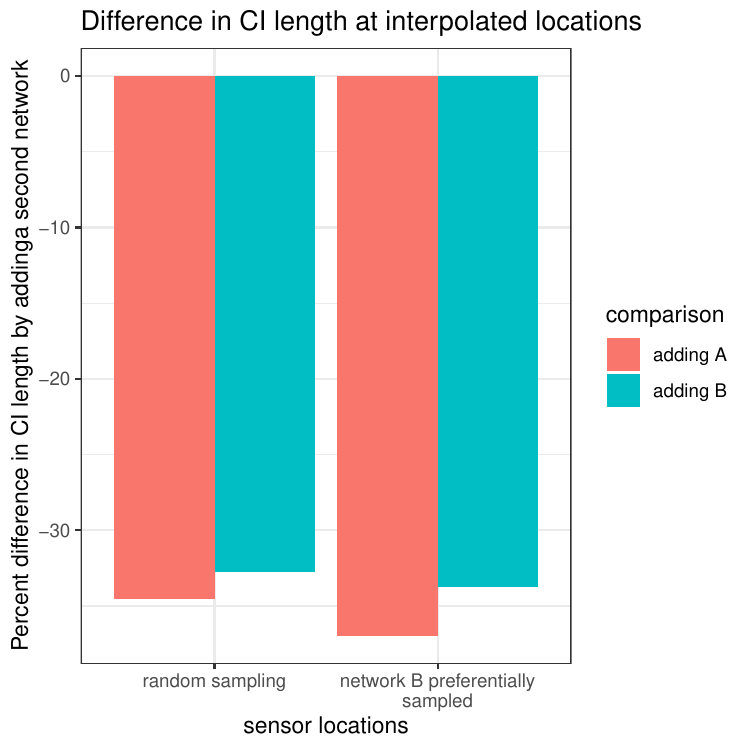}
\includegraphics[width=4in]{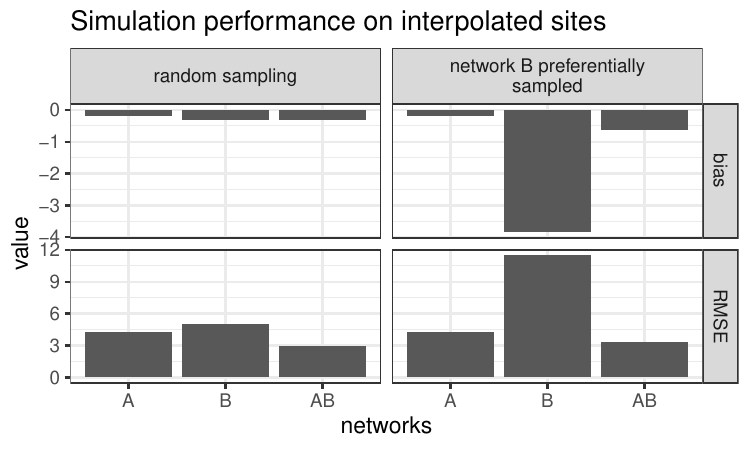}
\caption{(Top left) Percent difference in CI lengths using 2 networks or 1 network, at device sites. (Top right) Percent difference in CI lengths across all interpolated locations in the sampling region. (Bottom) RMSE and bias across all interpolated locations in the sampling region. }\label{fig:sim} 
\end{figure}

\end{document}